\documentclass[a4paper,12pt]{article}
\usepackage[margin=0.5in]{geometry}
\usepackage[utf8]{inputenc}
\usepackage{amsmath,amssymb,amsfonts,textcomp}
\usepackage[font=footnotesize]{caption}
\usepackage{hyperref}
\usepackage{fancyref}
\usepackage[nottoc]{tocbibind}
\usepackage{makecell,adjustbox,tikz}
\usepackage{indentfirst}

\usepackage[authoryear]{natbib}

\let\textcite\citept

\usepackage{xcolor}
\definecolor{greencyan}{rgb}{0.5, 1.0, 0.83}
\definecolor{marine}{rgb}{0.0, 0.5, 1.0}
\definecolor{firebrick}{rgb}{0.7, 0.13, 0.13}

\usepackage{listings,lstautogobble}
\title{
  CP4101 -- BComp Dissertation\\
  Final Year Project\\
  \large Literature Review\\
  Continuous Assessment Report\\
}
\author{Koo Zhengqun}
\date{November 5, 2020}

\begin{document}
\clearpage\setcounter{page}{1}
{\centering
  \maketitle

  \color{gray}
  Department of Computer Science\\
  School of Computing\\
  National University of Singapore\\
  \bigskip
  Project Title: Towards Bug-Free Distributed Go Programs\\
  Project ID: H018740\\
  Project Supervisor: Prof Chin Wei Ngan\\
  Project Co-Supervisor: Dr Andreea Costea\\
}
\clearpage

\setcounter{tocdepth}{10}
\renewcommand\contentsname{Table of Contents}
\tableofcontents
\clearpage

\section{Project Objectives}

\subsection{Motivation}
Distributed software is increasingly important in the digital world, with their footprint firmly anchored from intelligent appliances to large-scale cloud systems. Java and C used to be favorite languages for writing such software, but safer languages such as Go have become more popular, due to its more efficient and safer concurrency.

While Go does a good job in keeping programming bugs away, what about the logical bugs? How can Go programs be rigorously specified?

\subsection{Definitions}
We first define the terms in the motivation.
\begin{description}
\item[Distributed software.] A program that executes on distributed systems, and a distributed system has these characteristics \citep{vanSteen2017distsys}:
  \begin{enumerate}
  \item Is a concurrently executing collection of autonomous computing element. Each computing element is called a node. Since nodes are autonomous, we cannot always assume that there is a global clock that synchronizes the actions of all nodes. The nodes:
    \begin{enumerate}
    \item\label{nodes are geographically dispersed} Are geographically dispersed, so that sending and receiving takes a long time, and may be arbitrarily delayed.
    \item Can join and leave the system, and the connection network between nodes continuously changes its topology and performance.
    \end{enumerate}
  \item Appears as a single coherent system: distribution should be transparent to the user. Failure of some nodes should also be transparent to the user.
  \end{enumerate}
\item[Programming bugs.] Bugs revealed through some way that the implementation does not satisfy the specification. For example, type errors are revealed by the type checker that the program does not satisfy the typing rules.
\item[Logical bugs.] Bugs in the specification. For example, an error in the typing rules.
\end{description}

For the rest of this report, we use the terminology:
\begin{description}
\item[Bugs.] Logical bugs.
\item[Event.] Either a send or a receive. Events occur on a channel.
\item[Ordering.] Lamport's happens-before ordering \citep{lamport1978time}, unless otherwise stated.
\item[Concurrent events.] Two events are concurrent if they are not ordered \citep{go/ref/mem}.
\item[Goroutines.] Programming constructs that order events:
  \begin{enumerate}
  \item Within a goroutine, the order between events is the order expressed by the program.
  \item Between goroutines, assuming there is no synchronization, there is no order between events.
  \end{enumerate}
  This event ordering given by goroutines is similar to the read and write orderings given by goroutines in \citep{go/ref/mem}.
\item[User.] The user of our project's software. To avoid being ambiguous, whenever mentioning clients as in ``client-server protocol'', we use the technical term ``client'' instead of ``user''.
\end{description}

For the rest of this report, we assume:
\begin{enumerate}
\item All channels are first in, first out (FIFO) channels.
\item All channels have a type, and the channel type matches the type of events on this channel.
\end{enumerate}

\subsection{General Objectives}
In this project we aim to:
\begin{enumerate}
\item Investigate the requirements of writing rigorous specifications for distributed programs written in Go, and
\item Use these findings to design a specification language to help users avoid logical bugs specific to concurrent programs.
\end{enumerate}

\subsection{Specific Objectives}
\begin{description}
\item[Avoid communication races.]\label{Avoid communication races.} When a channel has multiple concurrent sends and multiple concurrent receives, where each send corresponds to a specific receive, how can we ensure that every receive gets the data only from the corresponding send, and not from any other send?

  A communication race looks like \ref{fig:communication race}.
  \begin{figure}
    {\centering
      \begin{tikzpicture}[x=.6in]
        \node (S_0) at (0,0) {$S_0$};
        \node (S_1) at (1,0) {$S_1$};
        \node (S_2) at (2,0) {$S_2$};
        \node (C) at (1,-1) {$C$};
        \node (R_0) at (0,-2) {$R_0$};
        \node (R_1) at (1,-2) {$R_1$};
        \node (R_2) at (2,-2) {$R_2$};
        \draw[dashed,->] (S_0) to (C);
        \draw[dashed,->] (S_1) to (C);
        \draw[dashed,->] (S_2) to (C);
        \draw[dashed,->] (C) to (R_0);
        \draw[dashed,->] (C) to (R_1);
        \draw[dashed,->] (C) to (R_2);
      \end{tikzpicture}
      \caption{\label{fig:communication race}Communication races happen when many sends and many receives share a common channel $C$, where $S_i$ are sends, and $R_i$ are receives, and dashed arrows to $C$ represent sending data to $C$, while dashed arrows from $C$ represent receiving data from $C$. Without loss of generality, assume for each $i$, $S_i$ should correspond to $R_i$. Since all $S_i$ are concurrent and all $R_i$ are concurrent, a realization of this concurrent execution may arbitrarily order $S_i$ and arbitrarily order $R_i$, e.g. $S_0, S_2, S_1, R_1, R_0, R_2$, where all sends correspond to a receive that it should not correspond to.}
    }
  \end{figure}
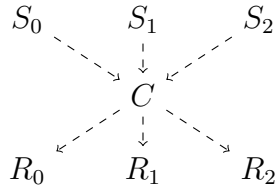
\item[Find a suitable representation of Go programs.] Which representation of Go programs best suits the objectives above?
\end{description}

\subsection{Out-of-scope Objectives}
The field of distributed systems is very broad. The following will likely be out of our project's scope, but are good to keep in mind:
\begin{description}
\item[Connectors in software architecture.] A distributed system is logically organized into components that are modular and replaceable, and these components are connected with connectors, which can be remote procedure calls, message passing, or streaming data \citep{vanSteen2017distsys}. Our project is ambivalent to the choice of connector, and our project will only verify aspects common to all connectors: all connectors send and receive between nodes.
\item[Dynamic analysis.]\label{Dynamic analysis.} Dynamic analysis verifies that certain properties hold in executing distributed system. Examples are Go's global deadlock detector (verifies that there is no global deadlock) and Go's data race detector (verifies that there is no data race) \citep{go/articles/racedetector}. It is a good idea to avoid analyzing a distributed system by executing it, since:
  \begin{enumerate}
  \item Analysis that spans many sends and receives takes a long time due to the characteristic of distributed systems \ref{nodes are geographically dispersed}.
  \item Due to the characteristic of distributed systems \ref{nodes are geographically dispersed}, sends and receives may be arbitrarily ordered. Analyzing a single execution deals with only one out of many possible orderings of sends and receives, so the other orderings are not analyzed. In general, dynamic analysis cannot guarantee that certain properties hold of these other orderings, because not all possible orderings are analyzed.
  \end{enumerate}
  So, we are focused on static analysis that does not require the distributed system to be executed.
\item[Consistency policies.] Replicated stores can have read policies and write policies that relate reading and writing on the local store to all other stores \citep{nagar2020semantics}. The distributed program can rely on these guarantees that the replicated store provides. We choose a different approach: we assume no consistency policy is provided. Rather, we leave to the programmer the burden of implementing protocols that enforce consistency policies. These protocols can then be passed to us for verification.
\item[Software Transactional Memory (STM).] In concurrent programming, lock-based synchronizations places the onus of correct behavior on writers, whereas STM places this onus on readers, who verifies that other threads have not concurrently made changes to memory that it accessed in the past. Notably, lock-based synchronizations are not composable, whereas STMs are composable \citep{harris2005composable}. In multithreaded programming, STM has been extended to Open Transactional Memory by relaxing isolation between transactions while still ensuring atomicity \citep{miculan2020software} (these loosely-coupled interactions remind of interactions in distributed systems), and STM has been extended to the setting of distributed systems \citep{Saad2011SupportingSI}.
\end{description}

\section{Literature Review}

\subsection{Background}
We first survey key historical developments in verifying sequential programs.

Floyd formally defines program semantics by using invariants on control flow graphs \citep{Floyd1967AssigningMT}. Hoare allows some program semantics to be undefined, by providing a choice of axioms (to be decided by the implementor), and explains the theoretical and practical benefits of Floyd-Hoare logic \citep{Hoare1969AnAB}. While Hoare takes axioms of program semantics to be ``self-evident'', Floyd derives these axioms from the soundness and completeness of the underlying deductive system.

Now, we survey key historical developments in verifying concurrent programs.

In parallel programs (a specific kind of concurrent programs), parallel threads interact through shared memory, and shared variables refer to shared memory. Races on shared variables can be avoided by making all assignments on shared variables atomic. The interference-allowing Owicki-Gries \citep{Owicki2004AnAP}:
\begin{enumerate}
\item Avoids races even when assignments on shared variables are not atomic (so, only assuming atomicity of memory reference, so that there are no data races), by following this convention: each expression in a thread only refers at most once to at most one shared variable that, during the evaluation of the expression, may be changed by another thread.
\item Verifies concurrent programs by first verifying sequential threads, then checks that a thread $T$ does not interfere with a \textit{proof} of a concurrent thread $S$, where interference is defined as: the execution of $T$ leaves the postcondition of $S$ invariant, and for each statement $S'$ in $S$, the execution of $T$ leaves the precondition of $S'$ invariant.
\end{enumerate}

Generalized Hoare Logic (GHL) derives invariants of a concurrent program from the invariants of atomic actions \citep{lamport1984the}, thus generalizing all three historical developments above \citep{Floyd1967AssigningMT, Hoare1969AnAB, Owicki2004AnAP}. GHL describes the control state \textit{pc} of a concurrent program with three predicates: \textit{at}, \textit{in}, \textit{after}, which the author claimed is sufficient to express the control flow semantics of any known concurrent programming language.

According to Brookes and O'Hearn in a retrospective paper \citep{Brookes2016ConcurrentSL}: due to a lack of a tractable semantic model for pointer-free shared-memory concurrency, semantic models were unable to prove race-freedom. Logics such as race-free Owicki-Gries \citep{owicki1976verifying} (not covered in our survey) shift the burden of proving race-freedom from semantics to logic, which only works when races are statically detectable. Concurrent separation logic (CSL) uses the semantics of separation logic (intuitively: ownership of pointers) as a semantic foundation for pointer-manipulating shared-memory concurrency, so race-freedom can be proved in richer settings where the existence of a potential race cannot merely be deduced from program syntax. An example of such a richer setting is parallel mergesort, where assignments to elements of an array are viewed by \citep{owicki1976verifying} as assignments to the same array variable and thus cannot be verified.

Brookes and O'Hearn \citep{Brookes2016ConcurrentSL} also note that message passing extensions to CSL \citep{Bell2010ConcurrentSL, Lei2014TraceBasedTV, villard2009proving} support reasoning about history, and histories can be composed by modular reasoning \citep{Fu2010ReasoningAO, Sergey2015MechanizedVO}. Reasoning about the history of events that a node observed is relevant to distributed systems, because due to the characteristic of distributed systems \ref{nodes are geographically dispersed}, each node in a distributed system may observe a different history than what other nodes observed.

\subsection{Preliminary Design}
Regarding specific objectives:

\subsubsection{Avoid communication races}
Communication races can be avoided by ordering events. The ordering on sends must be isomorphic to the ordering on receives: given an ordering on sends, there must exist a relabelling from send to receives that results in an ordering on receives.

Furthermore, this relabelling is the correspondence of sends to receives mentioned in \ref{Avoid communication races.}.

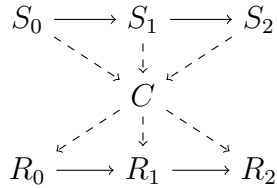
\begin{figure}
  {\centering
    \begin{tikzpicture}[x=.6in]
      \node (S_0) at (0,0) {$S_0$};
      \node (S_1) at (1,0) {$S_1$};
      \node (S_2) at (2,0) {$S_2$};
      \node (C) at (1,-1) {$C$};
      \node (R_0) at (0,-2) {$R_0$};
      \node (R_1) at (1,-2) {$R_1$};
      \node (R_2) at (2,-2) {$R_2$};
      \draw[dashed,->] (S_0) to (C);
      \draw[dashed,->] (S_1) to (C);
      \draw[dashed,->] (S_2) to (C);
      \draw[dashed,->] (C) to (R_0);
      \draw[dashed,->] (C) to (R_1);
      \draw[dashed,->] (C) to (R_2);
      \draw[->] (S_0) to (S_1);
      \draw[->] (S_1) to (S_2);
      \draw[->] (R_0) to (R_1);
      \draw[->] (R_1) to (R_2);
    \end{tikzpicture}
    \caption{\label{fig:no communication race}Communication races no longer happen when no sends are concurrent, no receives are concurrent, and sends have the same ordering as receives. A solid arrow from $A$ to $B$ means $A$ happens-before $B$. Assuming all sends happen before all receives, there is only one way this execution may be realized: $S_0, S_1, S_2, R_0, R_1, R_2$, where for each $i$, $S_i$ corresponds to $R_i$.}
  }
\end{figure}

A communication race that has been avoided by ordering events looks like \ref{fig:no communication race}. We have converted the concurrent execution into a deterministic, non-concurrent execution. This solution is very similar to the approach of avoiding data races in \citep{adve1991detecting} under a sequentially consistent memory model.

Now we consider a distributed program that the user is writing. To verify that this program is free of communication races, all events on a shared channel must be deterministic. To help the user avoid the logical bug of communication races, we might consider the solution of having those events be deterministic by default, and nondeterminism introduced judiciously only where needed \citep{Lee2006ThePW}.

However, this solution has shortcomings:
\begin{enumerate}
\item In a distributed system, events are almost always nondeterministic: not only is it possible for each event to be in a different concurrent goroutine, but it is also possible for each goroutine to be in a different autonomous node.
\item When each event is on a different node, the events must be ordered by communicating between nodes, because the goroutine in each node needs to know when it can continue execution. Due to the characteristic of distributed systems \ref{nodes are geographically dispersed}, communication between nodes takes a long time, so ordering events takes a long time. Ordering all such events everywhere in the distributed program may lower performance below acceptable levels. This could defeat the purpose of executing a distributed program, if the distributed program is expected to be more performant than its non-distributed variant.
\end{enumerate}

On the other hand, increased performance is not the only purpose of distributed programs. A program may be made distributed to achieve the following purposes too \citep{vanSteen2017distsys}:
\begin{enumerate}
\item Share resources.
\item Make distribution transparent.
\item Be open.
\item Be scalable.
\end{enumerate}
Ordering events on shared channels should not defeat these other purposes (being scalable may be defeated, but here only being geographically scalable is defeated, whereas being size-scalable or administratively scalable are purposes that are not defeated). So, it could make sense to have events be deterministic by default.

Even so, having these events be deterministic by default is not a good idea for some distributed programs, where these programs make use of the nondeterminism in a distributed system. For such distributed programs, a communication race is not a bug: these programs restrict the operations on communicated data to only commutative operations. Although such a program has nondeterministic events on a shared channel, this is not a programming bug if these events on a shared channel are declared by the specification to be nondeterministic: the program satisfies the specification. Examples of data types with commutative operations are CRDTs \citep{preguiça2018conflictfree} and LVars \citep{Kuper2013LVarsLD}. Allowing nondeterminism increases speedup and performance. Resolving that nondeterminism through commutative operations ensures that the program is deterministic, so that the program gives the same result when run again.

We suspect that programs that use commutative operators are common in distributed systems, so communication races are rarely bugs, and events on a shared channel should be nondeterministic by default.

Concurrent events can be ordered through these two classes of solutions:
\begin{enumerate}
\item Send and receive synchronizations (which we call \textit{implicit synchronizations}). Implicit synchronizations could be formalized like the read and write synchronizations in Definition 2.1 of \citep{adve1991detecting}, where there is no syntactic difference between non-synchronizing reads and writes, and implicit synchronizations.
\item Synchronizations on critical regions of shared memory (which we call \textit{explicit synchronizations}). Brookes and O'Hearn \citep{Brookes2016ConcurrentSL} interprets that Dijkstra's Principle advocates concurrent programs should not interfere except through explicit synchronization. Dijkstra's Principle \citep{dijkstra1968cooperating} says: ``... processes should be loosely connected; by this we mean that apart from the (rare) moments of explicit intercommunication, the individual processes are to be regarded as completely independent of each other''.
\end{enumerate}

Brookes and O'Hearns' interpretation of Dijkstra's Principle \citep{dijkstra1968cooperating} has some upsides:
\begin{enumerate}
\item An explicit synchronization can be easily statically located: a programmer looking through the program syntax knows where in the syntax processes synchronize. This eases verification that there are no programming bugs.
\item Detecting implicit synchronizations require dynamic analysis, especially if the channel on which the implicit synchronization occurs has nondeterministic runtime behavior. Following this interpretation helps us ignore all implicit synchronizations, and avoid the out-of-scope dynamic analysis objective \ref{Dynamic analysis.}.
\end{enumerate}

So, we would like to follow the principle of only externally synchronizing concurrent nodes. However, not all distributed systems have distributed shared memory, and distributed shared memory is less performant when the interconnection network is slow \citep{Ramesh2011IsIT}. Furthermore, in most distributed systems, nodes communicate (and therefore synchronize) by sending and receiving, so we should reason about implicit synchronizations. Even so, we can still follow the spirit of Dijkstra's Principle, by making all implicit synchronizations easily statically located.

One approach is to have the programmer annotate all implicit synchronizations in the implementation. However, this puts significant burden on the programmer, who should be focusing on the implementation instead of the specification. Our approach thus instead has the expert annotate all implicit synchronizations in the specification.

The relevance of inter-node communications does not mean that explicit synchronizations are not relevant. Explicit synchronizations are directly relevant to distributed programming in practice, which \citep{castro2019distributed} claims often involves local concurrency, since goroutines and channels are effective for dealing locally with the inherent asynchrony of distributed interactions.

So, our approach will support reasoning about explicit synchronizations, which order events. For each explicit synchronization primitive in the Go standard library \citep{go/pkg/sync}, we will formalize exactly what pattern of orderings are created. However, we plan to formalize only these primitives:
\begin{enumerate}
\item \texttt{func NewCond(l Locker) *Cond}
\item \texttt{func (c *Cond) Broadcast()}
\item \texttt{func (c *Cond) Signal()}
\item \texttt{func (c *Cond) Wait()}
\item \texttt{func (m *Mutex) Lock()}
\item \texttt{func (m *Mutex) Unlock()}
\item \texttt{func (o *Once) Do(f func())}
\item \texttt{func (rw *RWMutex) Lock()}
\item \texttt{func (rw *RWMutex) RLock()}
\item \texttt{func (rw *RWMutex) RLocker() Locker}
\item \texttt{func (rw *RWMutex) RUnlock()}
\item \texttt{func (rw *RWMutex) Unlock()}
\item \texttt{func (wg *WaitGroup) Add(delta int)}
\item \texttt{func (wg *WaitGroup) Done()}
\item \texttt{func (wg *WaitGroup) Wait()}
\end{enumerate}
so we do not plan to formalize the other primitives \texttt{type Map} and \texttt{type Pool} which behave more like concurrent data structures instead of behaving like explicit synchronization primitives.

Explicit synchronizations in the Go standard library \citep{go/pkg/sync} are only between goroutines that share memory. In most distributed systems, this can be interpreted as all these goroutines belonging to the same node. However, recognizing the relevance of inter-node communications, we intend to extend the interpretation of explicit synchronizations to synchronizations between goroutines in different nodes. We intend to find a distributed system topology that fits each explicit synchronization, and then use the abstract specification of the explicit synchronization (abstracted away from its shared memory model) in defining a communication synchronization usable in the context of distributed systems. For example, two geographically dispersed nodes could synchronize through \texttt{Cond}, whereas a geographically adjacent nodes could all synchronize by having the cluster of nodes define a \texttt{WaitGroup}. More research has to be conducted to see if this is a feasible idea.

\subsubsection{Find a suitable representation of Go programs}
The static analysis of a program can be divided into two general tasks \citep{oortwijn2019abstractions}:
\begin{enumerate}
\item Deductively verifying program states.
\item Reasoning about the control flow of a program.
\end{enumerate}

There are some general design choices to make when representing Go programs, somewhat related to deductively verifying program states:
\begin{description}
\item[Communicating Sequential Processes (CSP) or non-CSP.] According to \citep{go/blog/codelab-share}, Go's design was strongly influenced by Hoare's CSP \citep{hoare1978csp}, so the natural representation for Go programs is CSP. Indeed, \citep{Prasertsang2016formal} proposes translating Go programs to a machine-readable version of CSP, on which tools like \citep{fdrmanual} can be used. However, if we do not find other representations of Go programs, then in our analysis, we will not be able to use tools from these other representations.
\item[Step-wise or non-step-wise.] Representing Go programs can be done in a step-wise manner, where each step refines small parts of a representation towards the final representation (specification) \citep{kragl2018layered}. Although this has practical benefits: the step-wise approach makes the representation easier for the programmer to understand, this step-wise approach gives us unneeded implementation complexity. So, we represent Go programs with a single, big step.
\item[Influenced by or not influenced by concurrency.]\label{itm:concurrency influence} Although our project is focused on verifying distributed systems, we can make use of what we know from verifying concurrent programs. Program states of concurrent programs can be deductively verified with CSL \citep{Vafeiadis2011ConcurrentSL} or Concurrent Kleene Algebra (CKA) \citep{Hoare2009ConcurrentKA}. Still, verification techniques for concurrent programs are not enough to verify distributed systems, and these verification techniques need to be extended. Program states of distributed programs can be deductively verified with an extension to CSL \citep{ding2020formalizing}, or an extension to CKA \citep{Jaskolka2014EndowingCK}. Alternatively, we can skip deductively verifying concurrent programs, and directly formalize distributed systems like in \citep{sergey2017distributed}. We have not decided whether our project will be influenced by concurrency.
\end{description}

There are many ways to represent the control flow of Go programs:
\begin{description}
\item[Pomsets \citep{gischer1988pomsets}.] Pomsets are simple and properly understood through the characterization theorem and interpolation lemma. However, it is unclear how pomsets may be extended to include communication (sends and receives).
\item[Behavioral types \citep{ng2016deadlock, lange2018verification, gabet2020static, lange2017fencing}.] Although much work has been done on representing Go programs with behavioral types, behavioral types have a shortcoming: a representation with behavioral types is not the most comprehensive representation because behavioral types do not capture all information in a Go program (Go programs with infinitely occuring conditionals need to have some information discarded such that behavioral types correspond to these Go programs \citep{gabet2020static, lange2017fencing}). To address this shortcoming, we are considering a hybrid approach of behavioral types and CSL assertions \citep{oortwijn2019abstractions}. We believe this hybrid approach is the one of most comprehensive representations of Go programs, and will likely use this representation in our project.
\item[CKA \citep{Hoare2009ConcurrentKA}.] Like pomsets, CKA is formulated with axioms. The axioms of CKA come from well-known mathematical structures: quantales, ordered monoids, and lattices.
\item[Deep handlers of bidirectional algebraic effects \citep{zhang2020handling}.] We do not fully understand algebraic effects, so we will likely not represent Go programs with algebraic effects.
\end{description}

Here are some details on our preliminary design for representation of Go programs:
\begin{enumerate}
\item We will convert a Go program to its Single Static Assignment (SSA) intermediate representation \citep{go/tools/ssa} to ease our analysis, as Ng has done in \citep{ng2016deadlock}.
\item Inspired heavily by Ng's representation of goroutines as type graphs \citep{ng2016deadlock}, we will perform a similar representation in \ref{tab:happens-before representation}, where nodes are blocks of a Go program, and directed edges in the graph are Lamport's happens-before ordering \citep{lamport1978time}.
\end{enumerate}

\begin{table}
  {\centering
    \begin{tabular}{|c|c|c|}
      \hline 
      \textbf{SSA instructions} &\textbf{Happens-before graph} &\textbf{Go code}\\
      \hline\hline

      \adjustbox{valign=t}{
      \makecell[l]{
      Call\{Func, Method, Args\}\\
      Return
      }
      }
                                &\adjustbox{valign=t}{
                                  \begin{tikzpicture}[x=.6in]
                                    \node (P_0) at (0,0) {$P_0$};
                                    \node (Q) at (0,-1) {$Q$};
                                    \node (P_1) at (0,-2) {$P_1$};
                                    \draw[->] (P_0) to (Q);
                                    \draw[->] (Q) to (P_1);
                                  \end{tikzpicture}
                                  }
                                                               &\adjustbox{valign=t}{
\begin{lstlisting}
                                                                     func P() {
                                                                       // Block |\color{gray}$P_0$|
                                                                       Q()
                                                                       // Block |\color{gray}$P_1$|
                                                                     }
\end{lstlisting}
                                                                 }\\
      \hline 

      \adjustbox{valign=t}{
      Go\{Func, Method, Args\}
      }
                                &\adjustbox{valign=t}{
                                  \begin{tikzpicture}[x=.6in]
                                    \node (P_0) at (0,0) {$P_0$};
                                    \node (P_1) at (0,-1) {$P_1$};
                                    \node (Q) at (1,0) {new $Q$};
                                    \node (dots) at (1,-1) {$\ldots$};
                                    \draw[->] (P_0) to (P_1);
                                    \draw[->] (P_0) to (Q);
                                    \draw[->] (Q) to (dots);
                                  \end{tikzpicture}
                                  }
                                                               &\adjustbox{valign=t}{
\begin{lstlisting}
                                                                     func P() {
                                                                       // Block |\color{gray}$P_0$|
                                                                       go Q()
                                                                       // Block |\color{gray}$P_1$|
                                                                     }
\end{lstlisting}
                                                                 }\\
      \hline 

      \adjustbox{valign=t}{
      \makecell[l]{
      Defer\{Func, Method, Args\}\\
      RunDefers
      }
      }
                                &\adjustbox{valign=t}{
                                  \begin{tikzpicture}[x=.6in]
                                    \node (P_0) at (0,0) {$P_0$};
                                    \node (P_1) at (0,-1) {$P_1$};
                                    \node (P_2) at (0,-2) {$P_2$};
                                    \node (R) at (1,-3) {$R$};
                                    \node (Q) at (1,-4) {$Q$};
                                    \draw[->] (P_0) to (P_1);
                                    \draw[->] (P_1) to (P_2);
                                    \draw[->] (P_2) to (R);
                                    \draw[->] (R) to (Q);
                                  \end{tikzpicture}
                                  }
                                                               &\adjustbox{valign=t}{
\begin{lstlisting}
                                                                     func P() {
                                                                       // Block |\color{gray}$P_0$|
                                                                       defer Q()
                                                                       // Block |\color{gray}$P_1$|
                                                                       defer R()
                                                                       // Block |\color{gray}$P_2$|
                                                                     }
\end{lstlisting}
                                                                 }\\
      \hline 

    \end{tabular}
    \caption{\label{tab:happens-before representation}Notable control-flow SSA instructions and their happens-before graphs.}
  }
\end{table}

\section{Research Plan for the Next Semester}

\begin{enumerate}
\item Investigate the requirements of writing rigorous specifications for distributed programs written in Go. Specifically, find out what kind of common logical fallacies a specification of a distributed program usually has.
\item Investigate if it is feasible to extend the interpretation of explicit synchronizations to synchronizations between goroutines in different nodes.
\item Understand the following:
  \begin{enumerate}
  \item CSL: understand the operational semantics of CSL \citep{Gotsman2011PrecisionAT}.
  \item CKA: first \citep{Hoare2009ConcurrentKA}, then \citep{Hoare2011ConcurrentKA}, then \citep{Hoare2016DevelopmentsIC}, then finally how CKA is extended to distributed systems \citep{Jaskolka2014EndowingCK}.
  \item Read references in \ref{itm:concurrency influence}, and choose whether our project is influenced by concurrency.
  \item Abstract interpretation \citep{Cousot1977AbstractIA}, since abstract interpretation shows up many times in separation logic. Heaplets in the original formulation of separation logic are abstractions of many machine states, and may consist of many actual machine states \citep{Brookes2016ConcurrentSL}. Logical assertions in separation logic are abstracted away in \citep{Calcagno2007LocalAA}, and program semantics that a separation logic verifies are abstracted away in \citep{Hoare2009ConcurrentKA}. Related to abstract separation logic, Views \citep{DinsdaleYoung2013ViewsCR} has a similar logic, but its semantics is more relaxed and is worth understanding.
  \end{enumerate}
\item If time allows, understand the following:
  \begin{enumerate}
  \item Algebraic effects: \citep{zhang2020handling}.
  \end{enumerate}
\end{enumerate}

\clearpage
\bibliographystyle{apalike}
\bibliography{main}

\end{document}


\newpage
\renewcommand{\thepage}{\roman{page}}
\setcounter{page}{1}
{\centering
  \color{gray}
  B.Comp. Dissertation\\
  \begingroup
  \let\newpage\relax
  \color{black}
  \maketitle
  \endgroup
  Department of Computer Science\\
  School of Computing\\
  National University of Singapore (NUS)\\
  \bigskip
  Project ID: H018740\\
  Project Supervisor: Prof Chin Wei Ngan\\
  Project Co-Supervisor: Dr Andreea Costea\\
  Deliverables: Report: 1 Volume\\
}
\newpage

\section{Abstract}
Programmers of distributed systems need to reason about concurrency to avoid races. However, reasoning about concurrency is difficult, and unexpected races show up as bugs.

Data race detection in shared memory systems is well-studied (dynamic data race detection \parencite{fava2020ready}, behavioral types \parencite{gabet2020static}, dynamic race detection \parencite{Mathur2021OptimalPO}).

Similar to how a data race consists of reads and writes not related by happens-before at a shared memory location, a communication race consists of receives and sends not related by happens-before on a shared channel.
Communication races are problematic: a receiver expects a specific message from a specific sender, but with a communication race, the receiver can receive a message meant for another receiver, or not receive anything at all.

In this work, we describe a verification framework that can prove the absence of communication races for distributed programs that use a subset of the Go programming language, where synchronization is mainly achieved via message passing.
We statically reason about how a distributed program executes, using a happens-before order, extended to buffered and unbuffered channels.

CCS Concepts:
\begin{itemize}
\item Theory of computation \mytextsim Logic \mytextsim Logic and verification
\item \textbf{Theory of computation \mytextsim Logic \mytextsim Separation logic}
\item Theory of computation \mytextsim Logic \mytextsim Hoare logic
\item Theory of computation \mytextsim Semantics and reasoning \mytextsim Program reasoning \mytextsim Program verification
\item \textbf{Theory of computation \mytextsim Models of computation \mytextsim Concurrency \mytextsim Distributed computing models}
\end{itemize}
\newpage

\doublespacing
\section{Acknowledgements}
My sincere thanks to everyone from NUS Programming Languages \& Software Engineering Lab 2 for the wonderful conversations and interactions over the years.

Special thanks to Asst. Prof. Aquinas Hobor for introducing me to the logical fundamentals of formal verification, Dr. Andreea Costea for unwavering guidance across many research projects, and Assoc. Prof. Chin Wei Ngan for technical advice and discussions.
\newpage
\onehalfspacing

\setcounter{tocdepth}{10}
\renewcommand\contentsname{Table of Contents}
\tableofcontents
\newpage

\renewcommand{\thepage}{\arabic{page}}
\setcounter{page}{1}

\section{Introduction} \label{sec:Introduction}
Distributed software is important in the digital world, with their footprint firmly anchored in many areas, from intelligent appliances to large-scale cloud systems. Java and C used to be preferred languages for writing such software, but languages such as Go have become more popular, due to its more lightweight and supposedly safer concurrency.

Despite its careful design which relies heavily on message passing to achieve synchronization, empirical studies \parencites{Tu2019UnderstandingRC}{Liu2021AutomaticallyDA}{8668036} show that Go programs may still exhibit concurrency related bugs.
To ensure the safe and correct construction of Go programs, we propose a rigorous method to specify them, and demonstrate how to verify them using specifications.

In this work, we describe a formal verification framework that can prove that some sends or receives happen before other sends or receives.
We statically verify that a distributed program satisfies a protocol, by reasoning about a happens-before order, extended to buffered and unbuffered channels.
The happens-before order subsumes all possible sequential executions, therefore we can statically reason about all possible sequential executions.

Approaches to verify distributed programs can be categorised as follows:
\begin{enumerate}
\item Dynamic approaches need to execute the distributed program, and therefore need to wait for messages to travel across the network.
  Messages can take a long time to travel, or even longer if the messages are not delivered and must be resent.
  Dynamic approaches includes runtime testing of distributed programs, which can only ever cover a small subset of all possible inputs and all possible interleavings.

\item Model checking represents a distributed program as a state transition system.
  To verify a property, all states are explored.
  This does not scale with the number of interleavings of a distributed program.
  To mitigate this scalability issue, advances in model checking include the following:
  \begin{enumerate}
  \item Having less information in the model, by abstracting more information from the distributed program.
    Practical advances also increased the efficiency of model checking, allowing larger systems to be verified.
  \item The properties to be verified are weakened.
    This includes model checking only up to a bounded number of states, thus allowing infinite-state systems to be verified.
    These kinds of bounded model checking can verify less properties than usual.
  \end{enumerate}
  Although these advances have made model checking more attractive, the scalability issue is still prominent.

\item Deductive verification verifies a distributed program without executing it.
  Therefore, unlike dynamic approaches, deductive verification does not need to wait for messages to travel across the network.

  Deductive verification can universally quantify over all possible interleavings, and use a domain-specific theorem prover to deduce satisfiability of the universally quantified logical formula.
  Note that unlike model checking, deductive verification does not explore all possible interleavings, they merely reason about all possible interleavings.
  Therefore, deductive verification does not have the same kind of scalability issues that model checking has.
\end{enumerate}

Due to these differences of deductive verification with other approaches, we decided to verify Go programs deductively.

\begin{description}[labelwidth=0pt]
\item[Our main contribution.] This work includes an extension of \textcite{costea2017session} to deductively verify that implementations that use a subset of the Go programming language satisfy communication protocols.
  The verification specifically reasons about Go's memory model.
  In extending \citeauthor{costea2017session}'s work, we simplify their verification process by reducing the amount of formal machinery needed for verification.
  For example, unlike \textcite{costea2017session}, we do not need to formalize the dual of guards, referred to by \citeauthor{costea2017session} as ``assumptions''.
  Representing communication protocols as partial orders helped us come up with this simplified verification process.
\end{description}

The rest of this work is organized as follows.
Section \ref{sec:Overview} gives an overview of this work, and introduces basic concepts.
Section \ref{sec:Program Order, Events, Sequential Consistency, and Parties} introduces concepts central to our work, like events.
Section \ref{sec:HB Order} gives special attention to the central concept of the happens-before order.
Section \ref{sec:Global Protocols} motivates the use of a weaker kind of the happens-before order in specifications, by noting that executions can deviate from specifications.
Section \ref{sec:Sequential Executions} simplifies the events introduced in section \ref{sec:Program Order, Events, Sequential Consistency, and Parties}, so that discussions about executions are simplified.
Section \ref{sec:Verifications, Implementations, and Guards} describes how the verification of an implementation can prove guards.
Section \ref{sec:Formal Verification} defines the rules of verification, then demonstrates how programs are either verified to be safe, or verified to be unsafe.

\section{Overview} \label{sec:Overview}
Go is a programming language that concurrently executes lightweight threads, called goroutines, with shared memory.
A goroutine is a sequentially executing program that expresses an order on lines of code \parencite{go/ref/mem}.

We consider Go programs that execute on distributed systems.
According to \textcite[Chapter 4]{vanSteen2017distsys}:
\begin{quote}
  a distributed system is a concurrently executing collection of autonomous computing elements. Each computing element is called a node. Since nodes are autonomous, we cannot always assume that there is a global clock that synchronizes the actions of all nodes. The nodes:
  \begin{enumerate}
  \item are geographically dispersed, so that sending and receiving takes a long time, and may be arbitrarily delayed.
  \item can join and leave the system, and the connection network between nodes continuously changes topology and performance.
  \end{enumerate}
\end{quote}

The goal of this work is to describe a verification framework that can verify that an implementation satisfies a specification.
Users of our framework should follow the following software development methodology:
A specification designer writes a specification that the programmer's Go implementation is supposed to satisfy.
A programmer writes a distributed Go implementation.
If our framework then verifies that the implementation does not satisfy the specification, then both the programmer and the specification designer should rewrite their implementation and specification.
On the other hand, if our framework then verifies that the implementation satisfies the specification, then users of our framework can be confident that the implementation always has properties that the specification guarantees, no matter what kind of interleaving execution the implementation can have.

\subsection{Basics}
\textcite{Kurose1999ComputerN} gives the following informal dentition of protocols: ``a communication protocol defines the format and the order of messages exchanged between two or more communicating entities''.
In this work, the informal definition is made formal to have an unambiguous meaning, in order to for it to guide how we structure our verification framework:

\begin{definition}
  For distributed Go programs, specifications are global protocols.
\end{definition}

Furthermore, we say that a distributed Go program satisfies a global protocol if it communicates according to the protocol.
The notion of global protocols is formalized in section \ref{sec:Global Protocols}.

We now define ``parties'', the interacting elements of a concurrent system.
A party is a goroutine that has some line of code that is either a send or receive.
Parties are denoted in general by $\Party$. When referring to specific parties, we use the notation that ranges over $\{\party{A}, \party{B}, \party{C}, \party{D}, \ldots\}$.

Events are denoted in general by $\E{i}$, where $i$ is an index. Where an event belongs to party $\Party$, we use $\Ep{\Party}{i}$, where $i$ is an index.
Two events $\E{}$, $\E{'}$ may be ordered by a happens-before order, denoted by $\E{} \HB{} \E{'}$.
Our happens-before order $\HB{}$ order differs from Lamport's happens-before order \parencite{lamport1978time}, details deferred to section \ref{sec:HB Order}.

\begin{figure}
  {\centering
    \begin{tikzpicture}[]
      \node(A) at (0,0)  {$\party{A}$};
      \draw[line width=0.1mm] (-1.5,-0.5) -- (1.5,-0.5);

      \draw[line width=0.1mm,dotted] (-1.5,-0.95) -- (1.5,-0.95);
      \dop{AR1}{$\Ep{\party{A}}{R}$}{1}{-1}{left}
      \draw[line width=0.1mm,dotted] (-1.5,-1.35) -- (1.5,-1.35);
      \node(code) at (-1,-1.15) {\lstinline{<-c}};
    \end{tikzpicture}
    \caption{\label{fig:control flow and events}A line of code \lstinline{<-c} in party $\party{A}$ is abstracted away to become an event $\Ep{\party{A}}{R}$. Execution of the line of code starts at the upper dotted line, and completes at the bottom dotted line. The duration of the event is defined by these two dotted lines.}
  }
\end{figure}
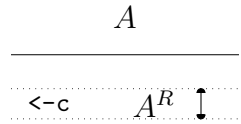

Figure \ref{fig:control flow and events} shows one party that has one event.
The figure looks like a sequence diagram that has only one participant: the lifeline is vertical, and time flows downwards.

We focus on applying our framework to verify that a Go program is communication race-free, for the first in, first out (FIFO) communication model.
In this work, we do not mention data races, except in reference to other works. Therefore, mentions of races refer only to communication races.
A usual definition of FIFO ordering in distributed systems is given by \textcite[Section 12.1]{Garg2004ConcurrentAD}, as follows:
\begin{quote}
  We use $e \prec f$ to denote that $e$ occured before $f$ in the same process.
  Let $s_1$ and $s_2$ be any two send events and $r_1$ and $r_2$ be the corresponding receive events. Then FIFO ordered computations is defined as $s_1 \prec s_2 \implies \neg(r_2 \prec r_1)$.
\end{quote}
This usual definition assumes that channels are not shared: if two processes are connected by a channel, then the channel connects only these two processes \parencite[Section 7.2]{Garg2004ConcurrentAD}.

However, this assumption is not valid in Go.
In Go, concurrently executing goroutines can access the same shared memory, as well as communicate with other concurrently executing goroutines over FIFO channels.
The sharing of a channel with concurrently executing goroutines presents a new problem not considered by the usual definition of FIFO ordering: despite the FIFO property of channels, it is possible that the messages sent end up at unexpectedly different receiving goroutines.
For example, even if a programmer writes the program \ref{lst:simple global protocol implementation} expecting \lstinline{B} to receive \lstinline{1} and \lstinline{C} to receive \lstinline{2}, this may not actually happen.
The program may execute with races, because communication does not proceed as specified by the global protocol designer.
It is possible for \lstinline{C} to receive \lstinline{1} and \lstinline{B} to not receive anything at all, because channel \lstinline{c} is shared between the goroutines that run \lstinline{go A()}, \lstinline{go B()}, and \lstinline{go C()}.

\begin{code}[caption={An implementation that satisfies the specification \eqref{eq:global protocol with races}.},label={lst:simple global protocol implementation},multicols=2]
package main

var c = make(chan int, 2)

func main() {
	go A()
	go B()
	go C()
}

func A() {
	c <- 1
}

func B() {
	<-c
	c <- 2
}

func C() {
	<-c
}
\end{code}

The invalid assumption of the usual definition of FIFO ordering motivates a different-than-usual definition of FIFO ordering of events for the Go programming language: we need to extend FIFO ordering to many concurrently executing goroutines.
Details in section \ref{sec:Transforming Global Protocols with Races into Race-free Global Protocols}.

\subsection{Transforming Global Protocols so that Satisfying Implementations are FIFO Race-free}
We call implementations that never execute with a race ``race-free implementations''.
On the other hand, we call implementations that may execute with a race ``implementations with races''.
If a global protocol may be satisfied by an implementation which may execute with races, then we call this a ``global protocol with races''.
Otherwise, we call this a ``race-free global protocol''.

\begin{figure}
  {\centering
    \begin{tikzpicture}[]
      \draw (0,0) node[draw,circle,minimum size=3cm,label=above:programs with races] (ImplsWithRaces) {};
      \draw (0,-7) node[draw,circle,minimum size=3cm,label=below:race-free programs] (ImplsRaceFree) {};
      \draw (7,0) node[draw,circle,minimum size=3cm,label=above:protocols with races] (ProtsWithRaces) {};
      \draw (7,-7) node[draw,circle,minimum size=3cm,label=below:race-free protocols] (ProtsRaceFree) {};

      \draw [-Implies,double distance=5pt] (ImplsWithRaces) -- (ProtsWithRaces) node [above,near start,fill=white,inner sep=1pt] {satisfies};
      \draw [-Implies,double distance=5pt] (ImplsRaceFree) -- (ProtsRaceFree) node [above,near start,fill=white,inner sep=1pt] {satisfies};
      \draw [-Implies,double distance=5pt] (ImplsRaceFree) -- (ProtsWithRaces) node [above,near start,fill=white,inner sep=1pt] {satisfies};

      \draw [-triangle 90] (ProtsWithRaces) -- (ProtsRaceFree) node [text width=3cm,right,near start,fill=white,inner sep=1pt] {transformation described in section \ref{sec:Transforming Global Protocols with Races into Race-free Global Protocols}};

      \draw (0,0) node[draw,circle,inner sep=1mm,fill=black,label=above:Program \ref{lst:simple global protocol implementation}] (ImplsWithRacesElem) {};
      \draw (0,-7) node[draw,circle,inner sep=1mm,fill=black,label=above:Program \ref{lst:simple global protocol guarded implementation}] (ImplsRaceFreeElem) {};
      \draw (7,0) node[draw,circle,inner sep=1mm,fill=black,label=above:Protocol \eqref{eq:global protocol with races}] (ProtsWithRacesElem) {};
      \draw (7,-7) node[draw,circle,inner sep=1mm,fill=black,label=above:Protocol \eqref{eq:global protocol race-free}] (ProtsRaceFreeElem) {};
    \end{tikzpicture}
    \caption{\label{fig:satisfaction with and without races}
      This figure shows two sets of programs, and two sets of protocols.
      Examples of protocol conformance and protocol transformation are concretely shown for one element of each set.
      Although our verification framework can verify protocol conformance (where protocol conformance is shown by a double arrow labelled ``satisfies'' from some set of programs to some set of protocols, and an example of conformance is that the program \ref{lst:simple global protocol implementation} in this figure satisfies the protocol with races \eqref{eq:global protocol with races} in this figure), users of our framework may want to verify that their Go program has more interesting properties than just protocol conformance.
      To this end, users should provide a richer specification which captures more interesting properties, such as race-freedom.
      Users can apply the transformation described in section \ref{sec:Transforming Global Protocols with Races into Race-free Global Protocols} to any protocol with races (e.g. \eqref{eq:global protocol with races}), to automatically obtain a new race-free protocol (e.g. \eqref{eq:global protocol race-free}) that communicates similarly to the old protocol with races.
      This transformation is shown by a single arrow from the set of protocols with races to the set of race-free protocols.
      Users can then make corrections to their original Go program, so that the program is race-free (e.g. \ref{lst:simple global protocol implementation} is corrected to \ref{lst:simple global protocol guarded implementation})
      Users can then use our verification framework to verify that the corrected race-free program (e.g. \ref{lst:simple global protocol guarded implementation}) satisfies a race-free protocol (e.g. \eqref{eq:global protocol race-free}).
      Furthermore, our verification framework can also verify that the corrected race-free program (e.g. \ref{lst:simple global protocol guarded implementation}) also satisfies the old protocols with races (e.g. \eqref{eq:global protocol with races}).
    }
  }
\end{figure}
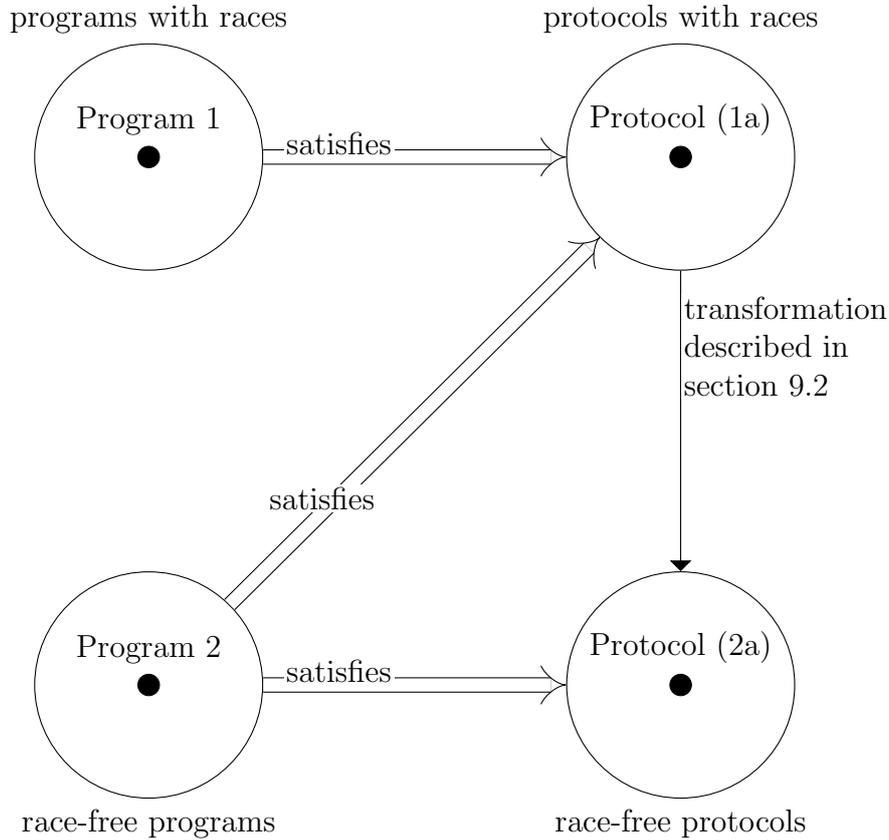

Figure \ref{fig:satisfaction with and without races} describes how global protocol designers can use our framework to convert a protocol with races to a race-free protocol, and how these protocols relate to implementations that a programmer writes.
The figure shows only the implementations that our verification framework can verify with success.

Our framework can verify that implementation \ref{lst:simple global protocol implementation} satisfies the following global protocol, where two transmissions are sequentially composed, and each transmission is of the form $\transmit{\Party}{\Channel}{i}{\Party'}$ (read as ``the $i^{th}$ transmission of the protocol specifies that party $\Party$ sends a message to party $\Party'$ over channel $\Channel$'', or ``the protocol specifies that event $\Ep{\Party}{i}$ sends a message to the receive event $\Ep{\Party'}{i}$ over channel $\Channel$''):
\begin{subequations}
  \begin{align}
    \transmit{\party{A}}{\Channel}{1}{\party{B}} \ssep \transmit{\party{B}}{\Channel}{2}{\party{C}} \label{eq:global protocol with races}
  \end{align}
\end{subequations}

However, our framework cannot guarantee that the implementation \ref{lst:simple global protocol implementation} is race-free, because the global protocol \eqref{eq:global protocol with races} does not conditionally guarantee the property of race-freedom.

Global protocols conditionally guarantee race-freedom by having certain guards at certain positions in the global protocol.
A guard $\guardP{\E{} \HB{} \E{'}}$ is an abstraction over how its happens-before element $\E{} \HB{} \E{'}$ is proven to be in the $\HB{}$ order.
Race-freedom is guaranteed on the condition that if verification of an implementation against that global protocol succeeds, then the verification shows that the implementation is race-free.

For example, the global protocol \eqref{eq:global protocol race-free} conditionally guarantees race-freedom, by having two guards in between its two transmissions.\footnote{If the guards were not there, these transmissions may race.}
\begin{subequations}
  \begin{align}
    \transmit{\party{A}}{\Channel}{1}{\party{B}} \ssep \guardP{\Ep{\party{A}}{1} \HB{} \Ep{\party{B}}{2}} \ssep \guardP{\Ep{\party{B}}{1} \HB{} \Ep{\party{C}}{2}} \ssep \transmit{\party{B}}{\Channel}{2}{\party{C}} \label{eq:global protocol race-free}
  \end{align}
\end{subequations}

As figure \ref{fig:satisfaction with and without races} shows, the race-free global protocol \eqref{eq:global protocol race-free} was obtained by applying the transformation described in section \ref{sec:Transforming Global Protocols with Races into Race-free Global Protocols} on the protocol with races \eqref{eq:global protocol with races}.
If users of our framework correct the original implementation with races to become a race-free implementation \ref{lst:simple global protocol guarded implementation}, then as shown by figure \ref{fig:satisfaction with and without races}, our framework can verify that this corrected implementation satisfies the race-free global protocol \eqref{eq:global protocol race-free}.

\begin{code}[caption={An implementation that satisfies the global protocol \eqref{eq:global protocol race-free}.},label={lst:simple global protocol guarded implementation},multicols=2]
package main

var c = make(chan int, 2)
var d = make(chan int, 0)
var e = make(chan int, 0)

func main() {
	go A()
	go B()
	go C()
}

func A() {
	c <- 1
	d <- 0
}

func B() {
	<-c
	<-d
	e <- 0
	c <- 2
}

func C() {
	<-e
	<-c
}
\end{code}

Indeed, the verification shows that the implementation is race-free, because the implementation \ref{lst:simple global protocol guarded implementation} transmits between $\party{A}$ and $\party{B}$ on an unbuffered channel $\channel{d}{0}$ to order the events $\Ep{\party{A}}{1} \HB{} \Ep{\party{B}}{2}$, and transmits between $\party{B}$ and $\party{C}$ on an unbuffered channel $\channel{e}{0}$ to order the events $\Ep{\party{B}}{1} \HB{} \Ep{\party{C}}{2}$.
Therefore, the implementation \ref{lst:simple global protocol guarded implementation} used two extra transmissions to fix a linear order of send events and a linear order of receive events.
These orderings are sufficient for the implementation to always execute such that $\party{B}$ receives data from $\party{A}$, and $\party{C}$ receives data from $\party{B}$, because Go channels are FIFO.
This means all executions of the implementation satisfy global protocol \eqref{eq:global protocol with races}, such that the implementation satisfies the global protocol.
In other words, the implementation always communicates according to the protocol: the implementation transmits according to $\transmit{\party{A}}{\Channel}{1}{\party{B}}$, then transmits according to $\transmit{\party{B}}{\Channel}{2}{\party{C}}$, and the implementation can never transmit in any other way.
Furthermore, the implementation also satisfies global protocol \eqref{eq:global protocol race-free}, because the two guard orders have been proven.

\subsection{Details of Verification}
In this work, we restrict how a Go program can be verified as race-free: Go programs must use Go's synchronization primitives to be verified as race-free.
Although Go's sync package \parencite{go/pkg/sync} defines many kinds of synchronization primitives, we require race-free Go programs to order their transmissions by synchronizing goroutines on an unbuffered channel.
We call such a synchronizing transmission on an unbuffered channel a ``rendezvous'' between the send and the receive of the transmission.
This restriction of synchronization to rendezvous simplifies verification, because the rendezvous is specified by the happens-before order over unbuffered channels, and our verification framework already models the happens-before order over unbuffered channels.

Although the given global protocol offers a global view of the communication, we opt for a modular verification where each channel endpoint in each party may be independently verified.
This means that verification can be parallelized.
To this purpose, the global protocol is projected onto parties, and then projected again onto channel endpoints.
Our framework then deductively verifies by using Hoare triples to consume the heads of each endpoint's protocol.
The verification succeeds, because the implementation has rendezvous on channels \lstinline{d} and \lstinline{e}.
Each rendezvous forces all events before the rendezvous to happen-before all events after the rendezvous, even if these events occur in different parties.
Therefore, the parties in the implementation must communicate in a race-free way.

\subsection{Channels in the Go Programming Language}
Go channels are FIFO: if there is an $i^{th}$ send and $i^{th}$ receive on the channel, then the $i^{th}$ receive on the channel receives data from the $i^{th}$ send on the channel. If there is no $i^{th}$ send on the channel, the $i^{th}$ receive blocks until there is an $i^{th}$ send on the channel.

A channel variable can be in shared memory, and accessed by concurrently executing goroutines.
We emphasize that despite the FIFO property of channels, a program may still send messages to unexpectedly different receiving parties.

Go channels are typed, and all messages on a channel must type check against the channel type. This may lead to runtime type errors if the channel type is something like a sum type, discussion of which is deferred to section \ref{sec:Avoid Runtime Type Errors}.
For brevity, in the rest of this work, unless necessary, we do not mention channel types, message types, and message values.
We still mention them in Go programs, because they are used by the Go compiler.
We also mention message values when we use the extended per-endpoint protocol syntax \ref{lst:extended per-endpoint protocol syntax}.

A channel buffer stores the messages provided by sends on the channel, that have not yet been received.
Channel buffers have a nonnegative size and a nonnegative capacity.
The size of a channel is the number of messages currently stored on the buffer.
The capacity of a channel is the number of messages that can be stored on the buffer, before the buffer is full.
A channel with nonnegative capacity $k$ is denoted by $\channel{c}{k}$.
An ``unbuffered channel'' is a channel with zero capacity, denoted by $\channel{c}{0}$.
A ``buffered channel'' is a channel with positive capacity, denoted by $\channel{c}{k}$ for some positive $k$.
If we do not care about the capacity of channel, we denote the channel by $\Channel$.
In this work, channels range over $\{\channel{c}{k}, \channel{d}{k'}, \channel{e}{k''}, \channel{f}{k'''}, \ldots\}$, for some nonnegative $k, k', k'', k'''$.

\section{Program Order, Events, Sequential Consistency, and Parties} \label{sec:Program Order, Events, Sequential Consistency, and Parties}
To ease our analysis, similar to the approach of \textcite{ng2016deadlock}, we convert a Go program to its Single Static Assignment (SSA) representation \parencite{go/tools/ssa}. Then we represent a goroutine as a control flow graph (CFG), where nodes are blocks of a Go program, and directed edges are the ordering of events given by program order.
We show only some graphs in table \ref{tab:control flow representation}, other graphs are defined as usual and not shown.

\begin{table}
  {\centering
    \begin{tabular}{|c|c|c|}
      \hline
      \textbf{SSA instructions} &\textbf{Control flow graph (CFG)} &\textbf{Go code}\\
      \hline\hline

      \adjustbox{valign=t}{
      \makecell[l]{
      Call\{Func, Method, Args\}\\
      Return
      }
      }
                                &\adjustbox{valign=t}{
                                  \begin{tikzpicture}[x=.6in]
                                    \node (P_0) at (0,0) {$P_0$};
                                    \node (Q) at (0,-1) {$Q$};
                                    \node (P_1) at (0,-2) {$P_1$};
                                    \draw[->] (P_0) to (Q);
                                    \draw[->] (Q) to (P_1);
                                  \end{tikzpicture}
                                  }
                                                               &\adjustbox{valign=t}{
\begin{lstlisting}
                                                                     func P() {
                                                                       // Block $P_0$
                                                                       Q()
                                                                       // Block $P_1$
                                                                     }
\end{lstlisting}
                                                                 }\\
      \hline

      \adjustbox{valign=t}{
      Go\{Func, Method, Args\}
      }
                                &\adjustbox{valign=t}{
                                  \begin{tikzpicture}[x=.6in]
                                    \node (P_0) at (0,0) {$P_0$};
                                    \node (P_1) at (0,-1) {$P_1$};
                                    \node (Q) at (1,0) {new $Q$};
                                    \node (dots) at (1,-1) {$\ldots$};
                                    \draw[->] (P_0) to (P_1);
                                    \draw[->] (P_0) to (Q);
                                    \draw[->] (Q) to (dots);
                                  \end{tikzpicture}
                                  }
                                                               &\adjustbox{valign=t}{
\begin{lstlisting}
                                                                     func P() {
                                                                       // Block $P_0$
                                                                       go Q()
                                                                       // Block $P_1$
                                                                     }
\end{lstlisting}
                                                                 }\\
      \hline

      \adjustbox{valign=t}{
      \makecell[l]{
      Defer\{Func, Method, Args\}\\
      RunDefers
      }
      }
                                &\hypertarget{cfg:go}{}\adjustbox{valign=t}{
                                  \begin{tikzpicture}[x=.6in]
                                    \node (P_0) at (0,0) {$P_0$};
                                    \node (P_1) at (0,-1) {$P_1$};
                                    \node (P_2) at (0,-2) {$P_2$};
                                    \node (R) at (1,-3) {$R$};
                                    \node (Q) at (1,-4) {$Q$};
                                    \draw[->] (P_0) to (P_1);
                                    \draw[->] (P_1) to (P_2);
                                    \draw[->] (P_2) to (R);
                                    \draw[->] (R) to (Q);
                                  \end{tikzpicture}
                                  }
                                                               &\adjustbox{valign=t}{
\begin{lstlisting}
                                                                     func P() {
                                                                       // Block $P_0$
                                                                       defer Q()
                                                                       // Block $P_1$
                                                                       defer R()
                                                                       // Block $P_2$
                                                                     }
\end{lstlisting}
                                                                 }\\
      \hline

    \end{tabular}
    \caption{\label{tab:control flow representation}Notable control flow SSA instructions and their CFGs.}
  }
\end{table}

Inside each block of code, control flows sequentially from one line of code to another.
For each line of code, we abstract away all information except the duration during which program control flows through the line of code. This information is called an ``event''.
We use the terminology in figure \ref{fig:start completion terminology} to refer to parts of an event.
\begin{figure}
  {\centering
    \begin{enumerate}
    \item The start of an event happens at the same time that the execution of the corresponding line of code starts.
    \item The completion of an event happens at the same time that the execution of the corresponding line of code completes.
    \end{enumerate}
    \caption{\label{fig:start completion terminology}Terminology describing the start and completion of an event.}
  }
\end{figure}

This terminology is adopted from \textcite{fava2020ready} where they rephrased the send and receive rules of the memory model of Go.
The rules of Go's memory model are restated in figure \ref{fig:send receive rules}.

Not only does $\HB{}$ relate events in program order, it also relates sends and receives according to the memory model of Go, given by rules \eqref{eq:send before receive} and \eqref{eq:receive before send}.

\begin{figure}
  {\centering
    \begin{subequations}
      \begin{align}
        &\text{The $i^{th}$ send on $\Channel$, $\E{S^i}$,} \notag \\
        &\text{happens-before the $i^{th}$ receive from $\Channel$, $\E{R^i}$, completes. In other words, $\E{S^i} \HB{\Channel} \E{R^i}$.} \label{eq:send before receive} \\
        \notag \\
        &\text{For some nonnegative $k$, the $i^{th}$ receive from $\channel{c}{k}$, $\E{R^i}$,} \notag \\
        &\text{happens-before the $(i + k)^{th}$ send on $\channel{c}{k}$, $\E{S^{i+k}}$, completes. In other words, $\E{R^i} \HB{} \E{S^{i+k}}$.} \label{eq:receive before send}
      \end{align}
    \end{subequations}
    \caption{\label{fig:send receive rules}Rules governing sends and receives, from the memory model of Go \parencite{go/ref/mem}.}
  }
\end{figure}

\begin{figure}
  \begin{subfigure}[t]{.45\textwidth}
    {\centering
      \begin{tikzpicture}[]
        \node(logic) at (0,1)  {$\transmit{\party{A}}{\channel{c}{0}}{1}{\party{B}}$};
        \node(A) at (-1,0)  {$\party{A}$};
        \node(B) at (1,0)  {$\party{B}$};
        \draw[line width=0.1mm] (-1.5,-0.5) -- (1.5,-0.5);

        \dop{AS1}{$\Ep{\party{A}}{1}$}{-1}{-2}{left}
        \doplength{BR1}{$\Ep{\party{B}}{1}$}{1}{-2}{right}{1}

        \draw [->,blue] (AS1s) -- (BR1e) node [near start,fill=white,inner sep=1pt] {$\channel{c}{0}$};
        \draw [->,red] (BR1s) -- (AS1e) node [] {};
      \end{tikzpicture}
      \caption{\label{fig:rendezvous with blocked receive execution}This execution demonstrates the blocking behavior of the receive event specified by rule \eqref{eq:send before receive}. Unlike execution \ref{fig:rendezvous with duration execution}, here we see that the receive event $\Ep{\party{B}}{1}$ starts early, but is blocked and does not complete until at least after the send event $\Ep{\party{A}}{1}$ starts. This is shown by the red arrow.}
    }
  \end{subfigure}\quad
  \begin{subfigure}[t]{.45\textwidth}
    {\centering
      \begin{tikzpicture}[]
        \node(logic) at (0,1)  {$\transmit{\party{A}}{\channel{c}{0}}{1}{\party{B}}$};
        \node(A) at (-1,0)  {$\party{A}$};
        \node(B) at (1,0)  {$\party{B}$};
        \draw[line width=0.1mm] (-1.5,-0.5) -- (1.5,-0.5);

        \doplength{AS1}{$\Ep{\party{A}}{1}$}{-1}{-2}{left}{1}
        \dop{BR1}{$\Ep{\party{B}}{1}$}{1}{-2}{right}

        \draw [->,blue] (AS1s) -- (BR1e) node [near start,fill=white,inner sep=1pt] {$\channel{c}{0}$};
        \draw [->,red] (BR1s) -- (AS1e) node [] {};
      \end{tikzpicture}
      \caption{\label{fig:rendezvous with blocked send execution}This execution demonstrates the blocking behavior of the send event specified by rule \eqref{eq:receive before send}. Unlike execution \ref{fig:rendezvous with duration execution}, here we see that the send event $\Ep{\party{A}}{1}$ starts early, but is blocked and does not complete until at least after the receive event $\Ep{\party{B}}{1}$ starts. This is shown by the blue arrow.}
    }
  \end{subfigure}
  \caption{\label{fig:rendezvous with blocked send and blocked receive executions}These executions demonstrate the blocking behavior of the send event, and the blocking behavior of the receive event.}
\end{figure}
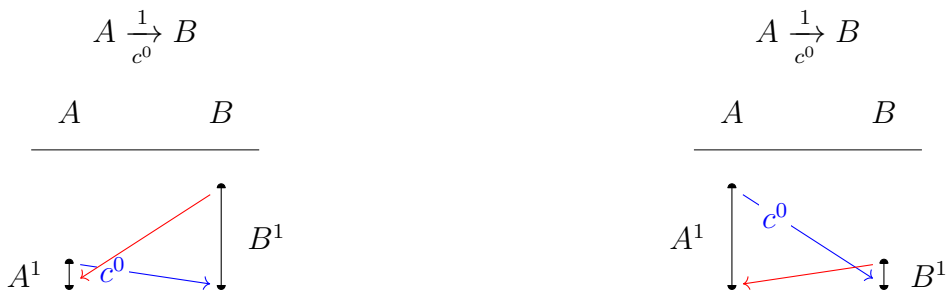

Rule \eqref{eq:send before receive} says that a receive must wait for some data to be written by a send, before it begins receiving data.
Note that the time at which the receive starts may be earlier than the time at which the receive begins receiving data. In this case, during the time in between, the receive blocks as in execution \ref{fig:rendezvous with blocked receive execution}.

Rule \eqref{eq:receive before send} says that if a channel buffer is full with $k$ messages, then a receive must begin receiving data from the channel so that the buffer is not full anymore, before a send begins sending data.
Note that the time at which the send starts may be earlier than the time at which the send begins sending data. In this case, during the time in between, the send blocks as in execution \ref{fig:rendezvous with blocked send execution}.

Parties are the interacting elements of a concurrent system.
A party is a goroutine that has some line of code that is either a send or receive.
It is not necessary for a goroutine to be a party.
For example, in program \ref{lst:simple global protocol implementation}, goroutines \lstinline{A}, \lstinline{B}, and \lstinline{C} are parties.
Even though the CFG of the goroutine \lstinline{main} has directed edges to the CFGs of \lstinline{A}, \lstinline{B}, and \lstinline{C} as indicated by the keyword \lstinline{go} (the CFG of goroutines with keyword \lstinline{go} was shown \hyperlink{cfg:go}{in this cell} in table \ref{tab:control flow representation}), \lstinline{main} does not have any line of code that is either a send or receive.
Therefore \lstinline{main} is not a party.
For the rest of this work, \lstinline{main} is never a party.

We want to show executions in the form of sequence diagrams, for better explanation.
However, the memory model of Go says:
\begin{weak} \label{weak:reorder}
The execution order observed by one goroutine may differ from the execution order observed by another goroutine.
\end{weak}
Therefore, it seems that for a single execution with $N$ goroutines, we need to show $N$ sequence diagrams, each as observed by one of the goroutines.
Despite this, only one sequence diagram is enough.
This is because sends and receives of one goroutine cannot be observed to execute differently by another goroutine.
Therefore, it is enough to consider only the execution where each goroutine sees itself executing according to its program order, and all other goroutines also see that this goroutine executes according to the same order.
In other words, it is enough to show only one order of events (the program order) on each lifeline in a sequence diagram.

An execution may be shown as a sequence diagram with many goroutines.
Each goroutine has a lifeline.
Events of each goroutine happen on the lifeline, and happen in program order.
All events happen at a specific time on the global clock, along a specific horizontal line in the sequence diagram.
This explains why the executions of Go programs are sequentially consistent if the Go programs have only sends and receives.

According to \textcite{10.1007/978-3-319-95582-7_15}, for a memory model \textit{inspired} by Go, executions of Go programs are sequentially consistent under data-race freedom (up to when an execution has a data race).
We believe this claim of sequential consistency also holds for the memory model of Go, for executions up to when the execution has a communication race.

\section{$\HB{}$ Order} \label{sec:HB Order}
We define our $\HB{}$ order by taking inspiration from Lamport's happens-before order.
Program order as mentioned in section \ref{sec:Program Order, Events, Sequential Consistency, and Parties} is a subset of $\HB{}$.
Events not ordered by $\HB{}$ may either happen in either order, or may happen at the same time.

However, in our work, we do not take inspiration from Lamport's other ideas. This simplifies our approach.
For example, although Lamport's happens-before order is related to Lamport's thread-local clocks, we do not define any local clocks.
Therefore, our approach is simplified because we do not have to synchronize any clocks.
In other words, we assume all events happen relative to the same global clock.
Although this assumption appears to contradict the characteristic of distributed systems that nodes are autonomous, this is assumption is sound in our work, because we focus only on verifying FIFO race-freedom.
To verify, we only need to prove that sends on one endpoint of a shared channel are ordered by $\HB{}$, and that receives on the other endpoint of the same shared channel are ordered by $\HB{}$.
It is reasonable to assume that events occuring on an endpoint happen relative to the same global clock.

Our $\HB{}$ order differs from Lamport's happens-before order \parencite{lamport1978time}.

Lamport's events happen instantaneously, whereas we use the terminology \ref{fig:start completion terminology} to refer to the start and completion of events with duration, in the same way that Gischer's tasks \parencite{gischer1988pomsets} happen over a duration.\footnote{Unlike Gischer's tasks, our events are unique, therefore we do not need a multiset of events; a set is enough.}

Lamport's happens-before order is a strict partial order over a set of events (irreflexive and transitive, therefore also asymmetric), whereas our happens-before order is also a strict partial order, but over a set containing both their starts and their completions. We must distinguish between the start of an event and its completion so that we can describe the events of a rendezvous with a $\HB{}$ order.

\subsection{Rendezvous} \label{sec:Rendezvous}
Recall the two rules \eqref{eq:send before receive} and \eqref{eq:receive before send} relating sends and receives.
The two rules allow for the send to start sending data at the same time that the receive starts receiving data (assuming a channel with no propogation delay). Similarly, the two rules allow for the send to complete sending data at the same time that the receive completes sending data.
Therefore, an unbuffered channel can be modelled by the two rules.
To apply rule \eqref{eq:receive before send} on unbuffered channels, we say that an unbuffered channel has capacity $0$.

A transmission on an unbuffered channel executes as in the rendezvous execution \ref{fig:rendezvous with duration execution}.
The rendezvous execution \ref{fig:rendezvous with duration execution} shows that there must be an overlap between the duration during which the send happens, and the duration during which the receive happens. The actual transmission happens instantaneously at some point during the overlapped durations. This corresponds to our intuitive notion of unbuffered channels: both the sender as well as the receiver must be ready to send and ready to receive, before the transmission can happen.
This assumes no propogation delay, because if the propogation delay is significant enough, then the completion of the send can happen-before the receive.
Significant propogation delay conflicts with the definition that the actual transmission happens instantaneously at some point during the overlapped durations.

We now compare how our $\HB{}$ order differs from Lamport's happens-before order, when representing a rendezvous.

\begin{enumerate}
\item By Lamport's definition, $\HB{}$ cannot contain both $\Ep{\party{A}}{1} \HB{\channel{c}{0}} \Ep{\party{B}}{1}$ and $\Ep{\party{B}}{1} \HB{} \Ep{\party{A}}{1}$ because $\HB{}$ is asymmetric. The execution \ref{fig:rendezvous without duration execution} captures this notion. Note that the two arrows corresponding to $\HB{}$ are horizontal arrows,\footnote{The arrows are horizontal if we require that all arrows are straight. Even though the arrows in the figure have been curved so that it is easier to tell them apart, we still consider these arrows to be horizontal.} indicating that $\Ep{\party{A}}{1}$ and $\Ep{\party{B}}{1}$ happen-at the same time.
\item By our definition, $\HB{}$ can contain both $\Ep{\party{A}}{1} \HB{\channel{c}{0}} \Ep{\party{B}}{1}$ and $\Ep{\party{B}}{1} \HB{} \Ep{\party{A}}{1}$, because the $\Ep{\party{A}}{1}$ in $\Ep{\party{A}}{1} \HB{\channel{c}{0}} \Ep{\party{B}}{1}$ refers to the start of the send event, whereas the $\Ep{\party{A}}{1}$ in $\Ep{\party{B}}{1} \HB{} \Ep{\party{A}}{1}$ refers to the completion of the send event. This does not violate asymmetry, since the two $\Ep{\party{A}}{1}$ refer to distinct parts of an event. Similarly, the two $\Ep{\party{B}}{1}$ refer to distinct parts of an event. The execution \ref{fig:rendezvous with duration execution} captures this notion. Note that all arrows corresponding to $\HB{}$ are diagonal arrows.
\end{enumerate}

\begin{figure}
  \begin{subfigure}[t]{.45\textwidth}
    {\centering
      \begin{tikzpicture}[]
        \node(logic) at (0,1)  {$\transmit{\party{A}}{\channel{c}{0}}{1}{\party{B}}$};
        \node(A) at (-1,0)  {$\party{A}$};
        \node(B) at (1,0)  {$\party{B}$};
        \draw[line width=0.1mm] (-1.5,-0.5) -- (1.5,-0.5);

        \pop{AS1}{$\Ep{\party{A}}{1}$}{-1}{-1}{left}
        \pop{BR1}{$\Ep{\party{B}}{1}$}{1}{-1}{right}

        \draw [->] (AS1) edge [bend left] node [near start,fill=white,inner sep=1pt] {$\channel{c}{0}$} (BR1);
        \draw [->] (BR1) edge [bend left] node [] {} (AS1);
      \end{tikzpicture}
      \caption{\label{fig:rendezvous without duration execution}The rendezvous execution where events happen instantaneously. An arrow from $\E{i}$ to $\E{j}$ means $\E{i}$ happens-before $\E{j}$. This is not possible, because the $\HB{}$ ordering is asymmetric.}
    }
  \end{subfigure}\quad
  \begin{subfigure}[t]{.45\textwidth}
    {\centering
      \begin{tikzpicture}[]
        \node(logic) at (0,1)  {$\transmit{\party{A}}{\channel{c}{0}}{1}{\party{B}}$};
        \node(A) at (-1,0)  {$\party{A}$};
        \node(B) at (1,0)  {$\party{B}$};
        \draw[line width=0.1mm] (-1.5,-0.5) -- (1.5,-0.5);

        \dop{AS1}{$\Ep{\party{A}}{1}$}{-1}{-1}{left}
        \dop{BR1}{$\Ep{\party{B}}{1}$}{1}{-1}{right}

        \draw [->,blue] (AS1s) -- (BR1e) node [near start,fill=white,inner sep=1pt] {$\channel{c}{0}$};
        \draw [->,red] (BR1s) -- (AS1e) node [] {};
      \end{tikzpicture}
      \caption{\label{fig:rendezvous with duration execution}The rendezvous execution where events happen over a duration. The time at which an event starts is its top half-circle, and the time at which an event completes is its bottom half-circle. A diagonal arrow from $\E{i}$ to $\E{j}$ means $\E{i}$ happens-before $\E{j}$ completes. This is possible, because this does not violate the asymmetric property of the $\HB{}$ ordering. Blue arrows correspond to $\HB{}$ elements given by rule \eqref{eq:send before receive}. Red arrows correspond to $\HB{}$ elements given by rule \eqref{eq:receive before send}.}
    }
  \end{subfigure}
  \caption{\label{fig:rendezvous without and with duration executions}Rendezvous executions between events $\Ep{\party{A}}{1}$ and $\Ep{\party{B}}{1}$ on channel $\channel{c}{0}$, one where events happen instantaneously, and one where events happen over a duration.}
\end{figure}

\section{Global Protocols} \label{sec:Global Protocols}
A global protocol is a specification of the events of parties, defined with global knowledge of what events all parties have.
For example, global protocol \eqref{eq:simple global protocol} specifies the events of parties $\party{A}$, $\party{B}$, and $\party{C}$.

\begin{subequations}
  \begin{align}
    \transmit{\party{A}}{\Channel}{1}{\party{B}} \ssep \transmit{\party{B}}{\Channel}{2}{\party{C}} \label{eq:simple global protocol}
  \end{align}
\end{subequations}

The syntax of global protocols \ref{fig:global protocol syntax} is explained as follows:
\begin{description}
\item[\protect{\parbox[t]{\widthof{\boldmath\bfseries \text{$\guardP{\Ep{\Party}{i} \HB{} \Ep{\Party'}{i}}$}}}{$\transmit{\Party}{\Channel}{i}{\Party'}$ where \\ $\Party \neq \Party'$}}] This is a transmission, read as ``the transmission uniquely identified by $i$ specifies that $\Party$ sends a message to $\Party'$ over channel $\Channel$''.
  The condition $\Party \neq \Party'$ syntactically prohibits self-transmissions.
\item[$\guardP{\Ep{\Party}{i} \HB{} \Ep{\Party'}{i}}$] This is a guard, read as ``it remains to be proven that event $\Ep{\Party}{i}$ happens before event $\Ep{\Party'}{i}$''. If a guard is proven, then the logical assertion $\Ep{\Party}{i} \HB{} \Ep{\Party'}{i}$ is known to be true. Further discussion of guards is deferred to section \ref{sec:Verifications, Implementations, and Guards}.
\item[$\emp$] The empty global protocol $\emp$ does nothing.
\item[$\GlobalProtocol \ssep \GlobalProtocol$] Global protocols may be sequentially composed by $\ssep$, such that one global protocol executes before the other.
\item[$\GlobalProtocol \asep \GlobalProtocol$] Global protocols may be concurrently composed by $\asep$, such that the two global protocols can execute at the same time.
  The concurrently composed global protocol does not impose any ordering between any event of the first global protocol and any event of the second global protocol.
\end{description}

This syntax is the syntax of session logic, but restricted to only transmissions, the empty protocol, sequential composition, and concurrent composition, as well as extended with guards.
Like Bocchi's session logic \parencite{Bocchi2012AMM}, our session logic is session types extended with logical assertions.
According to \textcite{10.1145/1328438.1328472}, ``session types model communications that often have a highly structured sequence of interactions, which as a whole form a natural unit of conversation. The structure of a conversation is abstracted as a session type through an intuitive syntax''.

We extend the usual session type syntax, by annotating transmissions with unique positive indexes.
Indexes are annotated by traversing the global protocol from left to right, and labelling transmissions with consecutive indexes starting from $1$.
For example, in global protocol \eqref{eq:simple global protocol}, $\transmit{\party{A}}{\Channel}{1}{\party{B}}$ has index $1$, and $\transmit{\party{B}}{\Channel}{2}{\party{C}}$ has index $2$.
Transmission indexes are defined so that we can refer to the events of a transmission.
For example, global protocol \eqref{eq:simple global protocol} has the events $\Ep{\party{A}}{1}$, $\Ep{\party{B}}{1}$, $\Ep{\party{B}}{2}$, and $\Ep{\party{C}}{2}$.

Our session logic has CSL as the underlying logic of its logical assertions.
Therefore, similar to \citeauthor{Bocchi2012AMM}'s session logic, our logical assertions can refer to the local state of each goroutine.
Moreover, separation logic allows, by the frame rule, a parallel decomposition that relates local computation of parallel programs to local computations of its components \parencite{BROOKES2007227}.

Other works that define session logics are \textcite{oortwijn2019abstractions} and \textcite{craciun2015towards}.
In both works, the session type and the logical assertions are able to share information.
For example, \textcite{oortwijn2019abstractions} defines a program annotation $\textbf{query } \text{b}$ that allows the logical assertion to assume a Boolean property $\text{b}$, which is proven to hold by process-algebraic analysis on the session type.

The syntax of global protocols include concurrent composition.
We explained that the two global protocols can execute at the same time.
This coincides with the usual notions of concurrent composition \parencites{gischer1988pomsets}{Hoare2009ConcurrentKA}, where composed parts execute concurrently.
However, we need to consider that in Go, parties must execute sequentially.
Therefore, there are cases where the concurrently composed global protocol imposes an ordering between some events of the first global protocol and some events of the second global protocol.

Consider the global protocol $\transmit{\party{A}}{\channel{c}{2}}{1}{\party{B}} \asep \transmit{\party{C}}{\channel{d}{2}}{2}{\party{A}}$.
In implementation \ref{lst:global protocol implementation concurrent}, the party $\party{A}$ that implements all of $\party{A}$'s events can order $\party{A}$'s send before $\party{A}$'s receive, as if the global protocol were sequentially composed instead of concurrently composed.
However, concurrent composition gives more freedom of implementation, in that it allows the party $\party{A}$ to instead order $\party{A}$'s receive before $\party{A}$'s send.
However, the party $\party{A}$ is not allowed to execute $\party{A}$'s concurrently with $\party{A}$'s receive.
In other words, parties are not allowed to spawn more concurrently executing parties.
Note that the \lstinline{main} goroutine is not a party, and therefore is allowed to spawn concurrently executing parties.
All concurrently executing parties must be determined in advance, and specified by the global protocol.

Other works \parencite[Section 4.4]{ng2016deadlock} similarly prohibit dynamic spawning of parties.

\begin{code}[caption={An implementation of the global protocol $\transmit{\party{A}}{\channel{c}{2}}{1}{\party{B}} \asep \transmit{\party{C}}{\channel{d}{2}}{2}{\party{A}}$, where $\party{A}$ is allowed to receive before sending.},label={lst:global protocol implementation concurrent},multicols=2]
package main

var c = make(chan int, 2)
var d = make(chan int, 2)

func main() {
	go A()
	go B()
	go C()
}

func A() {
	<-d
	c <- 1
}

func B() {
	<-c
}

func C() {
	d <- 2
}
\end{code}

The syntax of global protocols syntactically prohibits self-transmissions with the condition $\Party \neq \Party'$, to avoid the possibility of deadlock.
Deadlock is possible when self-transmissions are allowed, in the case where the self-transmission of a goroutine $\Party$ occurs on an unbuffered channel.
$\Party$ deadlocks because goroutines execute sequentially, so that there is never an overlap between the durations of the send and the receive of the self-transmission.
The rendezvous never happens.

Other works \parencite{DY2013} have similar syntactic conditions that prohibit self-sent messages in session types.

We also prohibit self-transmissions caused by the semantics of the concurrent composition operator $\asep$, to once again avoid the possibility of deadlock.
We achieve this by imposing \citeauthor{costea2017session}'s well-formedness condition of nondeterministic global protocols \parencite[Definition 12]{costea2017session}, which we will not restate in this work.
However, we will restate their explanation of why a non-well-formed global protocol $\transmit{\party{A}}{\channel{c}{2}}{1}{\party{B}} \asep \transmit{\party{C}}{\channel{c}{2}}{2}{\party{A}}$ will deadlock.
Their explanation refers to implementation \ref{lst:non-well-formed global protocol implementation deadlocked}.
The goroutine $\party{A}$ that implements all of $\party{A}$'s events in the global protocol is allowed to order $\party{A}$'s receive before $\party{A}$'s send, because the global protocol is non-well-formed, and concurrent composition does not fix the order of these events.
If a program that implements goroutine $\party{A}$ this way executes such that $\party{B}$ actually receives data from $\party{C}$'s send, then $\party{A}$'s receive will block forever, since no sends will ever execute ($\party{A}$'s receive must finish executing before $\party{A}$'s send can execute).

\begin{code}[caption={An implementation of the global protocol $\transmit{\party{A}}{\channel{c}{2}}{1}{\party{B}} \asep \transmit{\party{C}}{\channel{c}{2}}{2}{\party{A}}$ that deadlocks if $\party{A}$ is allowed to self-transmit.},label={lst:non-well-formed global protocol implementation deadlocked},multicols=2]
package main

var c = make(chan int, 2)

func main() {
	go A()
	go B()
	go C()
}

func A() {
	<-c
	c <- 1
}

func B() {
	<-c
}

func C() {
	c <- 2
}
\end{code}

Despite these syntactic and semantic prohibitions, we do not sacrifice expressiveness of our session logic, because it is still possible for parties to self-transmit.
For example, the global protocol $\transmit{\party{A}}{\Channel}{1}{\party{B}} \ssep \transmit{\party{B}}{\Channel}{2}{\party{A}}$, where all transmissions transmit the same message, simulates a self-transmission of party $\party{A}$.

\subsection{Partial Order Representation and Races}
Global protocol \eqref{eq:simple global protocol} specifies that party $\party{B}$'s events must happen in the order given by party $\party{B}$'s program order, because the global protocol orders $\party{B}$'s events with the $\HB{}$ relation such that $\party{B}$'s receive happens-before $\party{B}$'s send.
The global protocol also specifies that party $\party{B}$ \textit{should} receive data from party $\party{A}$, and party $\party{C}$ \textit{should} receive data from party $\party{B}$, because the global protocol orders events with the $\CB{}$ relation in that way.

We said $\party{B}$ \textit{should} (and not \textit{must}) receive data from $\party{A}$, where \textit{should} receive data means the protocol expects that $\party{B}$ receives data from $\party{A}$, and \textit{must} receive data means there is no execution where $\party{B}$ does not receive data from $\party{A}$.
The reason for this distinction between \textit{should} and \textit{must} is: an execution of the global protocol may be such that $\party{B}$ does not receive data from $\party{A}$, as in figure \ref{fig:simple global protocol sequential execution with races}.\footnote{The figure actually shows a sequential execution, but a sequential execution is a special kind of execution.}

To model the notion of ``\textit{should} receive data'', we define a communicates-before order $\CB{}$ as a set of triples $(\E{S}, \Channel, \E{R})$, where the global protocol expects $\E{R}$ to receive data from $\E{S}$ on channel $\Channel$, as expressed by the global subprotocol $\transmit{\E{S}}{\Channel}{i}{\E{R}}$.
Despite this expectation that $\E{R}$ receives data from $\E{S}$, an execution could have $\E{R}$ receive data from a different send, or have $\E{R}$ not receive any data at all. We call such deviations from the global protocol ``races''.
We consider an execution with races to be buggy, and we also consider the program that performed this execution to be buggy.
If a global protocol may be satisfied by an implementation which may execute with races, then we call this a ``global protocol with races''.
Otherwise, we call this a ``race-free global protocol''.
Unlike buggy executions with races, we do not consider a global protocol with races to be buggy, because the global protocol is the ultimate source of truth.
Therefore, we assume that the global protocol designer intended for the global protocol to have races.

A race-free global protocol always has race-free executions.
One possible race-free execution is the execution that replaces each $\E{S} \CB{\Channel} \E{R}$ with $\E{S} \HB{\Channel} \E{R}$, such that $\E{R}$ \textit{must} receive data from $\E{S}$.

We \textit{could} avoid defining $\CB{}$ by using $\E{S} \HB{\Channel} \E{R}$ for each $\E{S} \CB{\Channel} \E{R}$ in global protocols, with the provision that: under the context of global protocols, all $\E{S} \HB{\Channel} \E{R}$ actually means is that the receiver \textit{should} receive data. The problem with this is: under the context of executions, $\E{S} \HB{} \E{R}$ still means the receiver \textit{must} receive data. This double meaning of $\HB{}$ is potentially confusing. Introducing the $\CB{}$ relation avoids this confusion.

A global protocol can be represented with a partial order, over a set containing both the starts and the completions of events, by transforming the global protocol according to the following transformation:

\begin{description}[labelwidth=0pt]

\item[$\CB{\Channel}$ generation:] Each $\transmit{\Party}{\Channel}{i}{\Party'}$ is transformed into $\Ep{\Party}{i} \CB{\Channel} \Ep{\Party'}{i}$.
  For example, global protocol \eqref{eq:simple global protocol} generates $\Ep{\party{A}}{1} \CB{\Channel} \Ep{\party{B}}{1}$ and $\Ep{\party{B}}{2} \CB{\Channel} \Ep{\party{C}}{2}$.

\item[$\HB{}$ generation:]
  For any two transmissions ordered by sequential composition,\footnote{A formal definition of whether two transmissions are ordered by sequential composition is given by \citeauthor{costea2017session} \parencite[Definition 8]{costea2017session}. For transmissions $\transmit{\Party}{\Channel}{i}{\Party'}$ and $\transmit{\Party''}{\Channel}{i}{\Party'''}$, suppose that these transmissions are composed with many global subprotocols, like in the global protocol $(\transmit{\Party}{\Channel}{i}{\Party'} \asep \GlobalProtocol) \ssep (\GlobalProtocol' \asep \transmit{\Party''}{\Channel}{i}{\Party''')}$, where $\GlobalProtocol$ and $\GlobalProtocol'$ are global subprotocols. The formal definition still sequentially orders these transmissions, so that even these arbitrarily nested transmissions generate $\HB{}$ if party equality holds.\label{foot:formal two transmissions ordered sequentially}} such that $\transmit{\Party}{\Channel}{i}{\Party'}$ occurs before $\transmit{\Party''}{\Channel}{i}{\Party'''}$:
  \begin{enumerate}
  \item If $\Party = \Party''$, then $\Ep{\Party}{i} \HB{} \Ep{\Party''}{i}$.
  \item If $\Party = \Party'''$, then $\Ep{\Party}{i} \HB{} \Ep{\Party'''}{i}$.
  \item If $\Party' = \Party''$, then $\Ep{\Party'}{i} \HB{} \Ep{\Party''}{i}$.
  \item If $\Party' = \Party'''$, then $\Ep{\Party'}{i} \HB{} \Ep{\Party'''}{i}$.
  \end{enumerate}
  where party equality is defined in the obvious way, such that for example $\text{party } \party{A} = \text{party } \party{A}$.
  These generated $\HB{}$ elements are the program order of the respective parties.
  This $\HB{}$ ordering is similar to \citeauthor{10.1145/1328438.1328472}'s prefix ordering \parencite[Definition 3.3]{10.1145/1328438.1328472} for asynchronous communications, which imposes sequencing only on the actions of the same participant in ordered prefixes.

  As an example of generated $\HB{}$ elements, the global protocol \eqref{eq:simple global protocol} generates $\Ep{\party{B}}{1} \HB{} \Ep{\party{B}}{2}$.
\end{description}

As an example of this transformation, we represent the global protocol \eqref{eq:simple global protocol} with a partial order \eqref{eq:simple global protocol partial order}.
The partial order specifies the same things that the global protocol specifies.

The partial order \eqref{eq:simple global protocol partial order} is drawn as a sequence diagram \ref{fig:simple global protocol}.
A solid arrow from $\E{}$ to $\E{'}$ denotes $\E{} \HB{} \E{'}$, and a dotted arrow from $\E{}$ to $\E{'}$ denotes $\E{} \CB{\Channel} \E{'}$. Dotted arrows are labelled with the channel on which communication happens.
The vertical arrow $\Ep{\party{B}}{1} \HB{} \Ep{\party{B}}{2}$ is given by $\party{B}$'s program order.

Global protocols are defined up to party renaming, where if a party is replaced with another party, it is replaced everywhere in that global protocol.
For example, the global protocol \eqref{eq:simple global protocol} with $\party{B}$ renamed to $\party{D}$ is global protocol \eqref{eq:simple global protocol party renamed}.

\begin{subequations}
  \begin{align}
    \transmit{\party{A}}{\Channel}{1}{\party{D}} \ssep \transmit{\party{D}}{\Channel}{2}{\party{C}} \label{eq:simple global protocol party renamed}
  \end{align}
\end{subequations}

which is defined to be the same as the global protocol \eqref{eq:simple global protocol}.

\subsection{Indistinguishability of Global Protocols and Partial Orders}
We can change which party each event in the partial order belongs to. The partial order \eqref{eq:simple global protocol partial order} is the same as the partial order \eqref{eq:simple global protocol partial order different party}, but with event $\Ep{\party{B}}{2}$ renamed to event $\Ep{\party{D}}{2}$ everywhere. Note that we are renaming events, but not renaming parties.

\begin{subequations}
  \begin{align}
    \Ep{\party{A}}{1} \CB{\Channel} \Ep{\party{B}}{1} \land \Ep{\party{B}}{2} \CB{\Channel} \Ep{\party{C}}{2} \land \Ep{\party{B}}{1} \HB{} \Ep{\party{B}}{2} \label{eq:simple global protocol partial order} \\
    \Ep{\party{A}}{1} \CB{\Channel} \Ep{\party{B}}{1} \land \Ep{\party{D}}{2} \CB{\Channel} \Ep{\party{C}}{2} \land \Ep{\party{B}}{1} \HB{} \Ep{\party{D}}{2} \label{eq:simple global protocol partial order different party}
  \end{align}
\end{subequations}

An equivalence class of partial orders contains partial orders with the same structure, but different party information.
Therefore, an equivalence class of partial orders is the equivalent partial order that does not contain party information.
For example, both \eqref{eq:simple global protocol partial order} and \eqref{eq:simple global protocol partial order different party} are in the equivalence class \eqref{eq:simple global protocol partial order no party}:

\begin{subequations}
  \begin{align}
    \E{1} \CB{\Channel} \E{2} \land \E{3} \CB{\Channel} \E{4} \land \E{2} \HB{} \E{3} \label{eq:simple global protocol partial order no party}
  \end{align}
\end{subequations}

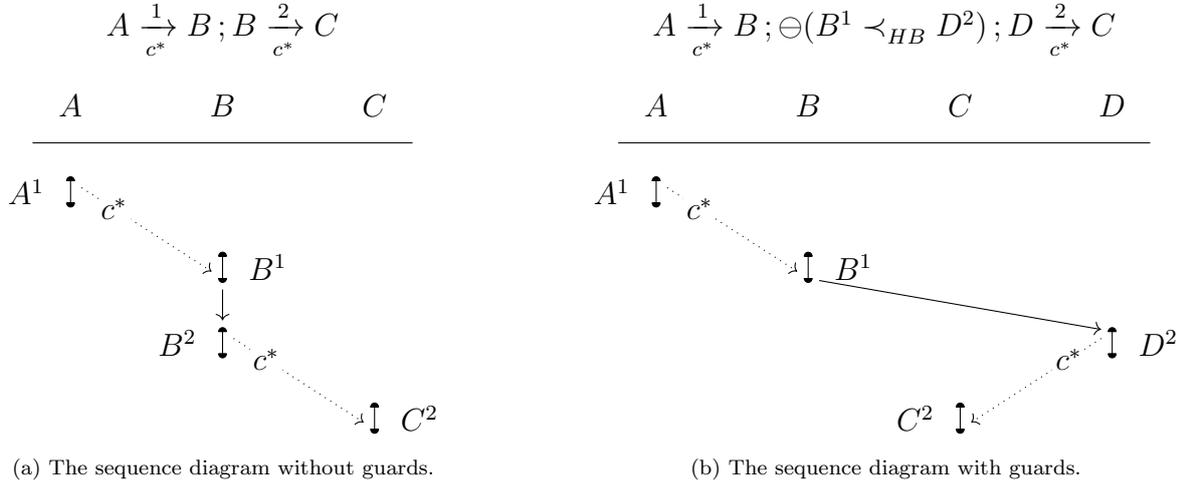
\begin{figure}
  \begin{subfigure}[t]{.45\textwidth}
    {\centering
      \begin{tikzpicture}[]
        \node(logic) at (0,1)  {$\transmit{\party{A}}{\Channel}{1}{\party{B}} \ssep \transmit{\party{B}}{\Channel}{2}{\party{C}}$};
        \node(A) at (-2,0)  {$\party{A}$};
        \node(B) at (0,0)  {$\party{B}$};
        \node(C) at (2,0)  {$\party{C}$};
        \draw[line width=0.1mm] (-2.5,-0.5) -- (2.5,-0.5);

        \dop{AS1}{$\Ep{\party{A}}{1}$}{-2}{-1}{left}
        \dop{BR1}{$\Ep{\party{B}}{1}$}{0}{-2}{right}
        \dop{BS1}{$\Ep{\party{B}}{2}$}{0}{-3}{left}
        \dop{CR1}{$\Ep{\party{C}}{2}$}{2}{-4}{right}

        \draw [->] (BR1e) -- (BS1s) node [] {};

        \draw [->,dotted] (AS1s) -- (BR1e) node [near start,fill=white,inner sep=1pt] {$\Channel$};
        \draw [->,dotted] (BS1s) -- (CR1e) node [near start,fill=white,inner sep=1pt] {$\Channel$};
      \end{tikzpicture}
      \caption{\label{fig:simple global protocol}The sequence diagram without guards.}
    }
  \end{subfigure}\quad
  \begin{subfigure}[t]{.45\textwidth}
    {\centering
      \begin{tikzpicture}[]
        \node(logic) at (0,1)  {$\transmit{\party{A}}{\Channel}{1}{\party{B}} \ssep \guardP{\Ep{\party{B}}{1} \HB{} \Ep{\party{D}}{2}} \ssep \transmit{\party{D}}{\Channel}{2}{\party{C}}$};
        \node(A) at (-3,0)  {$\party{A}$};
        \node(B) at (-1,0)  {$\party{B}$};
        \node(C) at (1,0)  {$\party{C}$};
        \node(D) at (3,0)  {$\party{D}$};
        \draw[line width=0.1mm] (-3.5,-0.5) -- (3.5,-0.5);

        \dop{AS1}{$\Ep{\party{A}}{1}$}{-3}{-1}{left}
        \dop{BR1}{$\Ep{\party{B}}{1}$}{-1}{-2}{right}
        \dop{DS1}{$\Ep{\party{D}}{2}$}{3}{-3}{right}
        \dop{CR1}{$\Ep{\party{C}}{2}$}{1}{-4}{left}

        \draw [->] (BR1e) -- (DS1s) node [] {};

        \draw [->,dotted] (AS1s) -- (BR1e) node [near start,fill=white,inner sep=1pt] {$\Channel$};
        \draw [->,dotted] (DS1s) -- (CR1e) node [near start,fill=white,inner sep=1pt] {$\Channel$};
      \end{tikzpicture}
      \caption{\label{fig:simple global protocol with guards}The sequence diagram with guards.}
    }
  \end{subfigure}
  \caption{\label{fig:simple global protocols without and with guards}Sequence diagrams of global protocol \eqref{eq:simple global protocol}, one without guards, and one with guards. The global protocol is written above each sequence diagram.}
\end{figure}

We can define the inverse transformation from a partial order back to the global protocol.
Therefore, each equivalence class of partial orders corresponds to their equivalence class of global protocols.
For example, the equivalence class \eqref{eq:simple global protocol partial order no party} corresponds to the set containing global protocol \ref{fig:simple global protocol} (corresponding to the partial order \eqref{eq:simple global protocol partial order}), global protocol \ref{fig:simple global protocol with guards} (corresponding to the partial order \eqref{eq:simple global protocol partial order different party}), and possibly other global protocols.
Therefore, global protocols \ref{fig:simple global protocol} and \ref{fig:simple global protocol with guards} are equal up to the equivalence relation that defines these equivalence classes.

The global protocol \ref{fig:simple global protocol with guards} has a guard.
In general, a global protocol may have guards.
Recall that the partial order \eqref{eq:simple global protocol partial order different party} that corresponds to the global protocol, is the same as partial order \eqref{eq:simple global protocol partial order}, but with event $\Ep{\party{B}}{2}$ renamed to event $\Ep{\party{D}}{2}$ everywhere.
In partial order \eqref{eq:simple global protocol partial order}, the order $\Ep{\party{B}}{1} \HB{} \Ep{\party{B}}{2}$ is given by program order of $\party{B}$.
In partial order \eqref{eq:simple global protocol partial order different party}, event $\Ep{\party{B}}{2}$ has been renamed to $\Ep{\party{D}}{2}$, so that the order $\Ep{\party{B}}{1} \HB{} \Ep{\party{B}}{2}$ no longer exists, and is no longer given by program order of $\party{B}$.
The guard $\guardP{\Ep{\party{B}}{1} \HB{} \Ep{\party{D}}{2}}$ in the global protocol \ref{fig:simple global protocol with guards} allows our verification framework to distinguish between the implementations that guarantee the ordering $\Ep{\party{B}}{1} \HB{} \Ep{\party{D}}{2}$, and the implementations that do not guarantee the ordering.
Our framework verifies with success only the implementations that guarantee the ordering $\Ep{\party{B}}{1} \HB{} \Ep{\party{D}}{2}$.
Further discussion on proving guards is deferred to section \ref{sec:Verifications, Implementations, and Guards}.

Not all partial orders correspond to a global protocol.
For example, the partial order $\Ep{\party{A}}{1} \CB{\Channel} \Ep{\party{B}}{1} \land \Ep{\party{A}}{2} \CB{\Channel} \Ep{\party{B}}{2} \land \Ep{\party{A}}{1} \HB{} \Ep{\party{A}}{2}$ does not correspond to a global protocol.
The $\HB{}$ element $\Ep{\party{A}}{1} \HB{} \Ep{\party{A}}{2}$ indicates that global protocol consists of sequentially composed transmissions $\transmit{\party{A}}{\Channel}{1}{\party{B}} \ssep \transmit{\party{A}}{\Channel}{2}{\party{B}}$, but then sequential composition indicates that the $\HB{}$ order must also have the element $\Ep{\party{B}}{1} \HB{} \Ep{\party{B}}{2}$, and yet the partial order does not have this element.
The inverse transformation from a partial order back to a global protocol maps these partial orders that do not correspond to a global protocol, to the empty set.

In this work, we only discuss partial orders that correspond to a global protocol.
Due to this correspondence, in this work, we do not distinguish between a global protocol and its partial order.
For example, in the figure \ref{fig:simple global protocol}, the partial order \eqref{eq:simple global protocol partial order} is written as a global protocol, and shown as a sequence diagram.

\section{Sequential Executions} \label{sec:Sequential Executions}
A sequential execution is a special kind of execution.

A sequential execution that refines a global protocol is a linear order, ordered only by the $\HB{}$ relation, on both the starts and the completions of events, for all events.
The linear order of a sequential execution that refines a global protocol is a linearization with the same underlying set of events as the set of events in the global protocol.
The linearization contains the global protocol's $\HB{}$ order, and maybe also contains some $\HB{\Channel}$ elements between sends and receives not in the global protocol, where each added $\HB{\Channel}$ can have different $\Channel$.
In a sequential execution, all events must be atomic, so that there is no event between the start and the completion of an event.
Each added $\E{S} \HB{\Channel} \E{R}$ for some $\Channel$ means the receive $\E{R}$ actually receives data from send $\E{S}$ on channel $\Channel$, and the completion of $\E{S}$ happens-before the start of $\E{R}$.
Each added $\E{} \HB{} \E{'}$ means the sequential execution executes in such a way that the completion of $\E{}$ happens-before the start of $\E{'}$.

Note that the linear order of a sequential execution does not include elements from the $\CB{}$ order of the global protocol: intuitively, a sequential execution does not care about how sends are specified to communicate with receives, it only cares about which receives actually receives data from certain sends (as given by $\HB{}$).

We are motivated to define sequential executions, because in most cases, sequential executions are easier to discuss about than concurrent executions.
Sequential execution events are atomic, so that we can define sequential execution events to be instantaneous without duration.
Therefore, when discussing sequential executions, we do not need to mention the start or completion of a sequential execution event.
We only need to mention the sequential execution event as a whole.
As another consequence, sequence diagrams of sequential executions become more concise.
For the rest of this work, the only executions we will consider are sequential executions with atomic events, except for when we must consider concurrent executions, or except for when we are citing another work.

A sequential execution simulates a concurrent execution.
We are allowed to represent a concurrent execution with a sequential execution, because this is similar to how true concurrency semantics and interleaved concurrency semantics are indistinguishable for process calculi.
Interleaved concurrency semantics simulates true concurrency semantics by arriving at the same conclusion in two reductions instead of one reduction. Process calculi do not distinguish the proof of a conclusion by the number of reductions.

A true concurrency rule that reduces process $P$ to process $P'$ and process $Q$ to process $Q'$:

$$\Rule{P \rightarrow P' \qquad Q \rightarrow Q'}{P \parallel Q  \rightarrow  P' \parallel Q'}$$

can be simulated by applying the two interleaved concurrency rules in sequence:

\begin{align*}
  \Rule{P \rightarrow P'}{P \parallel Q  \rightarrow  P' \parallel Q} & & \Rule{Q \rightarrow Q'}{P' \parallel Q  \rightarrow  P' \parallel Q'}
\end{align*}

Similarly, the sequential semantics for any two events that concurrently access a shared channel (send and send, send and receive, receive and send, receive and receive) is a linearization of this concurrent access, with the condition that no other events happen in-between these two events.
Intuitively, this means the two linearized concurrent events are regarded by all other events to happen at nearly the same time.
This simulates that the two linearized concurrent events happen at the same time.

However, there is one case where a sequential execution does not simulate a concurrent execution.
If $\E{S}$ and $\E{R}$ are events of a rendezvous, then the start of $\E{R}$ happens-before the completion of $\E{S}$, and by asymmetry of $\HB{}$, the completion of $\E{S}$ cannot happen-before the start of $\E{R}$.
There is a conflict between the definitions of rendezvous and sequential execution, because the completion of $\E{S}$ cannot happen-before the start of $\E{R}$, yet if we apply the definition of a sequential execution to a rendezvous, then the completion of $\E{S}$ must happen-before the start of $\E{R}$.
We resolve this case of conflict in the next section, so that by the end of that section, it should be clear that a sequential execution simulates a concurrent execution.

\subsection{Legal Sequential Executions}
\begin{figure}
  {\centering
    \begin{tikzpicture}[]
      \node(logic) at (0,1)  {$\transmit{\party{A}}{\Channel}{1}{\party{B}} \ssep \transmit{\party{B}}{\Channel}{2}{\party{C}}$};
      \node(A) at (-2,0)  {$\party{A}$};
      \node(B) at (0,0)  {$\party{B}$};
      \node(C) at (2,0)  {$\party{C}$};
      \draw[line width=0.1mm] (-2.5,-0.5) -- (2.5,-0.5);

      \pop{AS1}{$\Ep{\party{A}}{1}$}{-2}{-3}{left}
      \pop{BR1}{$\Ep{\party{B}}{1}$}{0}{-1}{right}
      \pop{BS1}{$\Ep{\party{B}}{2}$}{0}{-2}{left}
      \pop{CR1}{$\Ep{\party{C}}{2}$}{2}{-4}{right}

      \draw [->] (BR1) -- (BS1) node [] {};

      \draw [->,blue] (AS1) -- (CR1) node [near start,fill=white,inner sep=1pt] {$\Channel$};
      \draw [->] (BS1) -- (AS1) node [] {};
    \end{tikzpicture}
    \caption{\label{fig:simple global protocol sequential execution with races illegal}An illegal sequential execution of the global protocol \ref{fig:simple global protocol}.}
  }
\end{figure}
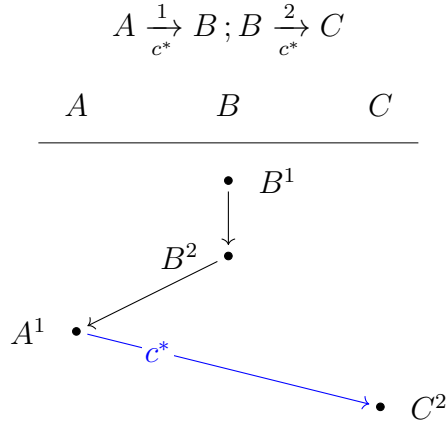

Not all sequential executions are legal. For example, the global protocol \ref{fig:simple global protocol} can never execute in the way given by the sequential execution \ref{fig:simple global protocol sequential execution with races illegal}.
This is because the sequential execution orders the receives such that $\Ep{\party{B}}{1}$ happens-before $\Ep{\party{C}}{2}$, and also says that the first send $\Ep{\party{A}}{1}$ sends data to the second receive $\Ep{\party{C}}{2}$.
However, by rule \eqref{eq:send before receive}, the first send $\Ep{\party{A}}{1}$ must send data to the first receive $\Ep{\party{B}}{1}$, not to the second receive.
This sequential execution is illegal because it violates rule \eqref{eq:send before receive}.

Using this observation, we can determine whether any given sequential execution is illegal.
\begin{lemma}
  There is an algorithm to determine whether any given sequential execution is illegal.
\end{lemma}
\begin{proof}
  Since the sequential execution linearizes concurrent sends, our verification knows which send the $i^{th}$ send is, and since the sequential execution linearizes concurrent receives, our verification knows which receive the $i^{th}$ receive is.
  Then, the rules \eqref{eq:send before receive} and \eqref{eq:receive before send} give a happens-before order, which we add to our $\HB{}$ order.
  If the resulting $\HB{}$ order has a cycle, then the sequential execution is illegal.
\end{proof}

In this work, we consider only legal sequential executions.

\subsection{Sequential Executions of Rendezvous}
Sequential executions must have atomic events, but this conflicts with the fact that the send and receive of a rendezvous must happen at the same time.
We resolve this conflict by redefining the sequential execution of a rendezvous.
A sequential execution must respect the semantics of a rendezvous:
\begin{subequations}
  \begin{align}
    \text{The send transmits data to the receive.} \label{eq:sequential execution rendezvous semantics transmit} \\
    \text{The send blocks at least until the receive starts.} \label{eq:sequential execution rendezvous semantics send blocks} \\
    \text{The receive blocks at least until the send starts.} \label{eq:sequential execution rendezvous semantics receive blocks}
  \end{align}
\end{subequations}

Semantics \eqref{eq:sequential execution rendezvous semantics transmit} is self-evident. Semantics \eqref{eq:sequential execution rendezvous semantics send blocks} is given by rule \eqref{eq:send before receive}, while semantics \eqref{eq:sequential execution rendezvous semantics receive blocks} is given by rule \eqref{eq:receive before send}.
Therefore, although atomic events have no notion of the ``start'' of an event or the ``completion'' of an event, these notions are still present in the memory model rules, and these (explicit) rules are now the (implicit) semantics that the sequential execution must respect.
In the following discussion about semantics, we show sequential executions such that its events are events with duration with a start and completion, because the discussion of semantics will mention the notions of ``start'' and ``completion''.

To respect semantics \eqref{eq:sequential execution rendezvous semantics transmit}, we have the rendezvous sequential execution \ref{fig:rendezvous sequential execution with events with duration} order the send before the receive: data has to be in-flight before it can be received.\footnote{We believe that it is also possible to respect semantics \eqref{eq:sequential execution rendezvous semantics transmit} if the sequential execution instead orders receive before send. However, we find this tougher to explain.} In other words, the sequential execution simulates that the data is stored in a buffered channel with capacity $1$.

The sequential execution \ref{fig:rendezvous sequential execution with events with duration} also respects semantics \eqref{eq:sequential execution rendezvous semantics receive blocks} (shown by the blue arrow), but does not respect semantics \eqref{eq:sequential execution rendezvous semantics send blocks} (shown by the red arrow), because the send happens-before the receive (shown by the black arrow); the send does not wait for the receive to start.

Another way of noticing that the sequential execution \ref{fig:rendezvous sequential execution with events with duration} does not respect semantics \eqref{eq:sequential execution rendezvous semantics send blocks} is, the sequential execution has a cycle formed by the red arrow and the black arrow, violating the assymetry condition of $\HB{}$.

To resolve this, we need to ensure the sequential execution satisfies some semantic conditions in figure \ref{fig:semantic conditions} that simulate the semantics \eqref{eq:sequential execution rendezvous semantics send blocks}.
We do not need to consider semantic conditions \eqref{eq:aft-recvs are aft-sends} and \eqref{eq:bef-sends are before-recvs}, because these are already satisfied by ordering the send before the receive.\footnote{If the rendezvous sequential execution instead ordered the receive before the send, then we would require the sequential execution to satisfy these other two semantic conditions.}
The sequential execution must satisfy condition \eqref{eq:bef-recvs are before-sends} such that no event of the sender happens-between the send and the receive, and the sequential execution must satisfy condition \eqref{eq:aft-sends are aft-recvs} such that no event of the receiver happens-between the send and the receive.

\begin{figure}
  {\centering
    \begin{subequations}
      \begin{align}
        &\E{S} \HB{\channel{c}{0}} \E{R} \land \E{R} \HB{} \E{} \implies \E{S} \HB{} \E{} \notag \\
        &\text{Each event that happens-after a receive must also happens-after the corresponding send.} \label{eq:aft-recvs are aft-sends} \tag{aft-recvs are aft-sends} \\
        &\E{} \HB{} \E{S} \land \E{S} \HB{\channel{c}{0}} \E{R} \implies \E{} \HB{} \E{R} \notag \\
        &\text{Each event that happens-before a send must also happens-before the corresponding receive.} \label{eq:bef-sends are before-recvs} \tag{bef-sends are before-recvs} \\
        &\E{S} \HB{\channel{c}{0}} \E{R} \land \E{S} \HB{} \E{} \implies \E{R} \HB{} \E{} \notag \\
        &\text{Each event that happens-after a send must also happens-after the corresponding receive.} \label{eq:aft-sends are aft-recvs} \tag{aft-sends are aft-recvs} \\
        &\E{S} \HB{\channel{c}{0}} \E{R} \land \E{} \HB{} \E{R} \implies \E{} \HB{} \E{S} \notag \\
        &\text{Each event that happens-before a receive must also happens-before the corresponding send.} \label{eq:bef-recvs are before-sends} \tag{bef-recvs are before-sends}
      \end{align}
    \end{subequations}
    \caption{\label{fig:semantic conditions}Semantic conditions for channel $\channel{c}{0}$.}
  }
\end{figure}

The semantic conditions \ref{fig:semantic conditions} must satisfy the requirement \ref{req:there are no other events on an unbuffered channel}, because if there is another receive $\E{R'}$ that happens concurrently with the receive $\E{R}$, then it is possible for a sequential execution to rendezvous $\E{R'}$ with $\E{S}$ instead.
In other words, in the sequential execution, we no longer have $\E{S} \HB{\channel{c}{0}} \E{R}$, therefore the conclusions of each semantic condition may no longer hold.

\begin{req} \label{req:there are no other events on an unbuffered channel}
  $\E{S}$ and $\E{R}$ are the only events occurring on channel $\channel{c}{0}$.
\end{req}

The semantic conditions \ref{fig:semantic conditions} and the requirement together ensure that no event of the global protocol happens-between the send and the receive.
A sequential execution with atomic events that satisfies the semantics of rendezvous is \ref{fig:rendezvous sequential execution with atomic events}.

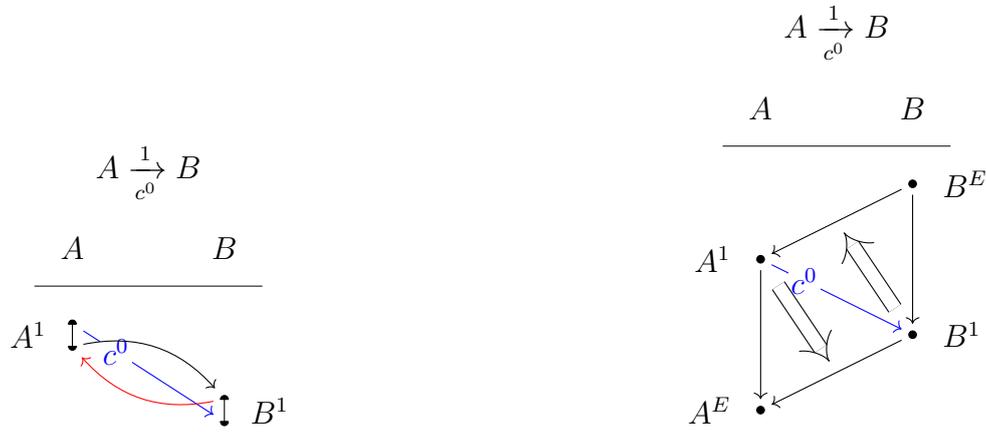
\begin{figure}
  \begin{subfigure}[t]{.2\textwidth}
    {\centering
      \begin{tikzpicture}[]
        \node(logic) at (0,1)  {$\transmit{\party{A}}{\channel{c}{0}}{1}{\party{B}}$};
        \node(A) at (-1,0)  {$\party{A}$};
        \node(B) at (1,0)  {$\party{B}$};
        \draw[line width=0.1mm] (-1.5,-0.5) -- (1.5,-0.5);

        \dop{AS1}{$\Ep{\party{A}}{1}$}{-1}{-1}{left}
        \dop{BR1}{$\Ep{\party{B}}{1}$}{1}{-2}{right}

        \draw [->,blue] (AS1s) -- (BR1e) node [near start,fill=white,inner sep=1pt] {$\channel{c}{0}$};
        \draw [->,red] (BR1s) edge [bend left] node [] {} (AS1e);

        \draw [->] (AS1e) edge [bend left] node [] {} (BR1s);
      \end{tikzpicture}
      \caption{\label{fig:rendezvous sequential execution with events with duration}The sequential execution with events with duration.}
    }
  \end{subfigure}\quad
  \begin{subfigure}[t]{.75\textwidth}
    {\centering
      \begin{tikzpicture}[]
        \node(logic) at (0,1)  {$\transmit{\party{A}}{\channel{c}{0}}{1}{\party{B}}$};
        \node(A) at (-1,0)  {$\party{A}$};
        \node(B) at (1,0)  {$\party{B}$};
        \draw[line width=0.1mm] (-1.5,-0.5) -- (1.5,-0.5);

        \pop{AS1}{$\Ep{\party{A}}{1}$}{-1}{-2}{left}
        \pop{AE}{$\Ep{\party{A}}{E}$}{-1}{-4}{left}
        \pop{BE}{$\Ep{\party{B}}{E}$}{1}{-1}{right}
        \pop{BR1}{$\Ep{\party{B}}{1}$}{1}{-3}{right}

        \draw [->] (AS1) -- node [] {} (AE);
        \draw [->] (BE) -- node [] {} (BR1);

        \draw [->,blue] (AS1) -- (BR1) node [near start,fill=white,inner sep=1pt] {$\channel{c}{0}$};

        \draw [->] (BE) -- node [](BEtoAS1) {} (AS1);
        \draw [-Implies,double distance=5pt,shorten <=7pt] (BR1) -- (BEtoAS1) node [below] {};

        \draw [->] (BR1) -- node [](BR1toAE) {} (AE);
        \draw [-Implies,double distance=5pt,shorten <=7pt] (AS1) -- (BR1toAE) node [above] {};
      \end{tikzpicture}
      \caption{\label{fig:rendezvous sequential execution with atomic events}The sequential execution with atomic events. The semantic conditions are shown as double arrows. A double arrow starts from a node (corresponding to the antecedent of the semantic condition, which is a conjunction of the two arrows (orders) leading in and out of the node), and ends at an arrow (corresponding to the consequent of the semantic condition). In this figure, the upward double arrow represents \eqref{eq:bef-recvs are before-sends}, and the downward double arrow represents \eqref{eq:aft-sends are aft-recvs}.}
    }
  \end{subfigure}
  \caption{\label{fig:rendezvous sequential execution with events with duration and with atomic events}Sequential executions of the rendezvous execution \ref{fig:rendezvous with duration execution}, one with events with duration, and one with atomic events.}
\end{figure}

To conclude the section, a concurrent execution of a rendezvous can be simulated as a sequential execution, if the sequential execution satisfies certain semantic conditions.
For the rest of the paper, any sequential execution of a rendezvous satisfies the semantic conditions.
We have only considered rendezvous, but similar considerations of semantics can be made for e.g. global protocols that communicate over buffered channels.

\subsection{Race-free Sequential Executions} \label{sec:Race-free Sequential Executions}
\begin{figure}
  \begin{subfigure}[t]{.45\textwidth}
    {\centering
      \begin{tikzpicture}[]
        \node(logic) at (0,1)  {$\transmit{\party{A}}{\Channel}{1}{\party{B}} \ssep \transmit{\party{B}}{\Channel}{2}{\party{C}}$};
        \node(A) at (-2,0)  {$\party{A}$};
        \node(B) at (0,0)  {$\party{B}$};
        \node(C) at (2,0)  {$\party{C}$};
        \draw[line width=0.1mm] (-2.5,-0.5) -- (2.5,-0.5);

        \pop{AS1}{$\Ep{\party{A}}{1}$}{-2}{-1}{left}
        \pop{BR1}{$\Ep{\party{B}}{1}$}{0}{-2}{right}
        \pop{BS1}{$\Ep{\party{B}}{2}$}{0}{-3}{left}
        \pop{CR1}{$\Ep{\party{C}}{2}$}{2}{-4}{right}

        \draw [->] (BR1) -- (BS1) node [] {};

        \draw [->,blue] (AS1) -- (BR1) node [near start,fill=white,inner sep=1pt] {$\Channel$};
        \draw [->,blue] (BS1) -- (CR1) node [near start,fill=white,inner sep=1pt] {$\Channel$};
      \end{tikzpicture}
      \caption{The race-free sequential execution.}
      \label{fig:simple global protocol sequential execution race-free}
    }
  \end{subfigure}\quad
  \begin{subfigure}[t]{.45\textwidth}
    {\centering
      \begin{tikzpicture}[]
        \node(logic) at (0,1)  {$\transmit{\party{A}}{\Channel}{1}{\party{B}} \ssep \transmit{\party{B}}{\Channel}{2}{\party{C}}$};
        \node(A) at (-2,0)  {$\party{A}$};
        \node(B) at (0,0)  {$\party{B}$};
        \node(C) at (2,0)  {$\party{C}$};
        \draw[line width=0.1mm] (-2.5,-0.5) -- (2.5,-0.5);

        \pop{AS1}{$\Ep{\party{A}}{1}$}{-2}{-1}{left}
        \pop{BR1}{$\Ep{\party{B}}{1}$}{0}{-3}{right}
        \pop{BS1}{$\Ep{\party{B}}{2}$}{0}{-4}{left}
        \pop{CR1}{$\Ep{\party{C}}{2}$}{2}{-2}{right}

        \draw [->] (BR1) -- (BS1) node [] {};

        \draw [->,blue] (AS1) -- (CR1) node [near start,fill=white,inner sep=1pt] {$\Channel$};
        \draw [->] (CR1) -- (BR1) node [] {};
      \end{tikzpicture}
      \caption{The sequential execution that has races.}
      \label{fig:simple global protocol sequential execution with races}
    }
  \end{subfigure}
  \caption{Sequential executions of global protocol \ref{fig:simple global protocol}, one race-free, and one with races.}
  \label{fig:simple global protocol sequential execution race-free and with races}
\end{figure}
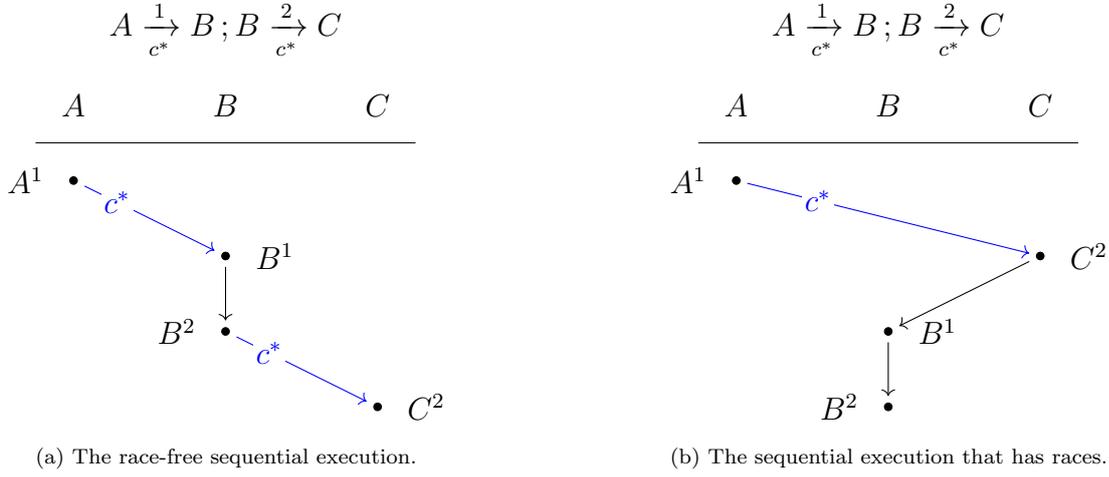

As another example of a sequential execution, global protocol \ref{fig:simple global protocol} can have the sequential execution \ref{fig:simple global protocol sequential execution race-free}.
This sequential execution is race-free, because:
\begin{enumerate}
\item Each receiver receives data from some sender.
\item Each receiver receives data from the correct sender, as specified by the global protocol.
\end{enumerate}
Intuitively, a race-free sequential execution communicates according to the global protocol.

The global protocol may also have a sequential execution with races \ref{fig:simple global protocol sequential execution with races}, such that $\party{C}$ receives data from $\party{A}$ instead of $\party{B}$ receiving data from $\party{A}$, and $\party{B}$ does not receive any data at all.

Combining these observations, we can conjecture this:
\begin{conj}
If the linear order of a sequential execution is a linearization of the order that is the union of the $\HB{}$ order of the global protocol \textit{and the $\CB{}$ order of the global protocol}, then the sequential execution is race-free. Otherwise, the sequential execution has races. \label{conj:sequential execution global protocol equivalence has races conjecture}
\end{conj}
Note that by the definition in section \ref{sec:Sequential Executions}, a sequential execution is already a linearization of the $\HB{}$ order of the global protocol.
Conjecture \ref{conj:sequential execution global protocol equivalence has races conjecture} says that if the linear order of the sequential execution furthermore contains the $\CB{}$ order of the global protocol, then the sequential execution is race-free.
Intuitively, a race-free sequential execution executes communications according to the global protocol.

\section{Verifications, Implementations, and Guards} \label{sec:Verifications, Implementations, and Guards}
We have described in section \ref{sec:Program Order, Events, Sequential Consistency, and Parties} how events are defined from a line of code in programs, and we use events to define global protocols in section \ref{sec:Global Protocols}.
We now describe how to verify if a given program, which is called an ``implementation'', satisfies a given global protocol.
We consider only verifications of implementations.

We now define guards.
A guard $\guardP{\Psi}$ is a proof obligation of proposition $\Psi$.
In other words, a guard is an abstraction over how proposition $\Psi$ is proven.
Only protocols may have guards.
If a guard is proven, then the protocol knows that proposition $\Psi$ is true.
Only verifications may prove guards.
A verification proves a guard $\guardP{\Psi}$ only if the verification can deduce via some proof rules that $\Psi$ is true.
An implementation satisfies a global protocol only if the implementation communicates according to the global protocol, and if the verification proves all the guards in the global protocol.

In global protocols, guards are sequentially composed like so: $\GlobalProtocol \ssep \guardP{\Psi} \ssep \GlobalProtocol'$, where $\GlobalProtocol$ is the global protocol before the guard, and $\GlobalProtocol'$ is the global protocol after the guard.
An implementation is verified for satisfaction of the global protocol.
We now describe each part of such a verification.
\begin{description}
\item[$\GlobalProtocol$] If $\GlobalProtocol$ does not contain another $\guardP{\Psi}$, then verification of $\GlobalProtocol$ proceeds without knowing whether $\Psi$ is true.
  This is because the sequential composition operator $\ssep$ requires that the verification verifies the global protocol $\GlobalProtocol$ before $\Psi$ is proven.
  If the implementation fails to satisfy $\GlobalProtocol$, then the verification terminates with failure.
  Otherwise, the verification continues, checking if the next part of the implementation satisfies the next part of the global protocol.
\item[$\guardP{\Psi}$]
  If the verification fails to prove the guard, then the verification terminates with failure.
  Otherwise, the verification continues, checking if the next part of the implementation satisfies the next part of the global protocol.
\item[$\GlobalProtocol'$] If verification reaches this part of the global protocol, then verification of the implementation against $\GlobalProtocol'$ proceeds with the knowledge that $\Psi$ is true.
  This is because $\guardP{\Psi}$ was already proven true by the verification, and the sequential composition operator $\ssep$ requires that the verification prove $\Psi$ before the rest of the global protocol $\GlobalProtocol'$ is verified.
\end{description}
For example, the global protocol \ref{fig:simple global protocol with guards} has a global protocol $\transmit{\party{A}}{\Channel}{1}{\party{B}}$ before the guard, the guard $\guardP{\Ep{\party{B}}{1} \HB{} \Ep{\party{D}}{2}}$, and a global protocol $\transmit{\party{D}}{\Channel}{2}{\party{C}}$ after the guard.

\subsection{Guards of $\HB{}$}
In this work, we consider only guards $\guardP{\Psi}$ where $\Psi$ is an assertion that an element is in $\HB{}$.
These guards have the form $\guardP{\E{} \HB{} \E{'}}$.
A verification proves $\guardP{\E{} \HB{} \E{'}}$ only if the verification proves that $\E{} \HB{} \E{'}$ indeed exists in $\HB{}$, which in turn only occurs if the implementation is implemented in a certain way.

There are at least two ways that an implementation can make a verification prove guards $\guardP{\E{} \HB{} \E{'}}$:
\begin{enumerate}
\item The program order of a party shows that $\guardP{\E{} \HB{} \E{'}}$ is true, where $\E{}$ and $\E{'}$ are events of the same party
\item Synchronization. Although there are many synchronization primitives in Go, for the rest of this work we will only consider synchronization by sending and receiving on an unbuffered channel.
\end{enumerate}
For example, the guard $\guardP{\Ep{\party{B}}{1} \HB{} \Ep{\party{D}}{2}}$ in the global protocol \ref{fig:simple global protocol with guards} may be proved through synchronizations by an implementation.

\subsection{Transforming Global Protocols with Races into Race-free Global Protocols} \label{sec:Transforming Global Protocols with Races into Race-free Global Protocols}
Guards can be used to prove arbitrary properties.
In this section, we use guards to prove that a global protocol is FIFO race-free.

A global protocol with races can be transformed into a FIFO race-free global protocol.
We first descriptively specify the transformation as follows:
\begin{enumerate}
\item Linearly order sends by $\HB{}$, so that there is only one way data can be written to the channel. We denote the linear order on sends by $<_S$. The linear order $<_S$ must not violate the program order of any party. The proof that the sends are linearly ordered is deferred as guards.
\item Ensure that channels are FIFO, so that data written to the channel is read in the same order from the channel. Note that FIFO channels do not fix the order of sends or receives.
\item Linearly order receives by $\HB{}$, so that there is only one way data can be read from the channel. We denote the linear order on receives by $<_R$. The linear order $<_R$ must not violate the program order of any party. The proof that the receives are linearly ordered is deferred as guards.
\end{enumerate}

Furthermore, the sending order is required to be equivalent to the receiving order. This requirement is rephrased as follows.
Recall that given a global protocol, for each transmission $\transmit{\E{S}}{\Channel}{i}{\E{R}}$ in that global protocol, $\CB{\Channel}$ relates its sender $\E{S}$ to its receiver $\E{R}$.
We define the left operation $\receiveOfSendP{\E{S}} = \E{R}$.
The requirement is that $\receiveOfSendP{\cdot}$ is order-preserving, given by rule \eqref{eq:CB is isotone}.

\begin{subequations}
  \begin{align}
    \E{S_1} <_S \E{S_2} \implies \receiveOfSendP{\E{S_1}} <_R \receiveOfSendP{\E{S_2}} \label{eq:CB is isotone} \tag{$\receiveOfSendP{\cdot}$ is isotone}
  \end{align}
\end{subequations}

This way of ensuring that access to the endpoint of a shared channel is linear is similar to the statement ``Programs that modify data being simultaneously accessed by multiple goroutines must serialize such access.'' \parencite{go/ref/mem}, as well as the approach of \parencite{adve1991detecting} to avoid data races for sequentially consistent executions.

We now operationally specify the transformation as follows.
Given a global protocol, e.g. global protocol \ref{fig:simple global protocol}, do the following for each pair of transmissions ordered by the $\ssep$ operator.\footnote{Recall from footnote \ref{foot:formal two transmissions ordered sequentially} that the formal definition of sequential order allows arbitrarily nested transmissions to still be sequentially ordered.}
If a pair of transmissions are on the same channel, e.g. $\Ep{\party{A}}{1} \CB{\Channel} \Ep{\party{B}}{1} \land \Ep{\party{B}}{2} \CB{\Channel} \Ep{\party{C}}{2}$ on channel $\Channel$ in partial order \eqref{eq:simple global protocol partial order}, then we prove that the send of the first transmission happens-before the send of the second transmission, e.g. $\Ep{\party{A}}{1} \HB{} \Ep{\party{B}}{2}$, and we prove that the receive of the first transmission happens-before the receive of the second transmission, e.g. $\Ep{\party{B}}{1} \HB{} \Ep{\party{C}}{2}$.
The proofs of these are abstracted as guards, e.g. $\guardP{\Ep{\party{A}}{1} \HB{} \Ep{\party{B}}{2}}$, and $\guardP{\Ep{\party{B}}{1} \HB{} \Ep{\party{C}}{2}}$.
The guards are inserted by sequential composition into the global protocol between the pair of transmissions.
The guards are inserted between the pair of transmissions, so that verification of the implementation proceeds to the second transmission only if the verification can show that the pair of transmissions alone is FIFO race-free (Intuitively, under the context of verifying FIFO race-freedom, there is no point in verifying a transmission that races. This is because if the verification already knows that the transmission will race, then the verification should immediately terminate with failure). Then, the guards are proved by verification of an implementation, e.g. implementation \ref{lst:simple global protocol guarded implementation}.
If the guards of all such pairs of transmissions are proven by verification, then verification would have shown that $\receiveOfSendP{\cdot}$ is order-preserving for the entirety of the global protocol, and therefore shown that the global protocol is FIFO race-free.

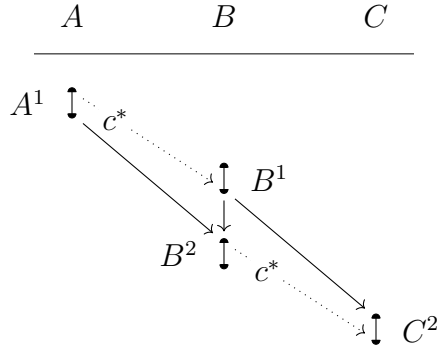
\begin{figure}
  {\centering
    \begin{tikzpicture}[]
      \node(logic) at (0,1)  {$\transmit{\party{A}}{\Channel}{1}{\party{B}} \ssep \guardP{\Ep{\party{A}}{1} \HB{} \Ep{\party{B}}{2}} \ssep \guardP{\Ep{\party{B}}{1} \HB{} \Ep{\party{C}}{2}} \ssep \transmit{\party{B}}{\Channel}{2}{\party{C}}$};
      \node(A) at (-2,0)  {$\party{A}$};
      \node(B) at (0,0)  {$\party{B}$};
      \node(C) at (2,0)  {$\party{C}$};
      \draw[line width=0.1mm] (-2.5,-0.5) -- (2.5,-0.5);

      \dop{AS1}{$\Ep{\party{A}}{1}$}{-2}{-1}{left}
      \dop{BR1}{$\Ep{\party{B}}{1}$}{0}{-2}{right}
      \dop{BS1}{$\Ep{\party{B}}{2}$}{0}{-3}{left}
      \dop{CR1}{$\Ep{\party{C}}{2}$}{2}{-4}{right}

      \draw [->] (AS1e) -- (BS1s) node [] {};
      \draw [->] (BR1e) -- (CR1s) node [] {};

      \draw [->] (BR1e) -- (BS1s) node [] {};

      \draw [->,dotted] (AS1s) -- (BR1e) node [near start,fill=white,inner sep=1pt] {$\Channel$};
      \draw [->,dotted] (BS1s) -- (CR1e) node [near start,fill=white,inner sep=1pt] {$\Channel$};
    \end{tikzpicture}
    \caption{\label{fig:simple global protocol guarded}The guarded global protocol converted from the global protocol with races \ref{fig:simple global protocol}, such that this protocol guarantees race-freedom.}
  }
\end{figure}

If the global protocol designer deems the global protocol with races \ref{fig:simple global protocol} to be a design flaw, then the designer can use our transformation to automatically obtain a FIFO race-free global protocol \ref{fig:simple global protocol guarded}.

The global protocol designer may insert arbitrary guards into a global protocol.
Despite this, $\HB{}$ remains a strict partial order.
It is impossible to prove that $\HB{}$ elements exist such that $\HB{}$ is not a strict partial order.
For example, $\guardP{\Ep{\party{C}}{2} \HB{} \Ep{\party{B}}{1}}$ may be inserted into global protocol \ref{fig:simple global protocol guarded} to get global protocol \eqref{eq:simple global protocol guarded cycle}.

\begin{subequations}
  \begin{align}
    \transmit{\party{A}}{\Channel}{1}{\party{B}} \ssep \guardP{\Ep{\party{A}}{1} \HB{} \Ep{\party{B}}{2}} \ssep \guardP{\Ep{\party{B}}{1} \HB{} \Ep{\party{C}}{2}} \ssep \transmit{\party{B}}{\Channel}{2}{\party{C}} \ssep \guardP{\Ep{\party{C}}{2} \HB{} \Ep{\party{B}}{1}} \label{eq:simple global protocol guarded cycle}
  \end{align}
\end{subequations}

This new global protocol has the cycle consisting of two elements, $\Ep{\party{B}}{1} \HB{} \Ep{\party{C}}{2}$ and $\Ep{\party{C}}{2} \HB{} \Ep{\party{B}}{1}$.
By transitivity, we have $\Ep{\party{B}}{1} \HB{} \Ep{\party{B}}{1}$ and $\Ep{\party{C}}{2} \HB{} \Ep{\party{C}}{2}$, violating asymmetry.
Such global protocols with cycles are called ``deadlocked global protocols'', because all implementations of such global protocols must deadlock, and all verifications must fail to prove some guard.

\subsection{Concurrent Executions and Order between Sends and Receives} \label{sec:Concurrent Executions and Order between Sends and Receives}
\begin{figure}
  \begin{subfigure}[t]{\textwidth}
    {\centering
      \begin{tikzpicture}[]
        \node(logic) at (0.5,1)  {$\transmit{\party{A}}{\Channel}{1}{\party{B}} \ssep \guardP{\Ep{\party{A}}{1} \HB{} \Ep{\party{B}}{2}} \ssep \guardP{\Ep{\party{B}}{1} \HB{} \Ep{\party{C}}{2}} \ssep \transmit{\party{B}}{\Channel}{2}{\party{C}} \ssep \guardP{\Ep{\party{B}}{2} \HB{} \Ep{\party{C}}{3}} \ssep \guardP{\Ep{\party{C}}{2} \HB{} \Ep{\party{A}}{3}} \ssep \transmit{\party{C}}{\Channel}{2}{\party{A}}$};
      \end{tikzpicture}
    }
  \end{subfigure}

  \begin{subfigure}[t]{.3\textwidth}
    {\centering
      \begin{tikzpicture}[]
        \node(A) at (-2,0)  {$\party{A}$};
        \node(B) at (0,0)  {$\party{B}$};
        \node(C) at (2,0)  {$\party{C}$};
        \draw[line width=0.1mm] (-2.5,-0.5) -- (2.5,-0.5);

        \dop{AS1}{$\Ep{\party{A}}{1}$}{-2}{-1}{left}
        \dop{BR1}{$\Ep{\party{B}}{1}$}{0}{-2}{right}
        \dop{BS1}{$\Ep{\party{B}}{2}$}{0}{-3}{left}
        \dop{CR1}{$\Ep{\party{C}}{1}$}{2}{-4}{right}
        \dop{CS1}{$\Ep{\party{C}}{2}$}{2}{-5}{right}
        \dop{AR1}{$\Ep{\party{A}}{2}$}{-2}{-6}{left}
        \dop{BS2}{$\Ep{\party{B}}{3}$}{0}{-7}{right}

        \draw [->] (AS1e) -- (BS1s) node [] {};
        \draw [->] (BS1e) -- (CS1s) node [] {};
        \draw [->] (CS1e) -- (BS2s) node [] {};
        \draw [->] (BR1e) -- (CR1s) node [] {};
        \draw [->] (CR1e) -- (AR1s) node [] {};

        \draw [->] (AS1e) -- (AR1s) node [] {};
        \draw [->] (BR1e) -- (BS1s) node [] {};
        \draw [->] (BS1e) -- (BS2s) node [] {};
        \draw [->] (CR1e) -- (CS1s) node [] {};

        \draw [->,dotted] (AS1s) -- (BR1e) node [near start,fill=white,inner sep=1pt] {$\channel{c}{2}$};
        \draw [->,dotted] (BS1s) -- (CR1e) node [near start,fill=white,inner sep=1pt] {$\channel{c}{2}$};
        \draw [->,dotted] (CS1s) -- (AR1e) node [near start,fill=white,inner sep=1pt] {$\channel{c}{2}$};
      \end{tikzpicture}
      \caption{\label{fig:send receive rules global protocol}The global protocol.}
    }
  \end{subfigure} \quad
  \begin{subfigure}[t]{.3\textwidth}
    {\centering
      \begin{tikzpicture}[]
        \node(A) at (-2,0)  {$\party{A}$};
        \node(B) at (0,0)  {$\party{B}$};
        \node(C) at (2,0)  {$\party{C}$};
        \draw[line width=0.1mm] (-2.5,-0.5) -- (2.5,-0.5);

        \dop{AS1}{$\Ep{\party{A}}{1}$}{-2}{-1}{left}
        \dop{BR1}{$\Ep{\party{B}}{1}$}{0}{-2}{right}
        \dop{BS1}{$\Ep{\party{B}}{2}$}{0}{-3}{left}
        \dop{CR1}{$\Ep{\party{C}}{1}$}{2}{-4}{right}
        \dop{CS1}{$\Ep{\party{C}}{2}$}{2}{-5}{right}
        \dop{AR1}{$\Ep{\party{A}}{2}$}{-2}{-6}{left}
        \dop{BS2}{$\Ep{\party{B}}{3}$}{0}{-7}{right}

        \draw [->] (AS1e) -- (BS1s) node [] {};
        \draw [->] (BS1e) -- (CS1s) node [] {};
        \draw [->] (CS1e) -- (BS2s) node [] {};
        \draw [->] (BR1e) -- (CR1s) node [] {};
        \draw [->] (CR1e) -- (AR1s) node [] {};

        \draw [->] (AS1e) -- (AR1s) node [] {};
        \draw [->] (BR1e) -- (BS1s) node [] {};
        \draw [->] (BS1e) -- (BS2s) node [] {};
        \draw [->] (CR1e) -- (CS1s) node [] {};

        \draw [->,blue] (AS1s) -- (BR1e) node [near start,fill=white,inner sep=1pt] {$\channel{c}{2}$};
        \draw [->,blue] (BS1s) -- (CR1e) node [near start,fill=white,inner sep=1pt] {$\channel{c}{2}$};
        \draw [->,blue] (CS1s) -- (AR1e) node [near start,fill=white,inner sep=1pt] {$\channel{c}{2}$};
        \draw [->,red] (BR1s) -- (CS1e) node [] {};
        \draw [->,red] (CR1s) -- (BS2e) node [] {};
      \end{tikzpicture}
      \caption{\label{fig:send receive rules happens-before}The untidy concurrent execution of the global protocol with the send-receive partial order, where the shared channel has buffer capacity $2$. Blue arrows correspond to $\HB{}$ elements given by rule \eqref{eq:send before receive}. Red arrows correspond to $\HB{}$ elements given by rule \eqref{eq:receive before send}.}
    }
  \end{subfigure} \quad
  \begin{subfigure}[t]{.3\textwidth}
    {\centering
      \begin{tikzpicture}[]
        \node(Ss) at (-1,0)  {Sends};
        \node(Rs) at (1,0)  {Receives};
        \draw[line width=0.1mm] (-1.5,-0.5) -- (1.5,-0.5);

        \dop{SsAS1}{$\Ep{\party{A}}{1}$}{-1}{-1}{left}
        \dop{RsBR1}{$\Ep{\party{B}}{1}$}{1}{-2}{right}
        \dop{SsBS1}{$\Ep{\party{B}}{2}$}{-1}{-3}{left}
        \dop{RsCR1}{$\Ep{\party{C}}{1}$}{1}{-4}{right}
        \dop{SsCS1}{$\Ep{\party{C}}{2}$}{-1}{-5}{left}
        \dop{RsAR1}{$\Ep{\party{A}}{2}$}{1}{-6}{right}
        \dop{SsBS2}{$\Ep{\party{B}}{3}$}{-1}{-7}{left}

        \draw [->] (SsAS1e) -- (SsBS1s) node [] {};
        \draw [->] (SsBS1e) -- (SsCS1s) node [] {};
        \draw [->] (SsCS1e) -- (SsBS2s) node [] {};
        \draw [->] (RsBR1e) -- (RsCR1s) node [] {};
        \draw [->] (RsCR1e) -- (RsAR1s) node [] {};

        \draw [->] (SsAS1e) -- (RsAR1s) node [] {};
        \draw [->] (RsBR1e) -- (SsBS1s) node [] {};
        \draw [->] (SsBS1e) -- (SsBS2s) node [] {};
        \draw [->] (RsCR1e) -- (SsCS1s) node [] {};

        \draw [->,blue] (SsAS1s) -- (RsBR1e) node [near start,fill=white,inner sep=1pt] {$\channel{c}{2}$};
        \draw [->,blue] (SsBS1s) -- (RsCR1e) node [near start,fill=white,inner sep=1pt] {$\channel{c}{2}$};
        \draw [->,blue] (SsCS1s) -- (RsAR1e) node [near start,fill=white,inner sep=1pt] {$\channel{c}{2}$};
        \draw [->,red] (RsBR1s) -- (SsCS1e) node [] {};
        \draw [->,red] (RsCR1s) -- (SsBS2e) node [] {};
      \end{tikzpicture}
      \caption{\label{fig:send receive rules happens-before tidy}The same concurrent execution of the global protocol with the send-receive partial order. Instead of parties being $\party{A}$, $\party{B}$, and $\party{C}$ like in \ref{fig:send receive rules happens-before}, we treat the collective of sends as a unit, and the collective of receives as a unit, and draw a sequence diagram with these collectives.}
    }
  \end{subfigure}
  \caption{\label{fig:send receive rules global protocol and concurrent executions}A global protocol and its concurrent execution with the send-receive partial order that subsumes all possible sequential executions. One concurrent execution figure is untidy and the other one is tidy. To avoid repetition, we write out the global protocol only once, above all three figures.}
\end{figure}
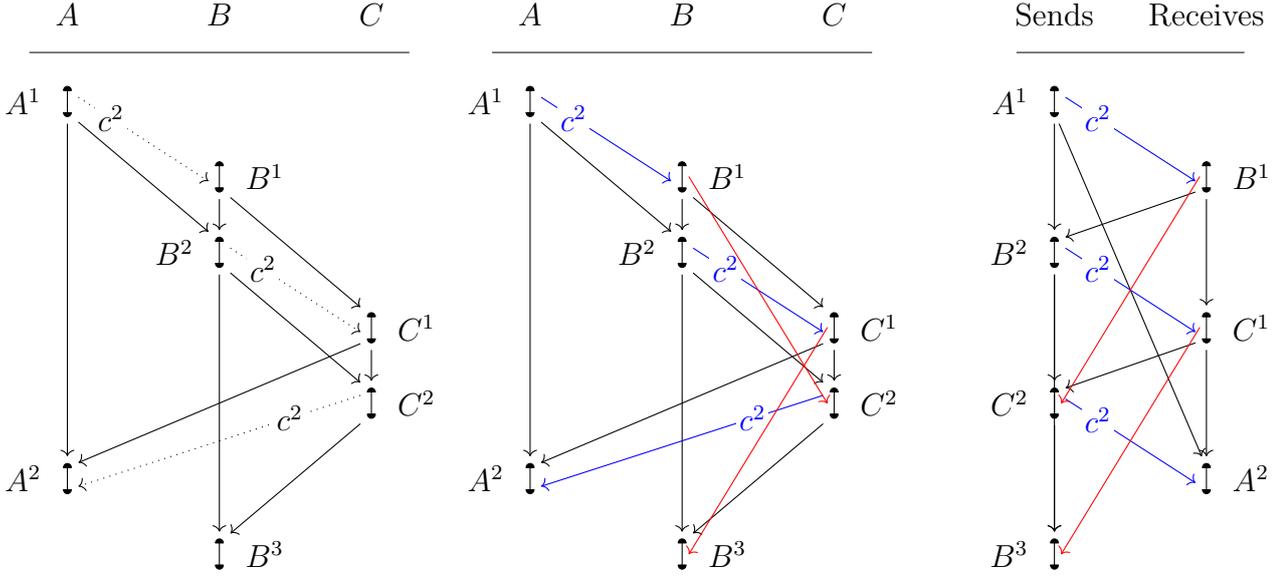

A race-free global protocol has linearized concurrent sends and linearized concurrent receives.
However, there is still more than one way to linearize the events consisting of both sends and receives.
For example, the global protocol \ref{fig:send receive rules global protocol} has three concurrent sends and three concurrent receives on a channel $\channel{c}{2}$ with buffer capacity $2$.
The following sequential executions of sends and receives of the global protocol are possible:
\begin{enumerate}
\item Two sends, then one receive, then one send, then two receives.
\item One send, then one receive, then two sends, then two receives.
\end{enumerate}
Instead of considering each possible sequential execution of sends and receives, we can consider only the concurrent execution with a partial $\HB{}$ order that subsumes all these sequential executions.
This is simply the $\HB{}$ given by the rules \eqref{eq:send before receive} and \eqref{eq:receive before send}.
The resulting $\HB{}$ order is shown in a concurrent execution \ref{fig:send receive rules happens-before}.
This figure can be made more tidy, as in figure \ref{fig:send receive rules happens-before tidy}.

Note that although this execution is concurrent, all sequential executions subsumed by this concurrent execution are race-free by conjecture \ref{conj:sequential execution global protocol equivalence has races conjecture}, therefore we also consider this concurrent execution to be race-free.

In previous sections, we have drawn figures that show the rules \eqref{eq:send before receive} and \eqref{eq:receive before send}, but those figures had either buffers of capacity $0$, or buffers of unspecified capacity.
There, the red arrows that correspond to $\HB{}$ elements given by rule \eqref{eq:receive before send} either point the $i^{th}$ receive to the $i^{th}$ send, or the red arrows do not exist at all.
This is the first time we draw figures that have a specific positive buffer capacity.
The red arrows in these figures showcase how the buffer capacity $k$ in rule \eqref{eq:receive before send} manifests itself in a $\HB{}$ partial order.

\section{Formal Verification} \label{sec:Formal Verification}
\begin{figure}
  \begin{subfigure}[t]{.45\textwidth}
    {\centering
      \begin{tikzpicture}[]
        \node(logic) at (0.5,1)  {$\transmit{\party{A}}{\Channel}{1}{\party{B}} \ssep \transmit{\party{A}}{\channel{d}{0}}{1.3}{\party{B}} \ssep \transmit{\party{B}}{\channel{e}{0}}{1.6}{\party{C}} \ssep \transmit{\party{B}}{\Channel}{2}{\party{C}}$};
        \node(A) at (-2,0)  {$\party{A}$};
        \node(B) at (0,0)  {$\party{B}$};
        \node(C) at (3,0)  {$\party{C}$};
        \draw[line width=0.1mm] (-2.5,-0.5) -- (3.5,-0.5);

        \dop{AS1}{$\Ep{\party{A}}{1}$}{-2}{-1}{left}
        \dop{AS2}{$\Ep{\party{A}}{1.3}$}{-2}{-3}{left}
        \dop{BR1}{$\Ep{\party{B}}{1}$}{0}{-2}{right}
        \dop{BR2}{$\Ep{\party{B}}{1.3}$}{0}{-3}{right}
        \dop{BS1}{$\Ep{\party{B}}{1.6}$}{0}{-4}{left}
        \dop{BS2}{$\Ep{\party{B}}{2}$}{0}{-5}{left}
        \dop{CR1}{$\Ep{\party{C}}{1.6}$}{3}{-4}{right}
        \dop{CR2}{$\Ep{\party{C}}{2}$}{3}{-6}{right}

        \draw [->] (AS1e) -- (AS2s) node [] {};
        \draw [->] (BR1e) -- (BR2s) node [] {};
        \draw [->] (BR2e) -- (BS1s) node [] {};
        \draw [->] (BS1e) -- (BS2s) node [] {};
        \draw [->] (CR1e) -- (CR2s) node [] {};

        \draw [->,dotted] (AS1s) -- (BR1e) node [near start,fill=white,inner sep=1pt] {$\Channel$};
        \draw [->,dotted] (AS2s) -- (BR2e) node [near start,fill=white,inner sep=1pt] {$\channel{d}{0}$};
        \draw [->,dotted] (BS1s) -- (CR1e) node [near start,fill=white,inner sep=1pt] {$\channel{e}{0}$};
        \draw [->,dotted] (BS2s) -- (CR2e) node [near start,fill=white,inner sep=1pt] {$\Channel$};
      \end{tikzpicture}
      \caption{\label{fig:simple global protocol race-free}The guarded global protocol \ref{fig:simple global protocol guarded} with all guards replaced by transmissions on different channels $\channel{d}{0}$ and $\channel{e}{0}$.}
    }
  \end{subfigure}\quad
  \begin{subfigure}[t]{.45\textwidth}
    {\centering
      \begin{tikzpicture}[]
        \node(logic) at (0.5,1)  {$\transmit{\party{A}}{\Channel}{1}{\party{B}} \ssep \transmit{\party{A}}{\channel{d}{0}}{1.3}{\party{B}} \ssep \transmit{\party{B}}{\channel{e}{0}}{1.6}{\party{C}} \ssep \transmit{\party{B}}{\Channel}{2}{\party{C}}$};
        \node(A) at (-2,0)  {$\party{A}$};
        \node(B) at (0,0)  {$\party{B}$};
        \node(C) at (3,0)  {$\party{C}$};
        \draw[line width=0.1mm] (-2.5,-0.5) -- (3.5,-0.5);

        \pop{AS1}{$\Ep{\party{A}}{1}$}{-2}{-1}{left}
        \pop{AS2}{$\Ep{\party{A}}{1.3}$}{-2}{-3}{left}
        \pop{BR1}{$\Ep{\party{B}}{1}$}{0}{-2}{right}
        \pop{BR2}{$\Ep{\party{B}}{1.3}$}{0}{-4}{right}
        \pop{BS1}{$\Ep{\party{B}}{1.6}$}{0}{-5}{left}
        \pop{BS2}{$\Ep{\party{B}}{2}$}{0}{-7}{left}
        \pop{CR1}{$\Ep{\party{C}}{1.6}$}{3}{-6}{right}
        \pop{CR2}{$\Ep{\party{C}}{2}$}{3}{-8}{right}

        \draw [->] (AS1) -- (AS2) node [] {};
        \draw [->] (BR1) -- node [] {} (BR2);
        \draw [->] (BR2) -- (BS1) node [] {};
        \draw [->] (BS1) -- node [] {} (BS2);
        \draw [->] (CR1) -- (CR2) node [] {};

        \draw [->] (AS1) -- (BR1) node [near start,fill=white,inner sep=1pt] {$\Channel$};
        \draw [->] (AS2) -- (BR2) node [near start,fill=white,inner sep=1pt] {$\channel{d}{0}$};
        \draw [->] (BS1) -- (CR1) node [near start,fill=white,inner sep=1pt] {$\channel{e}{0}$};
        \draw [->] (BS2) -- (CR2) node [near start,fill=white,inner sep=1pt] {$\Channel$};

        \draw [->] (BR1) -- node [](BR1toAS2) {} (AS2);
        \draw [-Implies,double distance=5pt,shorten <=7pt] (BR2) -- (BR1toAS2) node [below] {};

        \draw [->] (CR1) -- node [](CR1toBS2) {} (BS2);
        \draw [-Implies,double distance=5pt,shorten <=7pt] (BS1) -- (CR1toBS2) node [above] {};
      \end{tikzpicture}
      \caption{\label{fig:simple global protocol race-free sequential execution}The race-free sequential execution of the race-free global protocol \ref{fig:simple global protocol race-free}.}
    }
  \end{subfigure}
  \caption{\label{fig:simple global protocol race-free and its race-free sequential execution}The guarded global protocol \ref{fig:simple global protocol guarded} but race-free, and its race-free sequential execution.}
\end{figure}
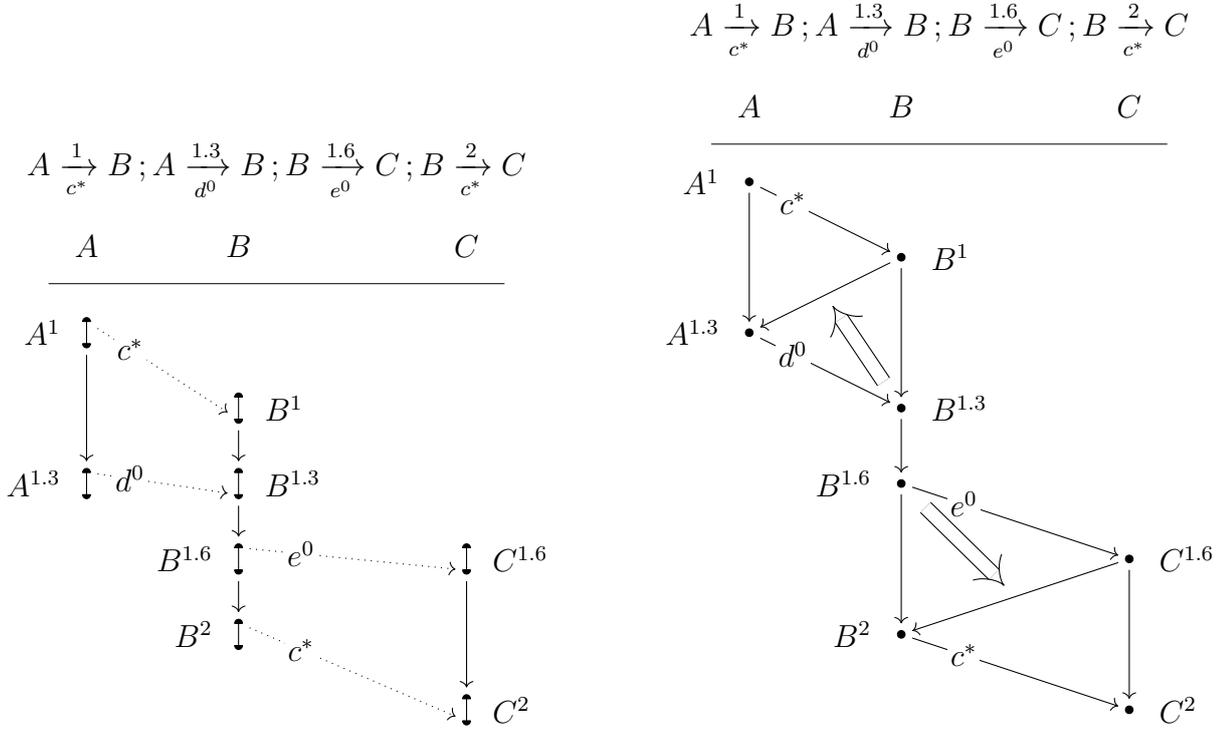

To easier understand why implementation \ref{lst:simple global protocol guarded implementation} satisfies global protocol \ref{fig:simple global protocol guarded}, we now describe a new global protocol \ref{fig:simple global protocol race-free} that has the same behavior as the implementation.\footnote{The sequence diagram of a global protocol also aids in understanding.}
We emphasize that only verifications can prove that $\HB{}$ elements exist, and global protocols cannot prove that $\HB{}$ elements exist.
We also emphasize that this new global protocol is not a part of our verification framework. This new global protocol is described only for easier understanding.

The new global protocol \ref{fig:simple global protocol race-free} is the same as the old global protocol \ref{fig:simple global protocol guarded}, but with a transmission $\Ep{\party{A}}{1.3} \CB{\channel{d}{0}} \Ep{\party{B}}{1.3}$ on channel $\channel{d}{0}$ inserted after event $\Ep{\party{A}}{1}$ and before event $\Ep{\party{B}}{2}$.
We show that the verification of the implementation that corresponds to this new protocol proves the guard $\guardP{\Ep{\party{A}}{1} \HB{} \Ep{\party{B}}{2}}$ in the old global protocol.
The inserted transmission means that the program order of $\party{A}$ gives $\Ep{\party{A}}{1} \HB{} \Ep{\party{A}}{1.3}$, and the program order of $\party{B}$ gives $\Ep{\party{B}}{1.3} \HB{} \Ep{\party{B}}{2}$.
As we have said in section \ref{sec:Rendezvous}, the send and the receive on channel $\channel{d}{0}$ must rendezvous, such that the $\HB{}$ order has these elements: $\Ep{\party{A}}{1.3} \HB{} \Ep{\party{B}}{1.3}$, and $\Ep{\party{B}}{1.3} \HB{} \Ep{\party{A}}{1.3}$.
Applying transitivity of $\HB{}$ to three elements $\Ep{\party{A}}{1} \HB{} \Ep{\party{A}}{1.3}$, $\Ep{\party{A}}{1.3} \HB{} \Ep{\party{B}}{1.3}$, and $\Ep{\party{B}}{1.3} \HB{} \Ep{\party{B}}{2}$, gives the desired element $\Ep{\party{A}}{1} \HB{} \Ep{\party{B}}{2}$, proving the guard.
A similar process inserts the new transmission $\Ep{\party{B}}{1.6} \CB{\channel{e}{0}} \Ep{\party{C}}{1.6}$ on channel $\channel{e}{0}$ (alternatively, transmission $\Ep{\party{B}}{1.16} \CB{\channel{e}{0}} \Ep{\party{C}}{1.16}$ can be inserted instead), applies transitivity of $\HB{}$, then concludes that the verification of the implementation that corresponds to this new global protocol proves the guard $\guardP{\Ep{\party{B}}{1} \HB{} \Ep{\party{C}}{2}}$.

Note that the unbuffered channels $\channel{d}{0}$ and $\channel{e}{0}$ must be unique, by requirement \ref{req:there are no other events on an unbuffered channel}.

To easier understand how the new global protocol \ref{fig:simple global protocol race-free} executes, we show a race-free sequential execution \ref{fig:simple global protocol race-free sequential execution} of this global protocol.

\subsection{Projections}
In this section, we show how to obtain local views of a given global protocol.
We do so by defining two projection functions to obtain ``per-party protocols'', and ``per-endpoint protocols''.
In a party, all channels actually refer to the channel endpoint that the party communicates on.
The resulting projected protocols guide the verification of an implementation.
We focus on applying our framework to verify that a Go program is FIFO race-free.

A given implementation is verified as follows.
A given global protocol is projected onto each party of the global protocol, to get per-party protocols.
Then, each party of the implementation is verified against the corresponding per-party protocol.
Each party of the implementation satisfies its corresponding per-party protocol only if its verification succeeds.
An implementation satisfies the global protocol only if all parties of the implementation satisfies their corresponding per-party protocol.

The syntax of per-party protocols \ref{fig:per-party protocol syntax} is somewhat familiar. We now explain the unfamiliar syntax:
\begin{description}
\item[$\Ep{\Party}{i} \CB{\Channel}$] As part of transmission $i$, party $\Party$ sends on endpoint $\Channel$.
\item[$\CB{\Channel} \Ep{\Party}{i}$] As part of transmission $i$, party $\Party$ receives on endpoint $\Channel$.
\item[$\guardP{\channel{d}{\anything}, \Ep{\Party}{i}}$] Endpoint $\channel{d}{\anything}$ has the burden of proving proposition $\Ep{\Party}{i}$.
  We added the endpoint parameter $\channel{d}{\anything}$ to guards so that we can project guards onto endpoints to obtain per-endpoint guards.
  The proposition in this guard is an event.
  The proposition is proved only if the verification of party $\Party$ knows that the event $\Ep{\Party}{i}$ already happened.
\end{description}
Concurrent composition $\asep$ is not defined in the syntax of per-party protocols, because a party is a goroutine that executes sequentially.

We do not use the familiar session type syntax of sends $\Channel!$ and receives $\Channel?$, because we want projection to preserve information about which event $\Ep{\Party}{i}$ occurs in the party.

Projection of a global protocol onto party $\Party$ is the function $\projectOntoP{\cdot}{\Party}$ in figure \ref{fig:global to per-party projection}.
If a transmission $\transmit{\Party'}{\Channel}{i}{\Party''}$ is projected onto party $\Party'$, then the per-party protocol is the send $\Ep{\Party'}{i} \CB{\Channel}$.
If the transmission is instead projected onto party $\Party''$, then the per-party protocol is the receive $\CB{\Channel} \Ep{\Party''}{i}$.
Projection preserves inaction $\emp$.
Projection is defined recursively on sequentially composed and concurrently composed global protocols.

We give special attention to how projection is defined on guards $\guardP{\Ep{\Party'}{i} \HB{} \Ep{\Party''}{j}}$.
Intuitively, projection on a guard inserts a new rendezvous between the two events ordered by $\HB{}$ in the guard, like how two rendezvous were inserted into the old global protocol \ref{fig:simple global protocol guarded} to obtain the new global protocol \ref{fig:simple global protocol race-free}.
In reality, a new rendezvous is not actually inserted into the global protocol.
Rather, the send of the rendezvous is inserted into one per-party protocol, and the receive of the rendezvous is inserted into the other per-party protocol.

If a guard $\guardP{\Ep{\Party'}{i} \HB{} \Ep{\Party''}{j}}$ is projected onto party $\Party'$, then the per-party protocol is the sequential composition of the guard $\guardP{\channel{d}{0}, \Ep{\Party'}{i}}$ and the send $\Ep{\Party'}{k} \CB{\channel{d}{0}}$, for some $\channel{d}{0}$ and some $k > i$.
If the guard is instead projected onto party $\Party''$, then the per-party protocol is the sequential composition of the receive $\CB{\channel{d}{0}} \Ep{\Party''}{h}$ and the guard $\guardP{\channel{d}{0}, \Ep{\Party''}{j}}$, for some $\channel{d}{0}$ and some $h < j$.
The send $\Ep{\Party'}{k} \CB{\channel{d}{0}}$ and the receive $\CB{\channel{d}{0}} \Ep{\Party''}{h}$ of the two projected protocols of a guard must rendezvous on the same unbuffered channel $\channel{d}{0}$.

The reason for defining projection on guards as such is as follows.
Consider an implementation that satisfies both guard-projected per-party protocols $\guardP{\channel{d}{0}, \Ep{\Party'}{i}} \ssep \Ep{\Party'}{k} \CB{\channel{d}{0}}$ and $\CB{\channel{d}{0}} \Ep{\Party''}{h} \ssep \guardP{\channel{d}{0}, \Ep{\Party''}{j}}$.
Our verification approach verifies that the implementation indeed satisfies the guard $\guardP{\Ep{\Party'}{i} \HB{} \Ep{\Party''}{j}}$, by:
\begin{enumerate}
\item Knowing from $\guardP{\channel{d}{0}, \Ep{\Party'}{i}} \ssep \Ep{\Party'}{k} \CB{\channel{d}{0}}$ that if verification reaches $\Ep{\Party'}{k} \CB{\channel{d}{0}}$, then verification of $\Ep{\Party'}{k} \CB{\channel{d}{0}}$ proceeds with the knowledge that $\Ep{\Party'}{i}$ has already happened.
  In other words, our verification knows that $\Ep{\Party'}{i} \HB{} \Ep{\Party'}{k}$.
\item Applying the semantic condition \eqref{eq:bef-sends are before-recvs} (instantiating $\E{}$ with $\Ep{\Party'}{i}$, $\E{S}$ with $\Ep{\Party'}{k}$, $\E{R}$ with $\Ep{\Party''}{h}$, and $\channel{c}{0}$ with $\channel{d}{0}$) to know that $\Ep{\Party'}{i} \HB{} \Ep{\Party''}{h}$.
\item Knowing from $\CB{\channel{d}{0}} \Ep{\Party''}{h} \ssep \guardP{\channel{d}{0}, \Ep{\Party''}{j}}$ that if verification reaches $\guardP{\channel{d}{0}, \Ep{\Party''}{j}}$, then verification of $\guardP{\channel{d}{0}, \Ep{\Party''}{j}}$ proceeds with the knowledge that $\CB{\channel{d}{0}} \Ep{\Party''}{h}$ has already happened.
  In other words, our verification knows that $\Ep{\Party''}{h} \HB{} \Ep{\Party''}{j}$.
\end{enumerate}
All in all, if all per-party protocols are verified with success, then our verification could apply transitivity of $\HB{}$ on the known orders to know that $\Ep{\Party'}{i} \HB{} \Ep{\Party''}{j}$, thus proving the guard $\guardP{\Ep{\Party'}{i} \HB{} \Ep{\Party''}{j}}$.
However, our verification does not need to actually apply transitivity of $\HB{}$, because to know that $\Ep{\Party'}{i} \HB{} \Ep{\Party''}{j}$, it is enough to ensure that both guard-projected per-party protocols are verified with success.
Therefore, applying transitivity of $\HB{}$ is not a verification step.

As an example of projection of a global protocol to per-party protocols, let global protocol \eqref{eq:simple global protocol guarded} be $\GlobalProtocol$.
The projected per-party protocols are as follows:
\begin{subequations}
  \begin{align} 
    \GlobalProtocol &= \transmit{\party{A}}{\Channel}{1}{\party{B}} &&\ssep \guardP{\Ep{\party{A}}{1} \HB{} \Ep{\party{B}}{2}} && &&\ssep \guardP{\Ep{\party{B}}{1} \HB{} \Ep{\party{C}}{2}} && &&\ssep \transmit{\party{B}}{\Channel}{2}{\party{C}} \label{eq:simple global protocol guarded} \\
    \projectOntoP{\GlobalProtocol}{\party{A}} &= \PartySend{\Channel}{\Ep{\party{A}}{1}} &&\ssep \guardP{\channel{d}{0}, \Ep{\party{A}}{1}} &&\ssep \PartySend{\channel{d}{0}}{\Ep{\party{A}}{1.3}} &&\ssep \emp && &&\ssep \emp \label{eq:simple global protocol with rendezvous guarded projected-per-party A} \\
    \projectOntoP{\GlobalProtocol}{\party{B}} &= \PartyReceive{\Channel}{\Ep{\party{B}}{1}} &&\ssep \PartyReceive{\channel{d}{0}}{\Ep{\party{B}}{1.3}} &&\ssep \guardP{\channel{d}{0}, \Ep{\party{B}}{2}} &&\ssep \guardP{\channel{e}{0}, \Ep{\party{B}}{1}} &&\ssep \PartySend{\channel{e}{0}}{\Ep{\party{B}}{1.6}} &&\ssep \PartySend{\Channel}{\Ep{\party{B}}{2}} \label{eq:simple global protocol with rendezvous guarded projected-per-party B} \\
    \projectOntoP{\GlobalProtocol}{\party{C}} &= \emp &&\ssep \emp && &&\ssep \PartyReceive{\channel{e}{0}}{\Ep{\party{C}}{1.6}} &&\ssep \guardP{\channel{e}{0}, \Ep{\party{C}}{2}} &&\ssep \PartyReceive{\Channel}{\Ep{\party{C}}{2}} \label{eq:simple global protocol with rendezvous guarded projected-per-party C}
  \end{align}
\end{subequations}

We only defined projection on guards $\guardP{\Psi}$ where $\Psi$ is an assertion that an element is in $\HB{}$.
This limited definition of projection is enough to show FIFO race-freedom.
If guards are generalized to other $\Psi$, then the projection on guards need to be defined for these other $\Psi$.

Although it is possible for implementations to be verified against per-party projections, we want finer-grained projections: we project the per-party protocol of party $A$ onto each endpoint $\Channel, \channel{d}{\anything}, \ldots$ used by $A$.

The syntax of per-endpoint protocols \ref{fig:per-endpoint protocol syntax} is somewhat familiar. We now explain the unfamiliar syntax:
\begin{description}
\item[$\Ep{\Party}{i} \CB{}$] A send event $\Ep{\Party}{i}$ occurs on the endpoint.
\item[$\CB{} \Ep{\Party}{i}$] A receive event $\Ep{\Party}{i}$ occurs on the endpoint.
\item[$\guardP{\Ep{\Party}{i}}$] The endpoint has the burden of proving proposition $\Ep{\Party}{i}$.
  The proposition is proved only if the verification of party $\Party$ knows that the event $\Ep{\Party}{i}$ already happened.
\end{description}
Concurrent composition $\asep$ is not defined in the syntax of per-endpoint protocols, because an endpoint sequentially executes the sends and receives that occur on it.

\begin{figure}
  {\centering
    \begin{subfigure}[t]{.1\textwidth}
      \begin{align*}
        \textit{Protocol} \\
        \vphantom{\transmit{\Party}{\Channel}{i}{\Party'}} \textit{Transmission} \\
        \vphantom{\Ep{\Party}{i} \CB{\Channel}} \textit{Send} \\
        \vphantom{\Ep{\Party}{i} \CB{\Channel}} \textit{Receive} \\
        \vphantom{\guardP{\Ep{\Party}{i} \HB{} \Ep{\Party'}{i}}} \textit{Guard} \\
        \textit{Inaction} \\
        \textit{Sequence} \\
        \textit{Concurrency}
      \end{align*}
    \end{subfigure}
    \begin{subfigure}[t]{.25\textwidth}
      \begin{align*}
      {\quad \GlobalProtocol} {~ ::=} &\\
      &\vphantom{\transmit{\Party}{\Channel}{i}{\Party'}} ~{\,\; \transmit{\Party}{\Channel}{i}{\Party'} \text{where $\Party \neq \Party'$}}  \\
      &\vphantom{\Ep{\Party}{i} \CB{\Channel}} \\
      &\vphantom{\Ep{\Party}{i} \CB{\Channel}} \\
      &\vphantom{\guardP{\Ep{\Party}{i} \HB{} \Ep{\Party'}{i}}} ~|~ {\guardP{\Ep{\Party}{i} \HB{} \Ep{\Party'}{i}}}\\
      & ~|~ {\emp}\\
      & ~|~ {\GlobalProtocol \ssep \GlobalProtocol}\\
      & ~|~ {\GlobalProtocol \asep \GlobalProtocol}
      \end{align*}
      \caption{$\GlobalProtocol$ is a global protocol.}
      \label{fig:global protocol syntax}
    \end{subfigure}
    \begin{subfigure}[t]{.25\textwidth}
      \begin{align*}
      {\quad \PerPartyProtocol} {~ ::=} &\\
      &\vphantom{\transmit{\Party}{\Channel}{i}{\Party'}} \\
      &\vphantom{\Ep{\Party}{i} \CB{\Channel}} ~{\,\; \Ep{\Party}{i} \CB{\Channel}} \\
      &\vphantom{\Ep{\Party}{i} \CB{\Channel}} ~|~ {\CB{\Channel} \Ep{\Party}{i}}  \\
      &\vphantom{\guardP{\Ep{\Party}{i} \HB{} \Ep{\Party'}{i}}} ~|~ {\guardP{\channel{d}{\anything}, \Ep{\Party}{i}}}\\
      & ~|~ {\emp}\\
      & ~|~ {\PerPartyProtocol \ssep \PerPartyProtocol}\\
      &
      \end{align*}
      \caption{$\PerPartyProtocol$ is a per-party protocol.}
      \label{fig:per-party protocol syntax}
    \end{subfigure}
    \begin{subfigure}[t]{.25\textwidth}
      \begin{align*}
      {\quad \PerEndpointProtocol} {~ ::=} &\\
      &\vphantom{\transmit{\Party}{\Channel}{i}{\Party'}} \\
      &\vphantom{\Ep{\Party}{i} \CB{\Channel}} ~{\,\; \Ep{\Party}{i} \CB{}} \\
      &\vphantom{\Ep{\Party}{i} \CB{\Channel}} ~|~ {\CB{} \Ep{\Party}{i}}  \\
      &\vphantom{\guardP{\Ep{\Party}{i} \HB{} \Ep{\Party'}{i}}} ~|~ {\guardP{\Ep{\Party}{i}}}\\
      & ~|~ {\emp}\\
      & ~|~ {\PerEndpointProtocol \ssep \PerEndpointProtocol}\\
      &
      \end{align*}
      \caption{$\PerEndpointProtocol$ is a per-endpoint protocol.}
      \label{fig:per-endpoint protocol syntax}
    \end{subfigure}
    \caption{The syntax of global protocols, per-party protocols, and per-endpoint protocols.}
    \label{fig:projected protocol syntax}
  }
\end{figure}

Projection of a per-party protocol onto endpoint $\Channel$ is the function $\projectOntoP{\cdot}{\Channel}$ in figure \ref{fig:per-party to per-endpoint projection}.
If a send $\Ep{\Party'}{i} \CB{\Channel}$ is projected onto endpoint $\Channel$, then the per-endpoint protocol is the send without endpoint information $\Ep{\Party'}{i} \CB{}$.
If a receive $\CB{\Channel} \Ep{\Party'}{i}$ is projected onto endpoint $\Channel$, then the per-endpoint protocol is the receive without endpoint information $\CB{} \Ep{\Party'}{i}$.
If a guard $\guardP{\channel{d}{\anything}, \Ep{\Party'}{i}}$ is projected onto endpoint $\channel{d}{\anything}$, then the per-endpoint protocol is the guard without endpoint information $\guardP{\Ep{\Party'}{i}}$.
Projection preserves inaction $\emp$.
Projection is defined recursively on sequentially composed and concurrently composed protocols.

As an example of projection of a per-party protocol to per-endpoint protocols, let global protocol \eqref{eq:simple global protocol guarded} be $\GlobalProtocol$.
The projected per-endpoint protocols of the per-party protocol $\projectOntoP{\GlobalProtocol}{\party{A}}$ are as follows:
\begin{subequations}
  \begin{align} 
    \projectOntoP{\GlobalProtocol}{\party{A}} &= \PartySend{\Channel}{\Ep{\party{A}}{1}} &&\ssep \guardP{\channel{d}{0}, \Ep{\party{A}}{1}} &&\ssep \PartySend{\channel{d}{0}}{\Ep{\party{A}}{1.3}} \label{eq:simple global protocol with rendezvous guarded projected-per-party A} \\
    \projectOntoP{\projectOntoP{\GlobalProtocol}{\party{A}}}{\Channel} &= \Ep{\party{A}}{1} \CB{} &&\ssep \emp &&\ssep \emp \label{eq:simple global protocol with rendezvous guarded projected-per-party A projected-per-endpoint c} \\
    \projectOntoP{\projectOntoP{\GlobalProtocol}{\party{A}}}{\channel{d}{0}} &= \emp &&\ssep \guardP{\Ep{\party{A}}{1}} &&\ssep \Ep{\party{A}}{1.3} \CB{} \label{eq:simple global protocol with rendezvous guarded projected-per-party A projected-per-endpoint d}
  \end{align}
\end{subequations}

Now that we have introduced the protocol syntax, we define structural congruence for these protocols.
Structural congruence is the smallest relation $\equiv$ on protocols that includes the relations in figure \ref{fig:protocol structural congruence}.

\begin{figure}
  {\centering
    \begin{subequations}
      \begin{align}
        \text{Protocol} \ssep \emp &\equiv \text{Protocol} \\
        \emp \ssep \text{Protocol} &\equiv \text{Protocol} \\
        \text{Protocol} \asep \emp &\equiv \text{Protocol} \\
        \text{Protocol}_1 \asep \text{Protocol}_2 &\equiv \text{Protocol}_2 \asep \text{Protocol}_1
      \end{align}
    \end{subequations}
    \caption{Protocol structural congruence, where Protocol is either $\GlobalProtocol$, $\PerPartyProtocol$, or $\PerEndpointProtocol$.}
    \label{fig:protocol structural congruence}
  }
\end{figure}

Projection results in concurrent composition of per-party protocols $\projectOntoP{\GlobalProtocol_1}{\Party} \asep \projectOntoP{\GlobalProtocol_2}{\Party}$ and concurrent composition of per-endpoint protocols $\projectOntoP{\GlobalProtocol_1}{\Channel} \asep \projectOntoP{\GlobalProtocol_2}{\Channel}$, even though concurrent composition $\asep$ is not defined in the syntax of both per-party protocols and per-endpoint protocols.
Here, concurrent composition $\asep$ partitions the global protocol into per-party protocols, like how state is partitioned into substates in the parallel composition rule of CSL.
To prove a concurrent composition of parties, we give each party a separate per-party protocol, and separately combine the postconditions for each party.
This separation of per-party protocols allows each party to be verified independently.
Similarly, the separation of per-endpoint protocols allows each endpoint to be verified independently.

Projection is faithful to the communications expressed by the session logic: the interactions of a projected protocol follow the interactions of the unprojected protocol. Projection fidelity was shown in \parencite[Proposition 1]{costea2017session} and is omitted in this work.

The notion of per-endpoint projections has some benefits. Let endpoint $\Channel$ used by party $A$ have the per-endpoint protocol $\PerEndpointProtocol_{\Channel}$. Then, the benefits are as follows:
\begin{enumerate}
\item Party $A$ can delegate execution of $\PerEndpointProtocol_{\Channel}$ to some other party, by sending both the endpoint $\Channel$ and its per-endpoint protocol $\PerEndpointProtocol_{\Channel}$ across another endpoint $\channel{c'}{\anything}$, by using the Hoare triple \eqref{eq:Endpoint Send}.
  Then, the other party receives on endpoint $\channel{c'}{\anything}$, by using the Hoare triple \eqref{eq:Endpoint Receive}.
  This endpoint protocol delegation is mentioned by \textcite{craciun2015towards} and \textcite{costea2017session}.
  This is unlike \citeauthor{10.1145/1328438.1328472}'s session delegation \parencite{10.1145/1328438.1328472}, where their session types use static party information, and the party that the session is delegated to must be specified.
  On the other hand, our session delegation can have many parties race to be delegated.\footnote{But only one party will be delegated the session.}
\item The number of implementations that satisfy the per-endpoint protocols is greater than the number of implementations that satisfy the corresponding per-party protocol.
  For example, the global protocol $\transmit{\party{A}}{\Channel}{1}{\party{B}} \ssep \transmit{\party{A}}{\channel{d}{\anything}}{2}{\party{C}}$ projected onto party $\party{A}$ gives the per-party protocol $\Ep{\party{A}}{1} \CB{\Channel} \ssep \Ep{\party{A}}{2} \CB{\channel{d}{\anything}}$.
  An implementation of party $\party{A}$ satisfying this per-party protocol must send on endpoint $\Channel$ before sending on endpoint $\channel{d}{\anything}$. There is only one possible implementation.
  On the other hand, an implementation of $\party{A}$ may implement the per-endpoint protocols $\Ep{\party{A}}{1} \CB{}$ of endpoint $\Channel$ and $\Ep{\party{A}}{2} \CB{}$ of endpoint $\channel{d}{\anything}$ in either order. There are two possible implementations: one sends on endpoint $\Channel$ before sending on endpoint $\channel{d}{\anything}$, and the other sends on endpoint $\channel{d}{\anything}$ before sending on endpoint $\Channel$.
\end{enumerate}

\begin{figure}
  {\centering
    \begin{subfigure}[t]{\textwidth}
      \begin{subequations}
        \begin{align}
          &\projectOntoP{\transmit{\Party'}{\channel{c'}{\anything}}{i}{\Party''}}{\Party} &&= \begin{cases}
            \Ep{\Party'}{i} \CB{\channel{c'}{\anything}} &\text{if $\Party = \Party'$} \\
            \CB{\channel{c'}{\anything}} \Ep{\Party''}{i} &\text{if $\Party = \Party''$} \\
            \emp &\text{otherwise} \\
          \end{cases} \tag{Transmission} \\
          &\text{For some fixed $\channel{d}{0}$,} \notag \\
          &\projectOntoP{\guardP{\Ep{\Party'}{i} \HB{} \Ep{\Party''}{j}}}{\Party} &&= \begin{cases}
            \guardP{\channel{d}{0}, \Ep{\Party'}{i}} \ssep \Ep{\Party'}{k} \CB{\channel{d}{0}} &\text{if $\Party = \Party'$, for some $k > i$} \\
              \CB{\channel{d}{0}} \Ep{\Party''}{h} \ssep \guardP{\channel{d}{0}, \Ep{\Party''}{j}} &\text{if $\Party = \Party''$, for some $h < j$} \\
                \emp &\text{otherwise} \\
              \end{cases} \tag{Guard} \\
          &\projectOntoP{\emp}{\Party} &&= \emp \tag{Inaction} \\
          &\projectOntoP{\GlobalProtocol_1 \ssep \GlobalProtocol_2}{\Party} &&= \projectOntoP{\GlobalProtocol_1}{\Party} \ssep \projectOntoP{\GlobalProtocol_2}{\Party} \tag{Sequence} \\
          &\projectOntoP{\GlobalProtocol_1 \asep \GlobalProtocol_2}{\Party} &&= \projectOntoP{\GlobalProtocol_1}{\Party} \asep \projectOntoP{\GlobalProtocol_2}{\Party} \tag{Concurrency}
        \end{align}
      \end{subequations}
      \caption{The projection function that projects global protocols onto party $\Party$ to obtain per-party protocols.}
      \label{fig:global to per-party projection}
    \end{subfigure}

    \begin{subfigure}[t]{\textwidth}
      \begin{subequations}
        \begin{align}
          \projectOntoP{\Ep{\Party'}{i} \CB{\channel{c'}{\anything}}}{\Channel} &= \begin{cases}
            \Ep{\Party'}{i} \CB{} &\text{if $\Channel = \channel{c'}{\anything}$} \\
            \emp &\text{otherwise} \\
          \end{cases} \tag{Send} \\
          \projectOntoP{\CB{\channel{c'}{\anything}} \Ep{\Party'}{i}}{\Channel} &= \begin{cases}
            \CB{} \Ep{\Party'}{i} &\text{if $\Channel = \channel{c'}{\anything}$} \\
            \emp &\text{otherwise} \\
          \end{cases} \tag{Receive} \\
          \projectOntoP{\guardP{\channel{d}{\anything}, \Ep{\Party'}{i}}}{\Channel} &= \begin{cases}
            \guardP{\Ep{\Party'}{i}} &\text{if $\Channel = \channel{d}{\anything}$} \\
            \emp &\text{otherwise} \\
          \end{cases} \tag{Guard} \\
          \projectOntoP{\emp}{\Channel} &= \emp \tag{Inaction} \\
          \projectOntoP{\GlobalProtocol_1 \ssep \GlobalProtocol_2}{\Channel} &= \projectOntoP{\GlobalProtocol_1}{\Channel} \ssep \projectOntoP{\GlobalProtocol_2}{\Channel} \tag{Sequence} \\
          \projectOntoP{\GlobalProtocol_1 \asep \GlobalProtocol_2}{\Channel} &= \projectOntoP{\GlobalProtocol_1}{\Channel} \asep \projectOntoP{\GlobalProtocol_2}{\Channel} \tag{Concurrency}
        \end{align}
      \end{subequations}
      \caption{The projection function that projects per-party protocols onto endpoint $\Channel$ to obtain per-endpoint protocols.}
      \label{fig:per-party to per-endpoint projection}
    \end{subfigure}
    \caption{The projection functions.}
    \label{fig:projection functions}
  }
\end{figure}

\subsection{Verifying Implementations}
Now we verify that implementation \ref{lst:simple global protocol guarded implementation} satisfies global protocol \ref{fig:simple global protocol guarded}.
For ease of reference, the global protocol is restated here.
\begin{subequations}
  \begin{align}
    \transmit{\party{A}}{\Channel}{1}{\party{B}} \ssep \guardP{\Ep{\party{A}}{1} \HB{} \Ep{\party{B}}{2}} \ssep \guardP{\Ep{\party{B}}{1} \HB{} \Ep{\party{C}}{2}} \ssep \transmit{\party{B}}{\Channel}{2}{\party{C}} \label{eq:simple global protocol guarded}
  \end{align}
\end{subequations}

The verified implementation is implementation \ref{lst:simple global protocol guarded implementation verified}.

\begin{code}[caption={A verified implementation that satisfies the global protocol \ref{fig:simple global protocol guarded}.},label={lst:simple global protocol guarded implementation verified}]
package main

var c = make(chan int, 2)
var d = make(chan int, 0)
var e = make(chan int, 0)

func main() {
	// $\hoareAssert{\transmit{\party{A}}{\Channel}{1}{\party{B}} \ssep \guardP{\Ep{\party{A}}{1} \HB{} \Ep{\party{B}}{2}} \ssep \guardP{\Ep{\party{B}}{1} \HB{} \Ep{\party{C}}{2}} \ssep \transmit{\party{B}}{\Channel}{2}{\party{C}}}$
	/* $\{\PartyProtocolPredicate{\party{A}}{\Ep{\party{A}}{1} \CB{\Channel} \ssep \guardP{\channel{d}{0}, \Ep{\party{A}}{1}} \ssep \Ep{\party{A}}{1.3} \CB{\channel{d}{0}} \ssep \emp \ssep \emp}$
	$\hphantom{/* \{}\asep \PartyProtocolPredicate{\party{B}}{\CB{\Channel} \Ep{\party{B}}{1} \ssep \CB{\channel{d}{0}} \Ep{\party{B}}{1.3} \ssep \guardP{\channel{d}{0}, \Ep{\party{B}}{2}} \ssep \guardP{\channel{e}{0}, \Ep{\party{B}}{1}} \ssep \Ep{\party{B}}{1.6} \CB{\channel{e}{0}} \ssep \Ep{\party{B}}{2} \CB{\Channel}}$
	$\hphantom{/* \{}\asep \PartyProtocolPredicate{\party{C}}{\emp \ssep \emp \ssep \CB{\channel{e}{0}} \Ep{\party{C}}{1.6} \ssep \guardP{\channel{e}{0}, \Ep{\party{C}}{2}} \ssep \CB{\Channel} \Ep{\party{C}}{2}}\}$ 
	go A()
	go B()
	go C()
	/* $\{\PartyProtocolPredicate{\party{A}}{\EndpointProtocolPredicate{\Channel}{\emp} \asep \EndpointProtocolPredicate{\channel{d}{0}}{\emp}} \land \Ep{\party{A}}{1} \land \Ep{\party{A}}{1.3}$
	$\hphantom{/* \{}\asep \PartyProtocolPredicate{\party{B}}{\EndpointProtocolPredicate{\Channel}{\emp} \asep \EndpointProtocolPredicate{\channel{d}{0}}{\emp} \asep \EndpointProtocolPredicate{\channel{e}{0}}{\emp}} \land \Ep{\party{B}}{1} \land \Ep{\party{B}}{1.3} \land \Ep{\party{B}}{1.6} \land \Ep{\party{B}}{2}$
	$\hphantom{/* \{}\asep \PartyProtocolPredicate{\party{C}}{\EndpointProtocolPredicate{\Channel}{\emp} \asep \EndpointProtocolPredicate{\channel{e}{0}}{\emp} \land \Ep{\party{C}}{1.6} \land \Ep{\party{C}}{2}}\}$ */
}

func A() {
	// $\hoareAssert{\PartyProtocolPredicate{\party{A}}{\Ep{\party{A}}{1} \CB{\Channel} \ssep \guardP{\channel{d}{0}, \Ep{\party{A}}{1}} \ssep \Ep{\party{A}}{1.3} \CB{\channel{d}{0}}}}$
	// $\hoareAssert{\EndpointProtocolPredicate{\Channel}{\Ep{\party{A}}{1} \CB{} \ssep \emp \ssep \emp} \asep \EndpointProtocolPredicate{\channel{d}{0}}{\emp \ssep \guardP{\Ep{\party{A}}{1}} \ssep \Ep{\party{A}}{1.3} \CB{}}}$ 
	c <- 1
	// $\hoareAssert{\EndpointProtocolPredicate{\Channel}{\emp \ssep \emp} \asep \EndpointProtocolPredicate{\channel{d}{0}}{\guardP{\Ep{\party{A}}{1}} \ssep \Ep{\party{A}}{1.3} \CB{}} \land \Ep{\party{A}}{1}}$ 
	// $\hoareAssert{\EndpointProtocolPredicate{\Channel}{\emp \ssep \emp} \asep \EndpointProtocolPredicate{\channel{d}{0}}{\Ep{\party{A}}{1.3} \CB{}} \land \Ep{\party{A}}{1}}$ 
	d <- 0
	// $\hoareAssert{\EndpointProtocolPredicate{\Channel}{\emp} \asep \EndpointProtocolPredicate{\channel{d}{0}}{\emp} \land \Ep{\party{A}}{1} \land \Ep{\party{A}}{1.3}}$ 
	// $\hoareAssert{\PartyProtocolPredicate{\party{A}}{\EndpointProtocolPredicate{\Channel}{\emp} \asep \EndpointProtocolPredicate{\channel{d}{0}}{\emp}} \land \Ep{\party{A}}{1} \land \Ep{\party{A}}{1.3}}$ 
}

func B() {
	// $\hoareAssert{\PartyProtocolPredicate{\party{B}}{\CB{\Channel} \Ep{\party{B}}{1} \ssep \CB{\channel{d}{0}} \Ep{\party{B}}{1.3} \ssep \guardP{\channel{d}{0}, \Ep{\party{B}}{2}} \ssep \guardP{\channel{e}{0}, \Ep{\party{B}}{1}} \ssep \Ep{\party{B}}{1.6} \CB{\channel{e}{0}} \ssep \Ep{\party{B}}{2} \CB{\Channel}}}$
	/* $\{\EndpointProtocolPredicate{\Channel}{\CB{} \Ep{\party{B}}{1} \ssep \Ep{\party{B}}{2} \CB{}}$
	$\hphantom{/* \{}\asep \EndpointProtocolPredicate{\channel{d}{0}}{\CB{} \Ep{\party{B}}{1.3} \ssep \guardP{\Ep{\party{B}}{2}}}$
	$\hphantom{/* \{}\asep \EndpointProtocolPredicate{\channel{e}{0}}{\guardP{\Ep{\party{B}}{1}} \ssep \Ep{\party{B}}{1.6} \CB{}}\}$ 
	<-c
	// $\hoareAssert{\EndpointProtocolPredicate{\Channel}{\Ep{\party{B}}{2} \CB{}} \asep \EndpointProtocolPredicate{\channel{d}{0}}{\CB{} \Ep{\party{B}}{1.3} \ssep \guardP{\Ep{\party{B}}{2}}} \asep \EndpointProtocolPredicate{\channel{e}{0}}{\guardP{\Ep{\party{B}}{1}} \ssep \Ep{\party{B}}{1.6} \CB{}} \land \Ep{\party{B}}{1}}$ 
	<-d
	// $\hoareAssert{\EndpointProtocolPredicate{\Channel}{\Ep{\party{B}}{2} \CB{}} \asep \EndpointProtocolPredicate{\channel{d}{0}}{\guardP{\Ep{\party{B}}{2}}} \asep \EndpointProtocolPredicate{\channel{e}{0}}{\guardP{\Ep{\party{B}}{1}} \ssep \Ep{\party{B}}{1.6} \CB{}} \land \Ep{\party{B}}{1} \land \Ep{\party{B}}{1.3}}$ 
	// $\hoareAssert{\EndpointProtocolPredicate{\Channel}{\Ep{\party{B}}{2} \CB{}} \asep \EndpointProtocolPredicate{\channel{d}{0}}{\guardP{\Ep{\party{B}}{2}}} \asep \EndpointProtocolPredicate{\channel{e}{0}}{\Ep{\party{B}}{1.6} \CB{}} \land \Ep{\party{B}}{1} \land \Ep{\party{B}}{1.3}}$ 
	e <- 0
	// $\hoareAssert{\EndpointProtocolPredicate{\Channel}{\Ep{\party{B}}{2} \CB{}} \asep \EndpointProtocolPredicate{\channel{d}{0}}{\guardP{\Ep{\party{B}}{2}}} \asep \EndpointProtocolPredicate{\channel{e}{0}}{\emp} \land \Ep{\party{B}}{1} \land \Ep{\party{B}}{1.3} \land \Ep{\party{B}}{1.6}}$ 
	c <- 2
	// $\hoareAssert{\EndpointProtocolPredicate{\Channel}{\emp} \asep \EndpointProtocolPredicate{\channel{d}{0}}{\guardP{\Ep{\party{B}}{2}}} \asep \EndpointProtocolPredicate{\channel{e}{0}}{\emp} \land \Ep{\party{B}}{1} \land \Ep{\party{B}}{1.3} \land \Ep{\party{B}}{1.6} \land \Ep{\party{B}}{2}}$ 
	// $\hoareAssert{\EndpointProtocolPredicate{\Channel}{\emp} \asep \EndpointProtocolPredicate{\channel{d}{0}}{\emp} \asep \EndpointProtocolPredicate{\channel{e}{0}}{\emp} \land \Ep{\party{B}}{1} \land \Ep{\party{B}}{1.3} \land \Ep{\party{B}}{1.6} \land \Ep{\party{B}}{2}}$ 
	// $\hoareAssert{\PartyProtocolPredicate{\party{B}}{\EndpointProtocolPredicate{\Channel}{\emp} \asep \EndpointProtocolPredicate{\channel{d}{0}}{\emp} \asep \EndpointProtocolPredicate{\channel{e}{0}}{\emp}} \land \Ep{\party{B}}{1} \land \Ep{\party{B}}{1.3} \land \Ep{\party{B}}{1.6} \land \Ep{\party{B}}{2}}$ 
}

func C() {
	// $\hoareAssert{\PartyProtocolPredicate{\party{C}}{\CB{\channel{e}{0}} \Ep{\party{C}}{1.6} \ssep \guardP{\channel{e}{0}, \Ep{\party{C}}{2}} \ssep \CB{\Channel} \Ep{\party{C}}{2}}}$
	// $\hoareAssert{\EndpointProtocolPredicate{\Channel}{\CB{} \Ep{\party{C}}{2}} \asep \EndpointProtocolPredicate{\channel{e}{0}}{\Ep{\party{C}}{1.6} \CB{} \ssep \guardP{\Ep{\party{C}}{2}}}}$ 
	<-e
	// $\hoareAssert{\EndpointProtocolPredicate{\Channel}{\CB{} \Ep{\party{C}}{2}} \asep \EndpointProtocolPredicate{\channel{e}{0}}{\guardP{\Ep{\party{C}}{2}}} \land \Ep{\party{C}}{1.6} \CB{}}$ 
	<-c
	// $\hoareAssert{\EndpointProtocolPredicate{\Channel}{\emp} \asep \EndpointProtocolPredicate{\channel{e}{0}}{\guardP{\Ep{\party{C}}{2}}} \land \Ep{\party{C}}{1.6} \land \Ep{\party{C}}{2}}$ 
	// $\hoareAssert{\EndpointProtocolPredicate{\Channel}{\emp} \asep \EndpointProtocolPredicate{\channel{e}{0}}{\emp} \land \Ep{\party{C}}{1.6} \land \Ep{\party{C}}{2}}$ 
	// $\hoareAssert{\PartyProtocolPredicate{\party{C}}{\EndpointProtocolPredicate{\Channel}{\emp} \asep \EndpointProtocolPredicate{\channel{e}{0}}{\emp} \land \Ep{\party{C}}{1.6} \land \Ep{\party{C}}{2}}}$ 
}
\end{code}

Verification proceeds like program verification with Hoare logic \parencite{Hoare1969AnAB}: using Hoare triples.
Verification starts with the given global protocol as the precondition of the first Hoare triple.
We now explain verification rules and lemmas in figure \ref{fig:verification rules and lemmas}.

Projection proceeds by the projection lemmas in figure \ref{fig:projection lemmas}.
The given global protocol is projected onto each party in the global protocol, and then projected again onto each endpoint of each party.
We define predicate $\PartyProtocolPredicate{\Party}{\PerPartyProtocol}$ to know that the per-party protocol $\PerPartyProtocol$ corresponds to party $\Party$ after projection, and predicate $\EndpointProtocolPredicate{\Channel}{\PerEndpointProtocol}$ to know that the per-endpoint protocol $\PerEndpointProtocol$ corresponds to endpoint $\Channel$ after projection.

Verification of sends and receives on endpoints proceeds by the Hoare triples in figure \ref{fig:Hoare triple rules}.
Each Hoare triple consumes the first event of the per-endpoint protocol in the precondition.
Then, the event appears in the postcondition, indicating that the party of the endpoint knows that the event has happened.
The Hoare triples that verify the sending and receiving of endpoints are slightly different.
These triples use an extended syntax of per-endpoint protocols, where sends and receives also specify the message:
\begin{description}
\item[$\ChannelSend{\Ep{\Party}{i}} \cdot \EndpointProtocolPredicate{\channel{c'}{\anything}}{\PerEndpointProtocol'}$] A send event $\Ep{\Party}{i}$ occurs on the endpoint, and the message sent is the endpoint $\channel{c'}{\anything}$ that has the per-endpoint protocol $\PerEndpointProtocol'$.
\item[$\ChannelReceive{\Ep{\Party}{i}} \cdot \EndpointProtocolPredicate{\channel{c'}{\anything}}{\PerEndpointProtocol'}$] A receive event $\Ep{\Party}{i}$ occurs on the endpoint, and the message received is the endpoint $\channel{c'}{\anything}$ that has the per-endpoint protocol $\PerEndpointProtocol'$.
\captionof{enumcnt}{Extended per-endpoint protocol syntax.} \label{lst:extended per-endpoint protocol syntax}
\end{description}
\eqref{eq:Endpoint Send} verifies that endpoint $\channel{c'}{\anything}$ is sent over endpoint $\Channel$.
The triple requires the sending party to have the endpoint predicate $\EndpointProtocolPredicate{\channel{c'}{\anything}}{\PerEndpointProtocol'}$, and ensures that the sent endpoint predicate is no longer in the sending party's state.
\eqref{eq:Endpoint Receive} verifies that endpoint $\channel{c'}{\anything}$ is received over endpoint $\Channel$.
The triple ensures that the received endpoint predicate $\EndpointProtocolPredicate{\channel{c'}{\anything}}{\PerEndpointProtocol'}$ is in the receiving party's state.

Guards are proven by the lemma in figure \ref{fig:prove guard lemma}.
Specifically, guards $\guardP{\Ep{\Party}{i}}$ in per-endpoint protocols are proven by having $\Ep{\Party}{i}$ in the precondition, which comes from the rules in figure \ref{fig:Hoare triple rules}.

If verification of a per-endpoint protocol succeeds, then the entire per-endpoint protocol is consumed by verification, and only the empty per-endpoint protocol $\emp$ remains, and the verification terminates.
If verification of a per-endpoint protocol fails, then only some prefix of the per-endpoint protocol is consumed by verification, and the unconsumed per-endpoint protocol $\PerEndpointProtocol$ or $\PerPartyProtocol$ remains, and the verification terminates.
If verification of some per-endpoint protocols fail, then these failed verifications terminate, but other independent verifications continue until termination.\footnote{In other words, verification of a Go program waits for verification of all goroutines to complete. Note that this is in stark contrast to executions, where execution of a Go program does not wait for execution of all goroutines to complete}.

We identify and informally describe two ways that the verification of a per-endpoint protocol may fail.
\begin{description}
\item[[FAIL: precondition]] Verification proceeds until a line of code, but the Hoare assertion of the verification does not satisfy the precondition of the Hoare triple corresponding to the line of code.
\item[[FAIL: unconsumed]] There are no more lines of code to be verified, but the unconsumed per-endpoint protocols are nonempty.
  In this way of failure, the verification does not consume the head of the per-endpoint protocol.
  The reason for not consuming depends on the type of head:
  \begin{description}
  \item[The head is a guard.] The verification cannot prove the guard.
  \item[The head is a send.] The implementation did not send on the channel.
  \item[The head is a receive.] The implementation did not receive on the channel.
  \end{description}
\end{description}

Eventually, all verifications terminate.
Now consider the set of successful verifications and the set of failed verifications.
Reverse of projection proceeds by the unprojection lemmas in figure \ref{fig:unprojection lemma}.
Successful verifications have an empty unconsumed projected protocol $emp$.
Failed verifications have a nonempty unconsumed projected protocol.
All the unconsumed per-endpoint protocols $\PerEndpointProtocol$ are unprojected by the unprojection lemma to become an unprojected per-party protocol, called a ``residue''.

In lemma \eqref{eq:Per-endpoint to Per-party}, we abuse the notation of $\PartyProtocolPredicate{\Party}{\PerPartyProtocol}$ to store the per-party residue where $\PerPartyProtocol$ belongs.
For example, $\party{A}$ has the per-party residue $\PartyProtocolPredicate{\party{A}}{\EndpointProtocolPredicate{\Channel}{\emp} \asep \EndpointProtocolPredicate{\channel{d}{0}}{\emp}}$.
This is an abuse of notation, because the per-party residue is made up of unconsumed per-endpoint protocols, combined with the $\asep$ operator.
The $\asep$ operator is not defined in the syntax of both per-party protocols, therefore the per-party residue may not conform to the syntax of $\PerPartyProtocol$.

After all parties are verified, the per-party residues are combined with concurrent composition $\asep$ into a global residue, which is shown to the user.
If the global residue contains only the empty protocol, then verification of the entire global protocol succeeded.
If the global residue contains a nonempty protocol, then it indicates exactly at which lines of code, on which endpoints, and in which parties the verification failed.
For example, the global residue provided by verification of implementation \ref{lst:simple global protocol guarded implementation verified failed} is nonempty, because the per-party residues of $\party{A}$ and $\party{B}$ are nonempty.
The verification as a whole fails because verifications of parties $\party{A}$ and $\party{B}$ fail.
Although the global protocol specifies that $\party{B}$ communicates with $\party{C}$, and in the implementation $\party{B}$ does not communicate with $\party{C}$, the verification of $\party{C}$ assumes that a message was sent on channel $\Channel$.
Verification of party $\party{C}$ continues independently of the verification failures of $\party{A}$ and $\party{B}$, and eventually $\party{C}$ succeeds with an empty per-party residue.
Intuitively, $\party{C}$ assumes that $\party{A}$ and $\party{B}$ are successfully verified.
Our approach to verification is such that only $\party{A}$ and $\party{B}$ need to prove the guards that were projected to them, while $\party{C}$ does not need to prove those guards.
This approach to verification is inspired by \citeauthor{costea2017session}'s cooperative proving, where only one party has the burden of proving a guard, while other parties can assume that the guard has been proven \parencite[Chapter 4]{costea2017session}.
This approach to verification is similar to \citeauthor{costea2017session}'s cooperative proving, because $\party{C}$ assumes that $\party{A}$ and $\party{B}$ are successfully verified, therefore $\party{C}$ assumes that the guards that were projected to $\party{A}$ and $\party{B}$ have been proven.

\begin{code}[caption={An implementation verified to \textbf{not} satisfy the global protocol \ref{fig:simple global protocol guarded}.},label={lst:simple global protocol guarded implementation verified failed}]
package main

var c = make(chan int, 2)
var d = make(chan int, 0)
var e = make(chan int, 0)

func main() {
	// $\hoareAssert{\transmit{\party{A}}{\Channel}{1}{\party{B}} \ssep \guardP{\Ep{\party{A}}{1} \HB{} \Ep{\party{B}}{2}} \ssep \guardP{\Ep{\party{B}}{1} \HB{} \Ep{\party{C}}{2}} \ssep \transmit{\party{B}}{\Channel}{2}{\party{C}}}$
	/* $\{\PartyProtocolPredicate{\party{A}}{\Ep{\party{A}}{1} \CB{\Channel} \ssep \guardP{\channel{d}{0}, \Ep{\party{A}}{1}} \ssep \Ep{\party{A}}{1.3} \CB{\channel{d}{0}} \ssep \emp \ssep \emp}$
	$\hphantom{/* \{}\asep \PartyProtocolPredicate{\party{B}}{\CB{\Channel} \Ep{\party{B}}{1} \ssep \CB{\channel{d}{0}} \Ep{\party{B}}{1.3} \ssep \guardP{\channel{d}{0}, \Ep{\party{B}}{2}} \ssep \guardP{\channel{e}{0}, \Ep{\party{B}}{1}} \ssep \Ep{\party{B}}{1.6} \CB{\channel{e}{0}} \ssep \Ep{\party{B}}{2} \CB{\Channel}}$
	$\hphantom{/* \{}\asep \PartyProtocolPredicate{\party{C}}{\emp \ssep \emp \ssep \CB{\channel{e}{0}} \Ep{\party{C}}{1.6} \ssep \guardP{\channel{e}{0}, \Ep{\party{C}}{2}} \ssep \CB{\Channel} \Ep{\party{C}}{2}}\}$ 
	go A()
	go B()
	go C()
	/* $\{\PartyProtocolPredicate{\party{A}}{\EndpointProtocolPredicate{\Channel}{\emp} \asep \EndpointProtocolPredicate{\channel{d}{0}}{\Ep{\party{A}}{1.3} \CB{}} \land \Ep{\party{A}}{1}}$
	$\hphantom{/* \{}\asep \PartyProtocolPredicate{\party{B}}{\EndpointProtocolPredicate{\Channel}{\Ep{\party{B}}{2} \CB{}} \asep \EndpointProtocolPredicate{\channel{d}{0}}{\guardP{\Ep{\party{B}}{2}}} \asep \EndpointProtocolPredicate{\channel{e}{0}}{\Ep{\party{B}}{1.6} \CB{}} \land \Ep{\party{B}}{1} \land \Ep{\party{B}}{1.3}}$
	$\hphantom{/* \{}\asep \PartyProtocolPredicate{\party{C}}{\EndpointProtocolPredicate{\Channel}{\emp} \asep \EndpointProtocolPredicate{\channel{e}{0}}{\emp} \land \Ep{\party{C}}{1.6} \land \Ep{\party{C}}{2}}\}$ */
}

func A() {
	// $\hoareAssert{\PartyProtocolPredicate{\party{A}}{\Ep{\party{A}}{1} \CB{\Channel} \ssep \guardP{\channel{d}{0}, \Ep{\party{A}}{1}} \ssep \Ep{\party{A}}{1.3} \CB{\channel{d}{0}}}}$
	// $\hoareAssert{\EndpointProtocolPredicate{\Channel}{\Ep{\party{A}}{1} \CB{} \ssep \emp \ssep \emp} \asep \EndpointProtocolPredicate{\channel{d}{0}}{\emp \ssep \guardP{\Ep{\party{A}}{1}} \ssep \Ep{\party{A}}{1.3} \CB{}}}$ 
	c <- 1
	// $\hoareAssert{\EndpointProtocolPredicate{\Channel}{\emp} \asep \EndpointProtocolPredicate{\channel{d}{0}}{\guardP{\Ep{\party{A}}{1}} \ssep \Ep{\party{A}}{1.3} \CB{}} \land \Ep{\party{A}}{1}}$ 
	// $\hoareAssert{\EndpointProtocolPredicate{\Channel}{\emp} \asep \EndpointProtocolPredicate{\channel{d}{0}}{\Ep{\party{A}}{1.3} \CB{}} \land \Ep{\party{A}}{1}}$ 
	e <- 0
	// $\hoareAssert{\EndpointProtocolPredicate{\Channel}{\emp} \asep \EndpointProtocolPredicate{\channel{d}{0}}{\Ep{\party{A}}{1.3} \CB{}} \land \Ep{\party{A}}{1}}$ [FAIL: precondition]
	// $\hoareAssert{\PartyProtocolPredicate{\party{A}}{\EndpointProtocolPredicate{\Channel}{\emp} \asep \EndpointProtocolPredicate{\channel{d}{0}}{\Ep{\party{A}}{1.3} \CB{}} \land \Ep{\party{A}}{1}}}$ 
}

func B() {
	// $\hoareAssert{\PartyProtocolPredicate{\party{B}}{\CB{\Channel} \Ep{\party{B}}{1} \ssep \CB{\channel{d}{0}} \Ep{\party{B}}{1.3} \ssep \guardP{\channel{d}{0}, \Ep{\party{B}}{2}} \ssep \guardP{\channel{e}{0}, \Ep{\party{B}}{1}} \ssep \Ep{\party{B}}{1.6} \CB{\channel{e}{0}} \ssep \Ep{\party{B}}{2} \CB{\Channel}}}$
	/* $\{\EndpointProtocolPredicate{\Channel}{\CB{} \Ep{\party{B}}{1} \ssep \Ep{\party{B}}{2} \CB{}}$
	$\hphantom{/* \{}\asep \EndpointProtocolPredicate{\channel{d}{0}}{\CB{} \Ep{\party{B}}{1.3} \ssep \guardP{\Ep{\party{B}}{2}}}$
	$\hphantom{/* \{}\asep \EndpointProtocolPredicate{\channel{e}{0}}{\guardP{\Ep{\party{B}}{1}} \ssep \Ep{\party{B}}{1.6} \CB{}}\}$ 
	<-c
	// $\hoareAssert{\EndpointProtocolPredicate{\Channel}{\Ep{\party{B}}{2} \CB{}} \asep \EndpointProtocolPredicate{\channel{d}{0}}{\CB{} \Ep{\party{B}}{1.3} \ssep \guardP{\Ep{\party{B}}{2}}} \asep \EndpointProtocolPredicate{\channel{e}{0}}{\guardP{\Ep{\party{B}}{1}} \ssep \Ep{\party{B}}{1.6} \CB{}} \land \Ep{\party{B}}{1}}$ 
	<-d
	// $\hoareAssert{\EndpointProtocolPredicate{\Channel}{\Ep{\party{B}}{2} \CB{}} \asep \EndpointProtocolPredicate{\channel{d}{0}}{\guardP{\Ep{\party{B}}{2}}} \asep \EndpointProtocolPredicate{\channel{e}{0}}{\guardP{\Ep{\party{B}}{1}} \ssep \Ep{\party{B}}{1.6} \CB{}} \land \Ep{\party{B}}{1} \land \Ep{\party{B}}{1.3}}$ 
	// $\hoareAssert{\EndpointProtocolPredicate{\Channel}{\Ep{\party{B}}{2} \CB{}} \asep \EndpointProtocolPredicate{\channel{d}{0}}{\guardP{\Ep{\party{B}}{2}}} \asep \EndpointProtocolPredicate{\channel{e}{0}}{\Ep{\party{B}}{1.6} \CB{}} \land \Ep{\party{B}}{1} \land \Ep{\party{B}}{1.3}}$ 
	// [FAIL: unconsumed]
	// $\hoareAssert{\PartyProtocolPredicate{\party{B}}{\EndpointProtocolPredicate{\Channel}{\Ep{\party{B}}{2} \CB{}} \asep \EndpointProtocolPredicate{\channel{d}{0}}{\guardP{\Ep{\party{B}}{2}}} \asep \EndpointProtocolPredicate{\channel{e}{0}}{\Ep{\party{B}}{1.6} \CB{}} \land \Ep{\party{B}}{1} \land \Ep{\party{B}}{1.3}}}$ 
}

func C() {
	// $\hoareAssert{\PartyProtocolPredicate{\party{C}}{\CB{\channel{e}{0}} \Ep{\party{C}}{1.6} \ssep \guardP{\channel{e}{0}, \Ep{\party{C}}{2}} \ssep \CB{\Channel} \Ep{\party{C}}{2}}}$
	// $\hoareAssert{\EndpointProtocolPredicate{\Channel}{\CB{} \Ep{\party{C}}{2}} \asep \EndpointProtocolPredicate{\channel{e}{0}}{\Ep{\party{C}}{1.6} \CB{} \ssep \guardP{\Ep{\party{C}}{2}}}}$ 
	<-e
	// $\hoareAssert{\EndpointProtocolPredicate{\Channel}{\CB{} \Ep{\party{C}}{2}} \asep \EndpointProtocolPredicate{\channel{e}{0}}{\guardP{\Ep{\party{C}}{2}}} \land \Ep{\party{C}}{1.6} \CB{}}$ 
	<-c
	// $\hoareAssert{\EndpointProtocolPredicate{\Channel}{\emp} \asep \EndpointProtocolPredicate{\channel{e}{0}}{\guardP{\Ep{\party{C}}{2}}} \land \Ep{\party{C}}{1.6} \land \Ep{\party{C}}{2}}$ 
	// $\hoareAssert{\EndpointProtocolPredicate{\Channel}{\emp} \asep \EndpointProtocolPredicate{\channel{e}{0}}{\emp} \land \Ep{\party{C}}{1.6} \land \Ep{\party{C}}{2}}$ 
	// $\hoareAssert{\PartyProtocolPredicate{\party{C}}{\EndpointProtocolPredicate{\Channel}{\emp} \asep \EndpointProtocolPredicate{\channel{e}{0}}{\emp} \land \Ep{\party{C}}{1.6} \land \Ep{\party{C}}{2}}}$ 
}
\end{code}

\begin{figure}
  \begin{subfigure}[t]{\textwidth}
    {\centering
      \begin{subequations}
        \begin{align}
          \GlobalProtocol &\implies \PartyProtocolPredicate{\Party}{\projectOntoP{\GlobalProtocol}{\Party}} \asep \PartyProtocolPredicate{\Party'}{\projectOntoP{\GlobalProtocol}{\Party'}} \asep \ldots &&\text{for some parties $\Party, \Party', \ldots$} \label{eq:Global to Per-party} \tag{Global to Per-party} \\
          \PerPartyProtocol &\implies \EndpointProtocolPredicate{\Channel}{\projectOntoP{\PerPartyProtocol}{\Channel'}} \asep \EndpointProtocolPredicate{\Channel}{\projectOntoP{\PerPartyProtocol}{\Channel'}} \asep \ldots &&\text{for some endpoints $\Channel, \Channel', \ldots$} \label{eq:Per-party to Per-endpoint} \tag{Per-party to Per-endpoint}
        \end{align}
      \end{subequations}
      \caption{Projection lemmas.}
      \label{fig:projection lemmas}
    }
  \end{subfigure}

  \begin{subfigure}[t]{\textwidth}
    {\centering
      \begin{subequations}
        \begin{align}
          \Rule
          {}
          {\triple
          {\EndpointProtocolPredicate{\Channel}{\ChannelSend{\Ep{\Party}{i}} \cdot \EndpointProtocolPredicate{\channel{c'}{\anything}}{\PerEndpointProtocol'} \ssep \PerEndpointProtocol} \asep \EndpointProtocolPredicate{\channel{c'}{\anything}}{\PerEndpointProtocol'}}
          {c <- c'}
          {\EndpointProtocolPredicate{\Channel}{\PerEndpointProtocol} \land \Ep{\Party}{i}}
          }&
             \label{eq:Endpoint Send} \tag{Endpoint Send} \\
          \Rule
          {}
          {\triple
          {\EndpointProtocolPredicate{\Channel}{\ChannelReceive{\Ep{\Party}{i}} \cdot \EndpointProtocolPredicate{\channel{c'}{\anything}}{\PerEndpointProtocol'} \ssep \PerEndpointProtocol}}
          {c' := <-c}
          {\EndpointProtocolPredicate{\Channel}{\PerEndpointProtocol} \asep \EndpointProtocolPredicate{\channel{c'}{\anything}}{\PerEndpointProtocol'} \land \Ep{\Party}{i}}
          }&
             \label{eq:Endpoint Receive} \tag{Endpoint Receive} \\
          \Rule
          {}
          {\triple
          {\EndpointProtocolPredicate{\Channel}{\ChannelSend{\Ep{\Party}{i}} \ssep \PerEndpointProtocol}}
          {c <- *}
          {\EndpointProtocolPredicate{\Channel}{\PerEndpointProtocol} \land \Ep{\Party}{i}}
          }&
             \label{eq:Data Send} \tag{Data Send} \\
          \Rule
          {}
          {\triple
          {\EndpointProtocolPredicate{\Channel}{\ChannelReceive{\Ep{\Party}{i}} \ssep \PerEndpointProtocol}}
          {<-c}
          {\EndpointProtocolPredicate{\Channel}{\PerEndpointProtocol} \land \Ep{\Party}{i}}
          }&
             \label{eq:Data Receive} \tag{Data Receive}
        \end{align}
      \end{subequations}
      \caption{Hoare triple rules.}
      \label{fig:Hoare triple rules}
    }
  \end{subfigure}

  \begin{subfigure}[t]{\textwidth}
    {\centering
      \begin{subequations}
        \begin{align}
          (\guardP{\Psi} \ssep \text{Protocol}) \land \Psi &\implies \text{Protocol} \land \Psi \label{eq:Prove Guard} \tag{Prove Guard}
        \end{align}
      \end{subequations}
      \caption{A lemma that proves a guard, where Protocol is either $\GlobalProtocol$, $\PerPartyProtocol$, or $\PerEndpointProtocol$.}
      \label{fig:prove guard lemma}
    }
  \end{subfigure}

  \begin{subfigure}[t]{\textwidth}
    {\centering
      \begin{subequations}
        \begin{align}
          \EndpointProtocolPredicate{\Channel}{\PerEndpointProtocol} \asep \EndpointProtocolPredicate{\Channel'}{\PerEndpointProtocol'} \asep \ldots &\implies \PartyProtocolPredicate{\Party}{\PerEndpointProtocol \asep \PerEndpointProtocol' \asep \ldots} \notag \\
          \text{for some endpoints $\Channel, \Channel', \ldots$ of some party $\Party$} \label{eq:Per-endpoint to Per-party} \tag{Per-endpoint to Per-party}
        \end{align}
      \end{subequations}
      \caption{Unprojection lemma, where residues of each endpoint are joined together.}
      \label{fig:unprojection lemma}
    }
  \end{subfigure}
  \caption{Verification rules and lemmas.}
  \label{fig:verification rules and lemmas}
\end{figure}

\section{Conclusion}
Our main contribution is the extension of \textcite{costea2017session} to deductively verify that implementations that use a subset of the Go programming language satisfy communication protocols, and we specifically reason about Go's memory model.

We achieved this by doing the following:
\begin{enumerate}
\item Represent Go programs as CFGs, thereby determining the program order of goroutines. Then, we combine the program orders with the send-receive order to obtain the $\HB{}$ order, which is different from Lamport's happens-before order in that we can use the $\HB{}$ to represent a rendezvous.
\item Recognize that a distributed Go program with only sends and receives is sequentially consistent up to races. This allows us to represent a global protocol or an execution with a sequence diagram.
\item Define sequential executions, which are easier to discuss about than concurrent executions, because sequential execution events are atomic. Therefore, when discussing sequential executions, we do not need to mention the start or completion of a sequential execution event. This simplifies our thought process when considering which executions a global protocol has, and this also simplifies and increases the help in visual understanding that sequence diagrams give us.
\item Define guards as proof obligations of whether an element exists in the $\HB{}$ order. This allows our verification framework to determine whether an implementation orders two events.
\item Use guards to define race-free global protocols. If our verification determines that an implementation satisfies a race-free global protocol, then users of our verification framework can be certain that the implementation is race-free.
\end{enumerate}

In doing so, we also simplify the verification process of \textcite{costea2017session} by reducing the amount of formal machinery needed for verification.

\section{Related Work}

\subsection{Other Verification Approaches}
We compare our approach to verification of Go programs, to other works with other approaches to similar problems.
\begin{enumerate}
\item Dynamic approaches need to execute the distributed program, whereas our static approach does not need to execute the distributed program.
  Despite the difference in approaches, our using of a happens-before order is similar to dynamic approaches that construct a happens-before order on-the-fly \parencite{10.1145/2103656.2103702}.
  Still, the dynamically constructed happens-before order depends on how the program executes.
  The large number of possible interleavings of a distributed programs makes it unlikely that a single execution exhibits a behavior representative of many other executions, therefore the dynamically constructed happens-before order could be imprecise.

\item Model checking approaches verify that a state representation of a program meets a specification.
  According to \textcite{go/blog/codelab-share}, the design of Go was influenced by Communicating Sequential Processes (CSP) \parencite{hoare1978csp}, therefore it is natural to represent Go programs with CSP.
  Indeed, \textcite{Prasertsang2016formal} proposes translating Go programs to a machine-readable version of CSP, on which tools like The CSP Refinement Checker \parencite{fdrmanual} can be used.
  Although we do not use a model checking approach, this representation with a process algebra suggests that Go programs are highly structured, and that we can use formal verification techniques on Go programs.
  CSP is similar to another process algebra, the $\pi$-calculus.
  \textcite{Caires2010SessionTA} has shown that the $\pi$-calculus and session types are related.
  Therefore, it is not surprising that there have been many works on verifying Go programs with session types \parencites{castro2019distributed}{DY2013}{lange2018verification}{gabet2020static}.
  There is also a work that model-checks up to a bound \textcite{lange2017fencing}.

\item Deductive verification approaches verify a distributed program without executing it.
  Our approach is also similar to the deductive verification approach of \textcite{oortwijn2019abstractions}, in that we also use concurrent separation logic (CSL) \parencite{Brookes2016ConcurrentSL} as the underlying logic.
  CSL is used to reason about local concurrency, which is relevant to distributed programs. According to \textcite[2]{castro2019distributed}, ``distributed programming in practice often involves local concurrency: goroutines and channels are effective for dealing locally with the inherent asynchrony of distributed interactions.''.
  Program states of concurrent programs can be deductively verified with CSL \parencite{Vafeiadis2011ConcurrentSL} or Concurrent Kleene Algebra (CKA) \parencite{Hoare2009ConcurrentKA}.
  However, verification techniques for local concurrency are not enough to verify distributed systems, and these verification techniques need to be extended.
  Program states of distributed programs can be deductively verified with an extension to CSL \parencite{ding2020formalizing}, or an extension to CKA \parencite{Jaskolka2014EndowingCK}.
  Our approach deductively verifies distributed Go programs by extending CSL with the Hoare triples \eqref{eq:Endpoint Send}, \eqref{eq:Endpoint Receive}, \eqref{eq:Data Send}, and \eqref{eq:Data Receive}.
  This is similar to how \citeauthor{ding2020formalizing} extends CSL with send and receive operations.
\end{enumerate}

\subsection{Global Protocol Syntax}
Our global protocol syntax is similar to the abstract syntax tree (AST) global types syntax of \textcite{10.1145/1328438.1328472}.
\citeauthor{10.1145/1328438.1328472} notice a limitation of expressiveness.
They identify that a safe process cannot be represented with abstract syntax tree (AST) global types \parencite[Section 6.1]{10.1145/1328438.1328472}.
The safe process they identified is the parallel composition of parties $P \equiv s!.t?$ and $Q \equiv t!.s?$ (each party sends before receiving).
Although we have a similar global protocol syntax, we do not have the limitation of expressiveness, because their ordering on events (definition 3.3 (prefix ordering)) differs from ours: in the global protocol $\transmit{\party{A}}{\Channel}{1}{\party{B}} \ssep \transmit{\party{A}}{\channel{d}{\anything}}{2}{\party{C}}$, they order the two sending actions of $\party{A}$, whereas we do not order the two sending actions of $\party{A}$ because they occur on different channels.

The safe process can be expressed as a global protocol in our syntax as $\transmit{\party{P}}{s}{1}{\party{Q}} \asep \transmit{\party{Q}}{t}{2}{\party{P}}$.
One limitation of our syntax is: the global protocol $\transmit{\party{P}}{s}{1}{\party{Q}} \asep \transmit{\party{Q}}{t}{2}{\party{P}}$ also expresses implementations that deadlock.
Using the syntax of \citeauthor{10.1145/1328438.1328472}, one such deadlocked implementation is the parallel composition of parties $P \equiv t?.s!$ and $Q \equiv s?.t!$ (each party receives before sending).
This limitation of our syntax can be mitigated by detecting implementations that deadlock with wait-for graphs \parencite[Section 5.3]{costea2017session}, such that only safe implementations remain.

\subsection{Considering All Sequential Executions}
A recent work by \textcite{fava2020ready} detects data races for programs that use channel communications to coordinate accesses to shared memory. The authors considered extending the definition of happens-before with the FIFO order of channels, but the authors decided against this extension, because race detection is performed locally by each goroutine, so that a receiver can learn about events that the sender does not know about. Our work does not have this limitation, because we perform a global static race detection, instead of local race detection.

Additionally, there is an example in \textcite[Section 6, Figure 13 and Figure 14]{fava2020ready} of how not extending the definition of happens-before with the FIFO order of channels results in a false positive (a data race is detected but there are actually no data races).
Their race detector detects a positive, because the write event and the read event are not related by happens-before.
This positive is false, because the FIFO order of the channel guarantees that the write event happens-before the read event.
If the definition of happens-before is extended with the FIFO order of channels, there are two possible sequential executions of concurrent receives (we use their notation for channels, therefore we do not indicate the channel capacity):
\begin{subequations}
  \begin{align}
    p_1\text{'s} \rvop{}{c}{} \quad \HB{} \quad p_2\text{'s} \rvop{}{c}{} \label{eq:conditional-race FIFO sequential execution 1} \\
    p_2\text{'s} \rvop{}{c}{} \quad \HB{} \quad p_1\text{'s} \rvop{}{c}{} \label{eq:conditional-race FIFO sequential execution 2}
  \end{align}
\end{subequations}

The false positive disappears if the receives linearize as in sequential execution \eqref{eq:conditional-race FIFO sequential execution 1}.

However, this example does not consider the other sequential execution \eqref{eq:conditional-race FIFO sequential execution 2}. This sequential execution is drawn as a sequence diagram \ref{fig:conditional-race partial order}.
The example has the program $\texttt{if r = 1 then load z}$, so that the read event $\texttt{load z}$ does not even exist if the sequential execution is \eqref{eq:conditional-race FIFO sequential execution 2}. There is one write event, but no read event for the write event to race with. In this sequential execution, a data race cannot occur, therefore there is no need to consider this sequential execution.

In general, for other programs, we cannot ignore such sequential executions.
If we modify the program in the example by replacing $\texttt{if r = 1 then load z}$ with $\texttt{load z}$, then the read event exists, and the write event happens concurrently with the read event, therefore a data race occurs if the sequential execution is \eqref{eq:conditional-race FIFO sequential execution 2}.
Even if we extend the definition of happens-before with the FIFO order of channels, a data race could still occur. This means we need something other than FIFO order to avoid data races. Particularly, we need a global protocol with guards that tells us which one of the two sequential executions we should force the execution to go along. This is the inspiration for our main result: the root problem of the data race is an execution where a communication race occurs, and our solution is a global protocol with guards that gets rid of all executions that contain a communication race.

Our solution offers some improvements for this modified example.
Firstly, our solution considers all possible sequential executions, because we represent a global protocol with a partial order, and a partial order subsumes all linearizations of it.
Secondly, our solution relates the two concurrent receives by happens-before, so that we avoid the data race caused by the communication race.

\section{Future Work}

\subsection{Extension to Other Communication Models}
In this work, we have only defined race-freeness for the FIFO communication model.
However, we believe we can formalize other communication models, for example one that orders messages in causal order, and another that orders messages in total order.
We believe this is possible by defining appropriate $\Psi$ for $\guardP{\Psi}$, and appropriate global-to-per-party projection and per-party-to-per-channel projection functions for $\guardP{\Psi}$.

\subsection{Removal of Guards Already Proven by the Global Protocol}
Recall our definition of projection on a guard $\guardP{\Ep{\Party'}{i} \HB{} \Ep{\Party''}{j}}$ in section \ref{sec:Projections}, where intuitively, a new rendezvous is inserted between the two events ordered by $\HB{}$ in the guard.
The inserted rendezvous ensures that if an implementation satisfies a race-free global protocol, then the implementation cannot execute with races.

However, for some global protocols, the inserted rendezvous is redundant, making an execution less performant.
For example, consider if the global protocol designer designs the following global protocol that has the rendezvous $\transmit{\party{A}}{\channel{f}{0}}{1.2}{\party{B}}$:
\begin{subequations}
  \begin{align}
    \transmit{\party{A}}{\Channel}{1}{\party{B}} \ssep \transmit{\party{A}}{\channel{f}{0}}{1.2}{\party{B}} \ssep \transmit{\party{B}}{\Channel}{2}{\party{C}} \label{eq:simple global protocol with rendezvous}
  \end{align}
\end{subequations}

but deems that the global protocol has races (because the receives $\Ep{\party{B}}{1}$ and $\Ep{\party{C}}{2}$ are not ordered), and therefore uses our transformation to automatically obtain the following race-free global protocol:
\begin{subequations}
  \begin{align}
    \transmit{\party{A}}{\Channel}{1}{\party{B}} \ssep \transmit{\party{A}}{\channel{f}{0}}{1.2}{\party{B}} \ssep \guardP{\Ep{\party{A}}{1} \HB{} \Ep{\party{B}}{2}} \ssep \guardP{\Ep{\party{B}}{1} \HB{} \Ep{\party{C}}{2}} \ssep \transmit{\party{B}}{\Channel}{2}{\party{C}} \label{eq:simple global protocol with rendezvous guarded}
  \end{align}
\end{subequations}

Let the race-free global protocol \eqref{eq:simple global protocol with rendezvous guarded} be $\GlobalProtocol$.
The following projected per-party protocols have a redundant rendezvous between $\party{A}$ and $\party{B}$:
\begin{subequations}
  \begin{align} 
    \projectOntoP{\GlobalProtocol}{\party{A}} &= \PartySend{\Channel}{\Ep{\party{A}}{1}} &&\ssep \PartySend{\channel{f}{0}}{\Ep{\party{A}}{1.2}} &&\ssep \guardP{\channel{d}{0}, \Ep{\party{A}}{1}} &&\ssep \PartySend{\channel{d}{0}}{\Ep{\party{A}}{1.3}} &&\ssep \emp && &&\ssep \emp \label{eq:simple global protocol with rendezvous guarded projected-per-party A} \\
    \projectOntoP{\GlobalProtocol}{\party{B}} &= \PartyReceive{\Channel}{\Ep{\party{B}}{1}} &&\ssep \PartyReceive{\channel{f}{0}}{\Ep{\party{B}}{1.2}} &&\ssep \PartyReceive{\channel{d}{0}}{\Ep{\party{B}}{1.3}} &&\ssep \guardP{\channel{d}{0}, \Ep{\party{B}}{2}} &&\ssep \guardP{\channel{e}{0}, \Ep{\party{B}}{1}} &&\ssep \PartySend{\channel{e}{0}}{\Ep{\party{B}}{1.6}} &&\ssep \PartySend{\Channel}{\Ep{\party{B}}{2}} \label{eq:simple global protocol with rendezvous guarded projected-per-party B} \\
    \projectOntoP{\GlobalProtocol}{\party{C}} &= \emp &&\ssep \emp &&\ssep \emp && &&\ssep \PartyReceive{\channel{e}{0}}{\Ep{\party{C}}{1.6}} &&\ssep \guardP{\channel{e}{0}, \Ep{\party{C}}{2}} &&\ssep \PartyReceive{\Channel}{\Ep{\party{C}}{2}} \label{eq:simple global protocol with rendezvous guarded projected-per-party C}
  \end{align}
\end{subequations}

where projection inserts rendezvous on unbuffered channels $\channel{d}{0}$ and $\channel{e}{0}$.

The guard $\guardP{\Ep{\party{A}}{1} \HB{} \Ep{\party{B}}{2}}$, when projected, inserts a redundant rendezvous $\transmit{\party{A}}{\channel{d}{0}}{1.5}{\party{B}}$.
This rendezvous is redundant because the rendezvous $\transmit{\party{A}}{\channel{f}{0}}{1.2}{\party{B}}$ is similar to it in two ways:
\begin{enumerate}
\item The unbuffered channel $\channel{f}{0}$ is unique, in that it satisfies requirement \ref{req:there are no other events on an unbuffered channel}.
\item The rendezvous $\transmit{\party{A}}{\channel{f}{0}}{1.2}{\party{B}}$ is positioned between the two events $\Ep{\party{A}}{1}$ and $\Ep{\party{B}}{2}$, so that it is possible to deduce that $\Ep{\party{A}}{1} \HB{} \Ep{\party{B}}{2}$.
\end{enumerate}

Future work can detect such similar rendezvous, and remove the global protocol guards that insert such redundant rendezvous, making an execution more performant.

\subsection{Channels between Distributed Nodes}
We understand that a send and a receive on a Go channel is not the same as a send and a receive between distributed nodes in a distributed Go program.
An unbuffered channel cannot communicate with other distributed nodes, whereas distributed sends and receives would for example be on Transmission Control Protocol sockets.
Nevertheless, the connections between distributed nodes can be modelled by buffered channels, which our framework analyzes.
Therefore, our work lays the foundation for future work, where messages on these distributed channels can be verified to satisfy certain properties.

\subsection{Other Synchronization Primitives}
Our approach of modelling only channels is similar to \textcite{Liu2021AutomaticallyDA}, where although they mainly model channels, their framework can be extended to other synchronization primitives.
This is because other synchronization primitives, e.g. mutexes and condition variables, can be implemented with channels, and thereby are covered under their model.
It remains to show that our framework can also be extended to mutexes and condition variables.

\textcite{Liu2021AutomaticallyDA} does not say whether other synchronization primitives, e.g. WaitGroups, can also be modelled with channels.
On the other hand, we can use CSL to model other synchronization primitives with their own set of specifications.
For example, we claim Go's WaitGroup is abstractly identical Java's CountDownLatch, which \textcite{Chin2019AutomatedVO} specifies with CSL.

\subsection{Avoid Runtime Type Errors} \label{sec:Avoid Runtime Type Errors}
We construct an example where a channel has a sum type, and a channel race results in a type error if received in the wrong order. However, there are no sum types in Go.

Still, it is possible to achieve the same functionality as sum types with the visitor pattern. The program \ref{lst:program sum type channel} has a channel where all messages on it have a sum type, and where channel \lstinline{f} is used only to prevent \lstinline{main} from exiting before any receives happen.

\begin{code}[caption={A program with communications on a channel with a sum type.},label={lst:program sum type channel},multicols=2]
package main

import "fmt"

type Int int
type Bool bool

type Expr interface {
	Visit(v ExprVisitor)
}

type ExprVisitor struct {
	VisitI func(Int)
	VisitB func(Bool)
}

func (i Int) Visit(v ExprVisitor) {
	v.VisitI(i)
}
func (b Bool) Visit(v ExprVisitor) {
	v.VisitB(b)
}

func main() {
	c := make(chan Expr, 2)
	f := make(chan bool, 0)
	c <- Int(0)
	c <- Bool(true)
	go A()
	go B()
	// Wait for both goroutines.
	<-f
	<-f
}

func A() {
	fmt.Println((<-c).(Int))
	// Tell main goroutine I'm done.
	f <- true
}

func B() {
	fmt.Println((<-c).(Bool))
	// Tell main goroutine I'm done.
	f <- true
}
\end{code}

Here the programmer sequentially orders the receives in the (supposed) send order, to try to avoid a runtime error. However, the programmer does not realize that the sends are not sequentially ordered, but instead concurrently ordered. Go may schedule execution of these concurrent send goroutines in the reverse order: first \lstinline{Bool(true)} is sent, then \lstinline{Int(0)} is sent. Indeed, running the program gives the runtime error \ref{lst:program sum type channel runtime error}.

\begin{code}[caption={A runtime error of a program with communications on a channel with a sum type.},label={lst:program sum type channel runtime error}]
panic: interface conversion: main.Expr is main.Int, not main.Bool

goroutine 7 [running]:
main.main.func2(0xc000052180, 0xc00005e060)
	/tmp/sandbox388283066/prog.go:30 +0x105
created by main.main
	/tmp/sandbox388283066/prog.go:29 +0x11f
panic: interface conversion: main.Expr is main.Bool, not main.Int

goroutine 6 [running]:
main.main.func1(0xc000052180, 0xc00005e060)
	/tmp/sandbox388283066/prog.go:26 +0x10a
created by main.main
	/tmp/sandbox388283066/prog.go:25 +0xf3

Program exited: status 2.
\end{code}

The way to fix this is to prevent Go from scheduling execution of these concurrent send goroutines in the reverse order. The programmer should make sure that the sends can only ever be scheduled in this order: first \lstinline{Int(0)} is sent, then \lstinline{Bool(true)} is sent. The fixed program  \ref{lst:program sum type channel synchronized} does this by synchronizing both sends with an unbuffered channel, \lstinline{d}.

\begin{code}[caption={A program with synchronized communications on a channel with a sum type.},label={lst:program sum type channel synchronized},multicols=2]
package main

import "fmt"

type Int int
type Bool bool

type Expr interface {
	Visit(v ExprVisitor)
}

type ExprVisitor struct {
	VisitI func(Int)
	VisitB func(Bool)
}

func (i Int) Visit(v ExprVisitor) {
	v.VisitI(i)
}
func (b Bool) Visit(v ExprVisitor) {
	v.VisitB(b)
}

func main() {
	c := make(chan Expr, 2)
	f := make(chan bool, 0)
	d := make(chan bool, 0)
	c <- Int(0)
	c <- Bool(true)
	go A()
	go B()
	// Wait for both goroutines.
	<-f
	<-f
}

func A() {
	fmt.Println((<-c).(Int))
	d <- true // rendezvous on channel d
	// Tell main goroutine I'm done.
	f <- true
}

func B() {
	<-d // rendezvous on channel d
	fmt.Println((<-c).(Bool))
	// Tell main goroutine I'm done.
	f <- true
}
\end{code}

This program can never give a runtime type (interface) conversion error.
Indeed, running the program gives output \ref{lst:program sum type channel synchronized output}.

\begin{code}[caption={The output of a program with synchronized communications on a channel with a sum type.},label={lst:program sum type channel synchronized output}]
0
true

Program exited.
\end{code}

\begin{figure}
  {\centering
    \begin{tikzpicture}[]
      \node (p0) at (0,0)  {$p_0$};
      \node (p1) at (-2,0) {$p_1$};
      \node (p2) at (2,0)  {$p_2$};
      \draw[line width=0.1mm] (-2.5,-0.5) -- (2.5,-0.5);

      \dop{p0sd0}{$\sdop{}{c}{0}$}{0}{-1}{right}
      \dop{p0sd1}{$\sdop{}{c}{1}$}{0}{-2}{right}
      \pop{p1z}{$\wrop{z}{}$}{-2}{-4}{left}
      \dop{p1rv}{$\rvop{}{c}{}$}{-2}{-5}{left}
      \dop{p2rv}{$\rvop{}{c}{}$}{2}{-3}{right}
      \pop{p2z}{$\rdop{z}{}$}{2}{-4}{right}

      \draw [->] (p0sd0e) -- (p0sd1s) node [near end,right]  {};
      \draw [->] (p0sd0s) -- (p1rve)  node [near end,right]  {};

      \draw [->] (p1z)   -- (p1rvs)  node [near end,left]  {};

      \draw [->] (p2rve)  -- (p2z)   node [near end,right] {};
      \draw [->] (p0sd1s) -- (p2rve)  node [near end,right]  {};

      \draw [->,dotted] (p2rve) -- (p1rvs)  node [near end,left]  {};
    \end{tikzpicture}
    \caption{\label{fig:conditional-race partial order}The unconsidered sequential execution in an example from \textcite{fava2020ready}, but with the program $\texttt{if r = 1 then load z}$ replaced with $\texttt{load z}$.}
  }
\end{figure}

\newpage
\singlespacing
\section{Appendix}

\subsection{Soundness or unsoundness of \eqref{eq:HB-CB} and \eqref{eq:CB-HB}}
The semantic conditions \eqref{eq:aft-recvs are aft-sends} and \eqref{eq:bef-sends are before-recvs} are derived respectively from \citeauthor{costea2017session}'s \eqref{eq:CB-HB} and \eqref{eq:HB-CB} propogation lemmas \parencite{costea2017session}, where communication occurs on a buffered channel $\channel{c}{\anything}$ instead of an unbuffered channel.
The propogation lemmas are restated in figure \ref{fig:costeas propogation lemmas}.

\begin{figure}
  {\centering
    \begin{subequations}
      \begin{align}
        \E{S} \CB{\channel{c}{\anything}} \E{R} \land \E{R} \HB{} \E{} &\implies \E{S} \HB{} \E{} \label{eq:CB-HB} \tag{CB-HB} \\
        \E{} \HB{} \E{S} \land \E{S} \CB{\channel{c}{\anything}} \E{R} &\implies \E{} \HB{} \E{R} \label{eq:HB-CB} \tag{HB-CB}
      \end{align}
    \end{subequations}
    \caption{\label{fig:costeas propogation lemmas}\citeauthor{costea2017session}'s propogation lemmas.}
  }
\end{figure}

A lemma is unsound if the lemma can be used to prove that a global protocol with races is race-free.
A lemma is sound if the lemma \textbf{cannot} be used to prove that a global protocol with races is race-free.

We want to know whether \citeauthor{costea2017session}'s propogation lemmas \ref{fig:costeas propogation lemmas} are sound or unsound.
We are interested because if a lemma is sound, then we can apply it on a partial order, resulting in a partial order with more elements.
A partial order with more elements has fewer sequential executions.
Intuitively, our verification can rule out more run-time sequential executions, and take less time to analyze.

\subsubsection{Unsoundness of \eqref{eq:HB-CB}}
We prove the unsoundness of propogation lemma \eqref{eq:HB-CB}.
\citeauthor{costea2017session} previously attempted this \parencite[Lemma 1]{costea2017session}, but \citeauthor{costea2017session}'s proof is incomplete under our definition of unsoundness.

\begin{lemma}
  The lemma \eqref{eq:HB-CB} is unsound.
\end{lemma}
\begin{proof}
  The global protocol \eqref{eq:simple global protocol} has races, because there are two sequential executions \ref{fig:simple global protocol sequential execution race-free} and \ref{fig:simple global protocol sequential execution with races}.
  Applying \eqref{eq:HB-CB} to \eqref{eq:simple global protocol} gives us a new element $\Ep{\party{B}}{1} \HB{} \Ep{\party{C}}{2}$, so that the two receives are ordered.
  To show that the global protocol is race-free, by \eqref{eq:CB is isotone} it remains to show that the two sends, $\Ep{\party{A}}{1}$ and $\Ep{\party{B}}{2}$, are ordered in the same way.
  It can be determined that the only two possible sequential executions of the global protocol are \ref{fig:simple global protocol sequential execution race-free} and \ref{fig:simple global protocol sequential execution with races}.
  In all sequential executions, $\Ep{\party{A}}{1} \HB{} \Ep{\party{B}}{2}$.
  If something is true in all sequential executions, then it must also be true for the global protocol.
  Therefore, $\Ep{\party{A}}{1} \HB{} \Ep{\party{B}}{2}$.
  Therefore, we have used \eqref{eq:HB-CB} to prove that the global protocol is race-free.
  Therefore, \eqref{eq:HB-CB} is unsound.
\end{proof}

\subsubsection{Intractability of proving the soundness of \eqref{eq:CB-HB}}
We suspect that the lemma \eqref{eq:CB-HB} is sound.
We considered a brute-force proof that firstly enumerates all global protocols.
However, we found out that the number of global protocols is larger than the number of sequentially composed global protocols, which follows the exponentially growing integer sequence A020556 \parencite{A020556} (as we will see shortly).
Therefore, the brute-force proof is intractable.

We are not aware of any other ways to prove that the lemma \eqref{eq:CB-HB} is sound.
We are also not aware of any proofs that the lemma \eqref{eq:CB-HB} is unsound.

Now, we count the number of sequentially composed global protocols with $n$ transmissions.
This number is lower than the number of global protocols with $n$ transmissions, because each sequential composition operator $\ssep$ may be replaced with the concurrent composition operator $\asep$.

For a given sequentially composed global protocol with any number of transmissions that already has $n$ old parties, the number of new sequentially composed global protocols formed by adding one transmission with two parties is shown in table \ref{tab:number of new sequentially composed global protocols formed by adding one transmission with two parties}, case splitting on whether each party already exists in previous transmissions (old) or not (new).

\begin{table}
  {\centering
    \begin{tabular}{|c|c|c|}
      \hline
      \textbf{Sender} &\textbf{Receiver} &\textbf{Number of new sequentially composed global protocols}\\
      \hline\hline

      old &old &$n \cdot (n-1)$\\
      \hline
      old &new &$n \cdot 1$\\
      \hline
      new &old &$1 \cdot n$\\
      \hline
      new &new &$1 \cdot 1$\\
      \hline
    \end{tabular}
    \caption{\label{tab:number of new sequentially composed global protocols formed by adding one transmission with two parties}Number of new sequentially composed global protocols formed by adding one transmission with two parties.}
  }
\end{table}

The number of sequentially composed global protocols with $n$ transmissions is the sum of the number of sequentially composed global protocols with $n$ transmissions and varying numbers of assigned parties.
The number of sequentially composed global protocols with $n$ transmissions and varying numbers of assigned parties is shown in table \ref{tab:number of sequentially composed global protocols with n transmissions}, where $(x, y, z, \ldots)$ is the number of sequentially composed global protocols with $(2, 3, 4, \ldots)$ assigned parties.
Multiplication distributes over tuples e.g. $2 \cdot (x, y, z) = (2 \cdot x, 2 \cdot y, 2 \cdot z)$, and tuple addition is defined pointwise e.g. $(u, v, w) + (x, y, z) = (u + x, v + y, w + z)$.

\begin{table}
  {\centering
    \begin{tabular}{|c|c|c|}
      \hline
      \textbf{n} &$\mathbf{(x, y, z, \ldots)}$ &\textbf{\thead{Number of sequentially composed global protocols\\with $n$ transmissions}}\\
      \hline\hline

      1
        &$\begin{aligned}
          (1,)
          \end{aligned}
          $
          &$1$\\
      \hline

      2
        &$\begin{aligned}
          & &&(2 \cdot 1, 2 \cdot 1 + 1 \cdot 2, 1 \cdot 1)\\
          &= &&(2, 4, 1)
          \end{aligned}
          $
          &$7$\\
      \hline

      3
        &$\begin{aligned}
          & && &&&2 \cdot (2 \cdot 1, 2 \cdot 1 + 1 \cdot 2, 1 \cdot 1)\\
          & &&+ &&&(0, 4 \cdot (3 \cdot 2, 3 \cdot 1+1 \cdot 3, 1 \cdot 1))\\
          & &&+ &&&(0, 0, 1 \cdot (4 \cdot 3, 4 \cdot 1+1 \cdot 4, 1 \cdot 1))\\
          &= && &&&(4, 32, 38, 12, 1)
          \end{aligned}
          $
          &$87$\\
      \hline

      4
        &$\begin{aligned}
          & && &&&4 \cdot (2 \cdot 1, 2 \cdot 1 + 1 \cdot 2, 1 \cdot 1)\\
          & &&+ &&&(0, 32 \cdot (3 \cdot 2, 3 \cdot 1+1 \cdot 3, 1 \cdot 1))\\
          & &&+ &&&(0, 0, 38 \cdot (4 \cdot 3, 4 \cdot 1+1 \cdot 4, 1 \cdot 1))\\
          & &&+ &&&(0, 0, 0, 12 \cdot (5 \cdot 4, 5 \cdot 1+1 \cdot 5, 1 \cdot 1))\\
          & &&+ &&&(0, 0, 0, 0, 1 \cdot (6 \cdot 5, 6 \cdot 1+1 \cdot 6, 1 \cdot 1))\\
          &= && &&&(8, 16, 4)\\
          & &&+ &&&(0, 192, 192, 32)\\
          & &&+ &&&(0, 0, 456, 304, 38)\\
          & &&+ &&&(0, 0, 0, 480, 120, 1)\\
          & &&+ &&&(0, 0, 0, 0, 30, 12, 1)\\
          &= && &&&(8, 208, 652, 576, 188, 24, 1)
          \end{aligned}
          $
          &$1657$\\
      \hline

    \end{tabular}
    \caption{\label{tab:number of sequentially composed global protocols with n transmissions}Number of sequentially composed global protocols with $n$ transmissions.}
  }
\end{table}

This is the integer sequence A020556 \parencite{A020556}, which is the number of oriented multigraphs on n labeled arcs (without loops) \parencite[Table 2 (Number of $(R;\Lambda)$-multigraphs with unlabelled nonisolated vertices and $k$ labelled links, Case 4: arcs no loops $(\lambda=0; \alpha=1; \epsilon=0)$, Ordinary multigraphs)]{LABELLE2000237}.

Indeed, sequentially composed global protocols of only sends or receives are multigraphs: each transmission is between unlabelled vertexes, and the edge is labelled, as in figure \ref{fig:multigraph of two-transmission sequentially composed global protocols}.

\begin{figure}
  {\centering
    \begin{tikzpicture}[->]
      \pop{A}{}{-2}{2}{};
      \pop{B}{}{2}{2}{};
      \pop{C}{}{-2}{-2}{};
      \pop{D}{}{2}{-2}{};

      \path
      (A)
      edge [black,bend left] node [above] {$\party{A} \text{ to } \party{B}$} (B)
      edge [purple] node [above] {$\party{A} \text{ to } \party{B}$} (B)
      edge [purple] node [right] {$\party{A} \text{ to } \party{C}$} (C)
      (B)
      edge [purple,bend left] node [above] {$\party{B} \text{ to } \party{A}$} (A)
      edge [purple,bend left] node [right,pos=0.25] {$\party{B} \text{ to } \party{C}$} (C)
      (C)
      edge [purple] node [left,pos=0.75] {$\party{C} \text{ to } \party{B}$} (B)
      edge [purple,bend left] node [left] {$\party{C} \text{ to } \party{A}$} (A)
      edge [purple] node [below] {$\party{C} \text{ to } \party{D}$} (D);
    \end{tikzpicture}
    \caption{\label{fig:multigraph of two-transmission sequentially composed global protocols}The one-transmission sequentially composed global protocol consists of only one arc from one of the vertexes to another vertex. For example, the black arc is a one-transmission sequentially composed global protocol. A two-transmission sequentially composed global protocol is a multigraph with two arcs between some number of vertexes, and such a sequentially composed global protocol consists of the black arc and one of the purple arcs. There are 7 possible two-transmission sequentially composed global protocols.}
  }
\end{figure}

The sequentially composed global protocol with one transmission consists of only the black arc from some vertex to some other vertex.
Then the sequentially composed global protocol with two transmissions is the multigraph with two arcs between some number of vertexes, and this sequentially composed global protocol consists of the black arc and some purple arc.
There are 7 possible two-transmission sequentially composed global protocols.

Normal graphs are not enough to represent sequentially composed global protocols.
We need multigraphs because a sequentially composed global protocol can have multiple edges between the same two nodes.

\subsection{Pitfalls to Avoid}

\subsubsection{Transitive Closure}
$\HB{}$ is transitive. Yet, we avoid taking the transitive closure, because this can lead to confusion about which send a receive receives data from.
For example, figure \ref{fig:simple global protocol race-free sequential execution transitive closure} shows that taking the transitive closure of the sequential execution \ref{fig:simple global protocol race-free sequential execution} creates the new element $\Ep{\party{A}}{1} \HB{\Channel} \Ep{\party{C}}{2}$. Now it is not immediately clear which send $\Ep{\party{C}}{2}$ corresponds to; is it $\Ep{\party{A}}{1}$, or $\Ep{\party{B}}{2}$?
This is a pitfall of representing sequential executions with the $\HB{}$ order.

Nevertheless, this pitfall is resolved by representing global protocols with session logic instead of with partial orders.
Session logic ensures that every receive corresponds to exactly one send.

\begin{figure}
  {\centering
    \begin{tikzpicture}[]
      \node(logic) at (0,1)  {$\transmit{\party{A}}{\Channel}{1}{\party{B}} \ssep \transmit{\party{B}}{\Channel}{2}{\party{C}}$};
      \node(A) at (-2,0)  {$\party{A}$};
      \node(B) at (0,0)  {$\party{B}$};
      \node(C) at (2,0)  {$\party{C}$};
      \draw[line width=0.1mm] (-2.5,-0.5) -- (2.5,-0.5);

      \pop{AS1}{$\Ep{\party{A}}{1}$}{-2}{-1}{left}
      \pop{BR1}{$\Ep{\party{B}}{1}$}{0}{-2}{right}
      \pop{BS1}{$\Ep{\party{B}}{2}$}{0}{-3}{left}
      \pop{CR1}{$\Ep{\party{C}}{2}$}{2}{-4}{right}

      \draw [->] (AS1) -- (BS1) node [] {};
      \draw [->] (BR1) -- (CR1) node [] {};

      \draw [->] (BR1) -- (BS1) node [] {};

      \draw [->,blue] (AS1) -- (BR1) node [near start,fill=white,inner sep=1pt] {$\Channel$};
      \draw [->,blue] (AS1) -- (CR1) node [near start,fill=white,inner sep=1pt] {$\Channel$};

      \draw [->,blue] (BS1) -- (CR1) node [near start,fill=white,inner sep=1pt] {$\Channel$};
    \end{tikzpicture}
    \caption{\label{fig:simple global protocol race-free sequential execution transitive closure}The transitive closure of sequential execution \ref{fig:simple global protocol race-free sequential execution}.}
  }
\end{figure}

\newpage
\bibliographystyle{apalike}
\bibliography{main}